\documentclass[fleqn,usenatbib]{mnras}

\usepackage{newtxtext,newtxmath}

\usepackage[T1]{fontenc}

\DeclareRobustCommand{\VAN}[3]{#2}
\let\VANthebibliography\thebibliography
\def\thebibliography{\DeclareRobustCommand{\VAN}[3]{##3}\VANthebibliography}

\usepackage{graphicx}	
\usepackage{subcaption} 
\usepackage{amsmath}	
\usepackage{acronym}    
\usepackage{tensor}     
\usepackage{braket}     

\usepackage{tikz}       
\usetikzlibrary{intersections}  
\usetikzlibrary{arrows.meta}    
\usetikzlibrary{shapes.geometric, arrows}
\tikzstyle{myNode} = [rectangle, rounded corners, minimum width=3.3cm, minimum height=2cm, text centered, draw=black, fill=black!10]
\tikzstyle{arrow} = [thick,->,>=stealth]

\newcommand{\dir}{{\mathrm{d}}}
\newcommand{\Ndir}{{N_{\text{dir}}}}

\newcommand{\eg}{e.g.,~}
\newcommand{\ie}{i.e.,~}
\usepackage{hyperref}
\hypersetup{
    colorlinks=true,
    linkcolor=blue,
    citecolor=blue,
    filecolor=magenta,
    urlcolor=cyan
    }
\usepackage{times}

\title[General-Relativistic LBM for Radiation
  Transport]{General-Relativistic Lattice-Boltzmann Method for Radiation Transport}

\author[Olsen and Rezzolla]{
Tom Olsen,$^{1,2}$
and Luciano Rezzolla,$^{2,1,3}$
\\
$^{1}$Frankfurt Institute for Advanced Studies, Ruth-Moufang-Str. 1,
D-60438 Frankfurt, Germany\\
$^{2}$Institut f\"ur Theoretische Physik, Goethe Universit\"at,
  Max-von-Laue-Str. 1, 60438 Frankfurt am Main, Germany\\
$^{3}$School of Mathematics, Trinity College, Dublin 2, Ireland}

\date{Accepted XXX. Received YYY; in original form ZZZ}

\pubyear{2024}

\begin{document}
\label{firstpage}
\pagerange{\pageref{firstpage}--\pageref{lastpage}}
\maketitle


\begin{abstract}
We present the first extension of the special-relativistic
Lattice-Boltzmann Method for radiative transport developed
by~\citet{Weih2020c}, to solve the radiative-transfer equation in curved
spacetimes. The novel approach is based on the streaming of carefully
selected photons along null geodesics and interpolating their final
positions, velocities, and frequency shifts to all photons in a given
velocity stencil. Furthermore, by transforming between the laboratory
frame, the Eulerian frame, and the fluid frame, we are able to perform
the collision step in the fluid frame, thus retaining the collision
operator of the special-relativistic case with only minor
modifications. As a result, with the new method we can model the
evolution of the frequency-independent (``grey'') radiation field as it
interacts with a background fluid via absorption, emission, and
scattering in a curved background spacetime. Finally, by introducing a
refined adaptive stencil, which is suitably distorted in the direction of
propagation of the photon bundle, we can reduce the computational costs
of the method while improving its performance in the optically-thin
regime. A number of standard and novel tests are presented to validate
the approach and exhibit its robustness and accuracy.
\end{abstract}


\section{Introduction}
\label{sec: Introduction}

In essentially all observations of astronomical sources, the radiation
that we receive is the result of a delicate and sometimes subtle
interaction between the radiation field and the matter field that emits
and absorbs it as it propagates. It is, therefore, of great importance
that this interaction, which is mathematically described by the
radiative-transfer equation (RTE). Given the nonlinear regimes normally
encountered in astrophysical scenarios and the complexity of the
radiative-transfer equation, the use of advanced numerical methods to
tackle the radiative-transfer problem becomes inevitable. Such methods
need to be combined with equally advanced approaches necessary to account
for the dynamics of plasmas often encountered in astrophysics and modeled
with the equations of magnetohydrodynamics (MHD). A perfect, but surely
not unique example is the modelling of astrophysical compact objects as
those involved in short gamma-ray bursts~\citep{Rezzolla:2011,
  Palenzuela2013a, Kiuchi2015}, core-collapse
supernovae~\citep{Mezzacappa01, Just2015b, Oconnor2015, Kuroda2016}, or
in the merger of binary systems of neutron stars (BNS). In all of these
scenarios, radiation fields composed of either photons or neutrinos, play
a fundamental role in shaping the dynamics of the compact objects and, of
course, in determining the astronomical observables~\citep[see,
  \eg][]{Rosswog2014a, Siegel2016a, Bovard2017, Dietrich2016, Perego2017,
  Siegel2017, Fujibayashi2018, Fernandez2018, Espino2024c}.

Several approaches are available in the literature for the inclusion of
the effects of neutrinos in general-relativistic hydrodynamical or
magnetohydrodynamical (GRMHD) simulations of BNS mergers. These range
from very simple and computationally efficient ``leakage-type''
schemes~\citep{Ruffert97, Galeazzi2013, Most2019b}, where the local
heating/cooling rates are directly estimated from the reaction
cross-sections corrected with a diffusion prescription, over to the
so-called ``moment schemes'', where a varying number of moments of the
Boltzmann equation for neutrinos is solved~\citep{Rezzolla1994,
  Foucart2015a, Just2015b, Kuroda2016, Skinner2019, MelonFuksman2019,
  Weih2020b, Radice2022, Sun:2022vri, Izquierdo2022}. The most advanced
approaches even consider the direct solution of the radiative transfer
equation via MonteCarlo or other methods~\citep{Radice2013, Foucart2020c,
  Roth2022, Izquierdo2024b}.

Among these different approaches, there is one that is closely related to
the content of our work, the Lattice-Boltzmann method (LBM) for radiative
transport recently developed by~\cite{Weih2020c} within a
special-relativistic context (SRLBM hereafter) and employed in
\texttt{BHAC}~\citep{Porth2017, Olivares2019}. The appealing aspects of
this approach are its low computation cost and its high adaptability to
optically intermediate and thick regimes. While it does not have issues
with beam crossing, like the M1 scheme~\citep{Weih2020b,
  Musolino2023, Izquierdo2024b}, its performance is less accurate in
optically thin regimes. More importantly, the method was developed for
flat spacetime and, therefore, is not applicable in some of the most
interesting scenarios described by {GRMHD} simulations of astrophysical
compact objects.

Given these prospects and limitations, we here present the first
implementation of the LBM for the solution of the general-relativistic
RTE in curved background spacetimes. The core of our approach is based on
the attempt to retain as much as possible of the logic of the SRLBM while
adapting to the more complex background geometry. Since in {LBM} schemes
``external forces'' are either treated as additional extra terms in the
collision operator or by altering the streaming step, we follow the
latter approach, as it is far more natural in a general-relativistic
context. In particular, during the streaming step, we solve only certain
null geodesics and interpolate the final photon positions, velocities,
and frequencies to the photons in our {LBM} velocity stencil with the
help of Fourier-transformations and spherical harmonics. In the collision
step, on the other hand, we retain the collision operator in its
special-relativistic form by transforming between multiple frames and
carefully adapting the definition of the discretized intensities.

Our paper is structured as follows: Sec. \ref{sec: Theoretical Basics}
presents the basic mathematical aspects of the LBM, the $3+1$ split of
spacetime, and the geodesic equation in $3+1$ form we employ. In
Sec. \ref{sec: SRLBMRT}, we review the recap the {SRLBM} as this will be
useful in Sec. \ref{sec:GRLBMRT}, where we illustrate the details of the
{GRLBM}. In Sec. \ref{sec: Numerics}, we discuss the numerical methods we
implemented to employ the GRLBM, while in Sec. \ref{sec: Implementation
  Tests} we present a long series of standard and novel tests to validate
the robustness and accuracy of our approach. Finally, conclusions and
future prospects of the LBM in general are presented in Sec. \ref{sec:
  Conclusion and Outlook}. Hereafter, we will adopt the $(-,+,+,+)$
signature and assume Greek indices to run from $0$ to $4$, and Latin
indices from $1$ to $3$.

\section{Mathematical setup}
\label{sec: Theoretical Basics}

\subsection{The Lattice-Boltzmann Method}
\label{subsec: LBM}

We recall that the LBM represent a numerical approach to model the
dynamics of fluids on a mesoscopic scale rather than on a macroscopic
one~\citet{Higuera1989, succi2001}. Thus, the starting point is the
classical Boltzmann equation~\citep[see, \eg][]{Rezzolla_book:2013}
\begin{align}
    \label{eq: Boltzmann eq}
    \frac{df(\vec x,\vec u, t)}{dt} = \partial_t f +
    \vec{u} \cdot \vec{\nabla}_{\vec x}f + \frac{\vec F}{\rho}
    \cdot \vec{\nabla}_{\vec u} f = \Gamma[f]\,,
\end{align}
where $\vec x$ and $\vec u$ are the position and velocity of a fluid
particle at time $t$, respectively, $\vec F$ is the force acting on the
fluid, and $\rho$ is the fluid density. The collision operator
$\Gamma[f]$ accounts for the interactions between the particles and is
responsible for the relaxation of the distribution function $f$ to the
local equilibrium distribution function $f^{\text{eq}}$.

Solving this set of partial differential equations is complicated by the
intrinsic seven-dimensionality and so it is convenient to retain the
total differential on the left-hand side and to integrate it directly in
time over an interval $\Delta t = t_{n+1} - t_n$
\begin{align}
  \label{eq:LB_int}
  \int_{t_n}^{t_{n+1}}\frac{df(\vec x,\vec u, t)}{dt} dt & =
    \int_{t_n}^{t_{n+1}}\Gamma[f] dt\,.
\end{align}
The left-hand side of Eq.~\eqref{eq:LB_int} can be integrated exactly,
while the right-hand side is approximated with a numerical integral,
assuming the collision operator $\Gamma[f]$ is known, \ie
\begin{align}
    \label{eq: integrated Boltzmann eq}
    f(\vec x_{n+1}, t_{n+1}) & \approx f(\vec x_n, t_n) + \Gamma[f]
    \,\Delta t \,,
\end{align}
where we use the compact notation $\vec x_{n+1} := \vec x(t_{n+1}) = \vec
x(t_n) + \vec u \, \Delta t$.

In its most general form, the collision operator $\Gamma[f]$ is a complex
multi-dimensional integral that cannot be solved analytically and poses
challenges even for those approaches that attempt to solve it
numerically. The Bhatnagar-Gross-Krook (BGK) collision operator
represents the simplest approximation that guarantees the conservation of
mass and momentum and is given by~\citep{Bhatnagar1954}
\begin{align}
    \label{eq:BGK operator}
    \Gamma[f] \approx -\frac{f - f^{\text{eq}}}{\tau}\,,
\end{align}
where $\tau$ is the ``relaxation time''. In particular, it forces the
particle distribution function $f$ to decay to the local equilibrium
$f^{\text{eq}}$ at an exponential rate of $e^{-t/\tau}$; for a classical
non-relativistic fluid, $f^{\text{eq}}$ is given by the Maxwell
equilibrium distribution 
\begin{align}
    \label{eq: Maxwell Equilibrium}
    f^{\text{eq}}(\rho,\vec v, \kappa,\vec u) & =
    \frac{\rho}{(2\pi\kappa)^{D/2}}\exp\left[-\frac{(\vec u-\vec
        v)^2}{2\kappa}\right]\,, 
\end{align}
where $T, m$ are the local temperature and the mass of the one-component fluid,
$\vec v$ the local fluid velocity (first moment), and $\kappa := {k_B T}/{m}$.

Within the Chapman-Enskog analysis~\citep{Chapman1970}, it is
possible to show that the BGK collision operator is sufficient to restore
the macroscopic properties of the fluid and, in particular, its
dissipative properties, over a timescale that is related to the shear
viscosity $\nu = c_s^2(\tau-\Delta t/2)$, where $c_s$ is the fluid's
sound speed.

The key point of the LBM is the discretization of the underlying
seven-dimensional phase space. More specifically, the $\Ndir$ discrete velocities
$\vec{u}_\dir$ are chosen such that the macroscopic first two moments can
be computed exactly with a Hermite-Gauss quadrature
\begin{align}
    \rho(\vec x, t) & = \int_{\mathbb{R}^D} f(\vec x,\vec u, t) d\vec u =
    \sum_{\dir=0}^{\Ndir-1} f_\dir(\vec x, t)\,, \\
    \vec v(\vec x,t) & =
    \frac{1}{\rho(\vec x, t)}\int_{\mathbb{R}^D} f(\vec x,\vec u, t)\vec
    u(\vec x,t) d \vec u\nonumber \\
    & = \frac{1}{\rho(\vec x,
      t)}\sum_{\dir=0}^{\Ndir-1} f_\dir(\vec x, t)\vec u_\dir(\vec x,t)\,.
\end{align}
This leads to a discretization of the distribution function
\begin{equation}
f_\dir(\vec x, t) = \frac{w_\dir}{\omega(\vec{u}_\dir)}f(\vec{x},\vec{u}_\dir,t)\,,
\end{equation}
with weight function
\begin{equation}
\omega(\vec{u}_\dir) = \frac{1}{(2\pi)^{D/2}} e^{-\vec{u}_\dir^2/2}\,,
\end{equation}
and quadrature weights $w_\dir$. Each population $f_\dir$ accounts for
the particles moving in the direction $\vec u_\dir$, and therefore, can
be interpreted as a pseudo-particle.

Several different velocity stencils $\{\vec u_\dir, w_\dir\}$, with $\dir
=\{0,...,\Ndir-1\}$, can be employed to guarantee the exact
reconstruction of the macroscopic moments. Common choices are velocity
stencils such that the velocities point from any source point to all
neighbouring points on a Cartesian grid. In this way, no interpolation is
needed and the LBM becomes a mass and momentum conservative scheme. The
most commonly encountered stencils in two- (2D) and three-dimensions (3D)
are the D2Q9 and D3Q27 stencils that we report in Fig.~\ref{fig: D2Q9 and
  D3Q27}.
\begin{figure}
    \centering
    \includegraphics[width=0.45\columnwidth]{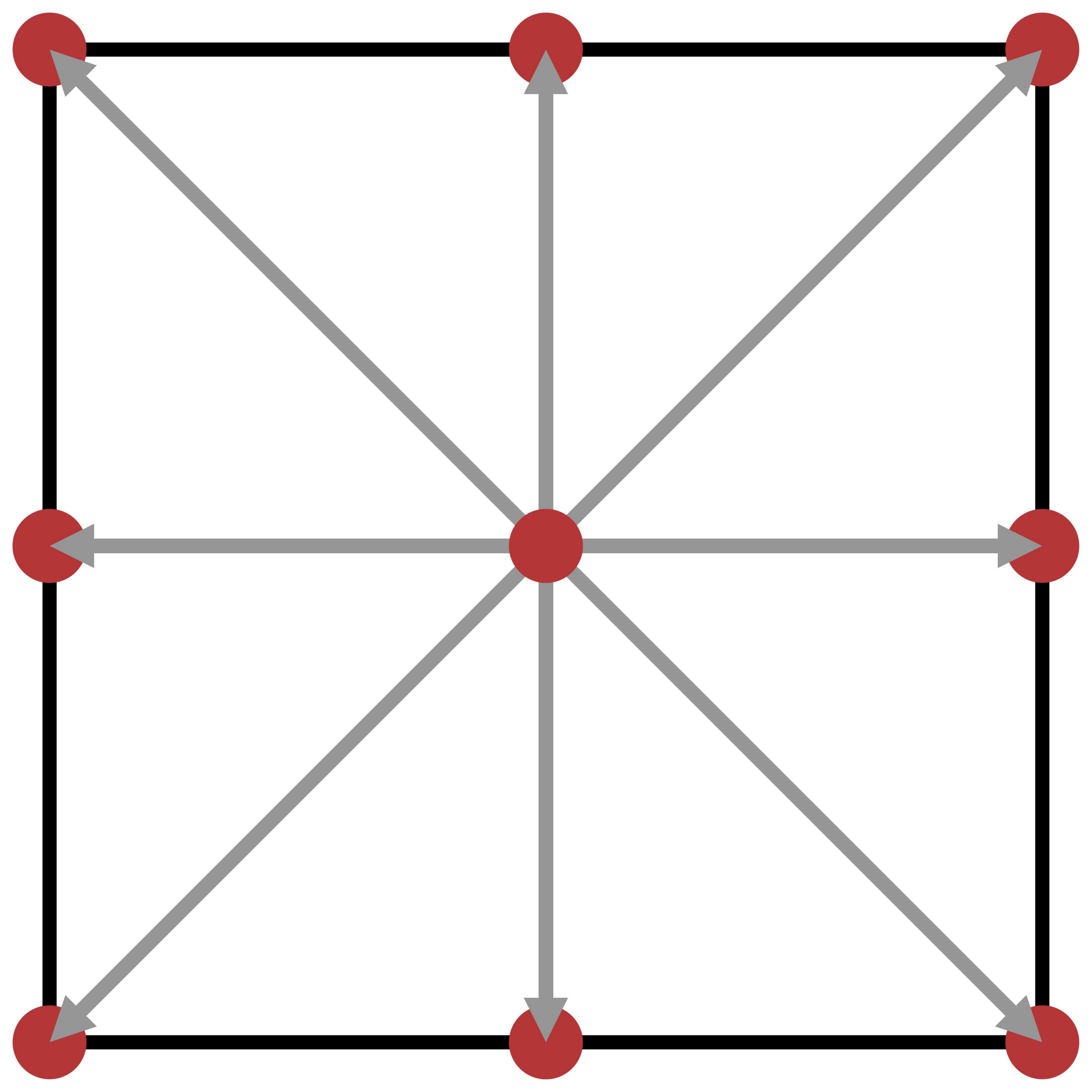}
    \hfill
    \includegraphics[width=0.5\columnwidth]{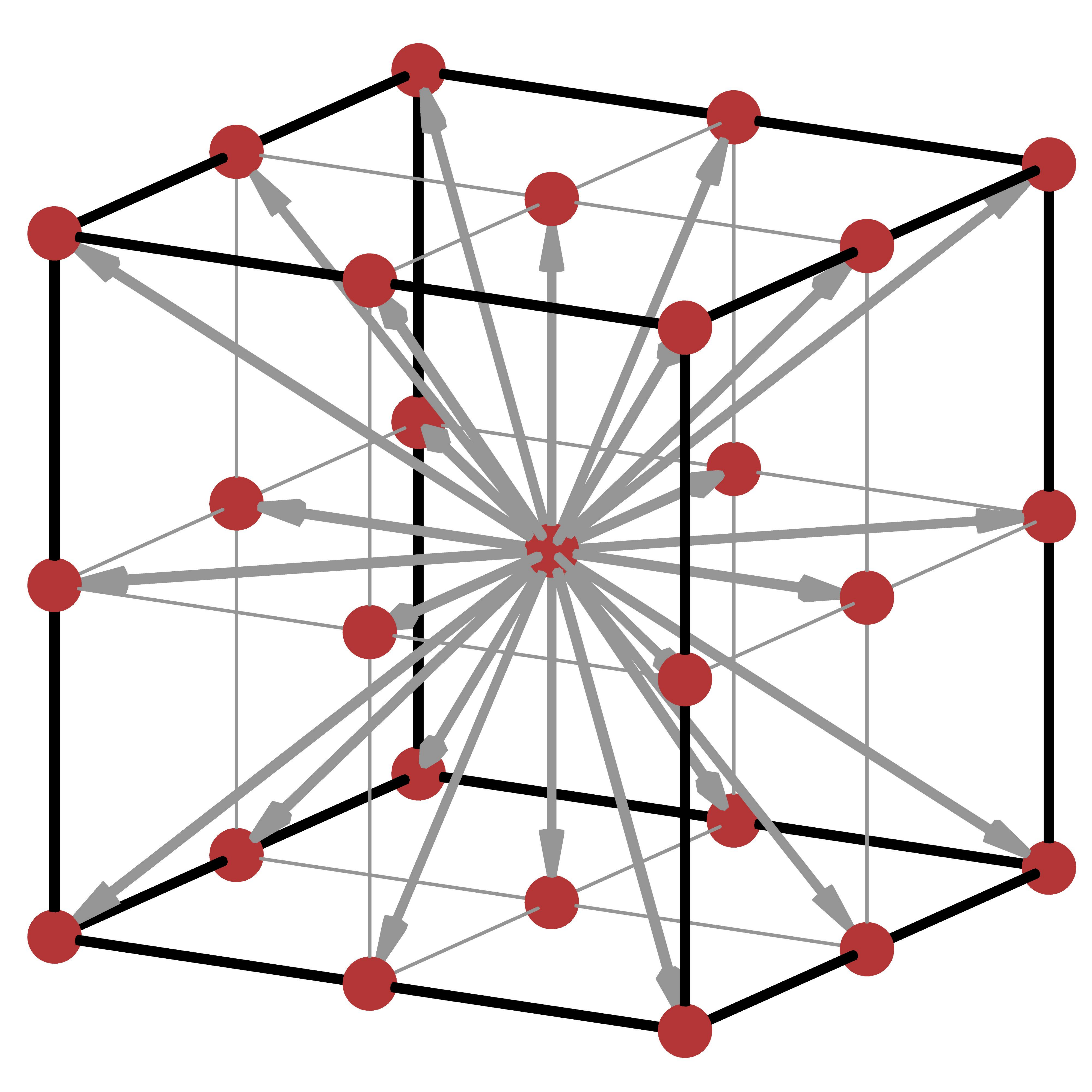}
    \caption{Discretisation stencils in 2D (left panel) and 3D (right
      panel). Red dots represent grid-points of the spatial
      discretization, while the grey arrows indicate the directions of
      the stencil. Besides the visible directions, the zero vector,
      pointing at the center, is also included in both velocity
      stencils.}
      \label{fig: D2Q9 and D3Q27}
\end{figure}
The equilibrium distribution function is discretized similarly to the
particle distribution function in a Hermite polynomial expansion,
\begin{align}
    f_\dir^{\rm eq}(\rho,\vec{u}_\dir,\vec v) = w_\dir\rho\left[ 1 +
      \frac{\vec{u}_\dir \cdot \vec v}{c_s^2} + \frac{(\vec{u}_\dir \cdot
        \vec v)^2}{2c_s^4} - \frac{\vec v^2}{2c_s^2}\right]\,,
\end{align}
where a second-order expansion is sufficient since we need to integrate
only the first three moments of the distribution (from zero to two).

The discretized form of Eq.~\eqref{eq: integrated Boltzmann eq} is given by
\begin{align}
    f_\dir(\vec x_{n+1}, t_{n+1}) & = f_\dir(\vec x_n, t_n) +
    \Gamma_\dir(\vec x_n, t_n)\, \Delta t
    \,,
\end{align}
and can then be split into a two-step procedure
\begin{align}
\label{eq: classical LBM collide}
    \textit{Collision:} &&
    f^\star_\dir(\vec x_n, t_n) &= f_\dir(\vec x_n, t_n) +
    \Gamma_\dir(\vec x_n, t_n)\, \Delta t\,,\\
\label{eq: classical LBM stream}
    \textit{Streaming:} &&
    f_\dir(\vec x_{n+1}, t_{n+1}) &=
    f^\star_\dir(\vec x_n, t_n)\,.
\end{align}
The first step is the so-called ``collision step'' and accounts for the
relaxation of the distribution function to the local equilibrium; it is
purely local, meaning it only depends on quantities at a single lattice point.
The second step is the so-called ``streaming step'', which simply moves populations
of particles from one lattice point to its neighbours, according to the
chosen velocity stencil. While the classical BGK collision operator
conserves the moments, more complicated operators,
\eg those accounting for external forces, may
not. Due to the nature of the collision process, the post-collision
intensities are closer to the equilibrium state and produce more accurate
moment integrals. Therefore, when using the moments for other operations,
like coupling to other codes, it is important
to use the moments computed directly after the collision step.

In summary, the lattice-Boltzmann method amounts to computing
Eq. \eqref{eq: classical LBM collide} for the collision step and
Eq. \eqref{eq: classical LBM stream} for the streaming step.  When
combined, these represent a system of $\Ndir$ coupled ODEs, where the
coupling is hidden in the moments $\rho$ and $\vec v$~\citep[see
  also][]{succi2001,Kruger2017} for additional details.

\subsection{3+1 Split}
\label{subsec: 3+1 Split}

A convenient way to handle the four-dimensional nature of spacetime as a
manifold in general relativity is to decompose it into timelike and
spacelike components by foliating it in terms of non-intersecting
spacelike hypersurfaces $\Sigma_t$ of constant coordinate time $x^0 =
t$~\citep[see, \eg][for additional details]{MTW1973,
  Rezzolla_book:2013}. As customary, we define the ``Eulerian'' observer
to be moving along a worldline orthogonal to $\Sigma_t$ and thus
tangent to the null-normalized local normal vector of $\Sigma_t$
\begin{align}
  n_\mu = -\alpha\nabla_\mu t = \left(-\alpha,\vec{0}\right)\,,
  \qquad n^\mu = \frac{1}{\alpha}\left(1,-\beta^k \right)\,,
\end{align}
where the lapse function $\alpha$ can be seen as the time dilation of the
Eulerian observer relative to a static observer at infinity, while the
shift vector $\beta^\mu$ is purely spatial and describes the coordinate
changes intrinsic to the curvature of the spacetime. The metric
associated with each hypersurface $\Sigma_t$ is given by the spatial
metric,
\begin{align}
    \gamma_{\mu\nu} = g_{\mu\nu} + n_\mu n_\nu,\qquad \gamma^{\mu\nu} =
    g^{\mu\nu} + n^\mu n^\nu\,.
\end{align}
The unit normal $n^\mu$ to a spacelike hypersurface $\Sigma_t$ does not
represent the direction along which the evolution needs to be carried out
to obtain coordinate synchronous events on a new spacelike hypersurface.
This is ensured by using the rescaled timelike vector
\begin{align}
    t^\mu = e_t^\mu = \alpha n^\mu + \beta^\mu\,.
\end{align}

Within this $3+1$ split of spacetime, the four-metric can be expressed generally as 
\begin{align}
    g_{\mu\nu} &= \left(\begin{array}{cc}
        -\alpha^2 + \beta_k\beta^k & \beta_j \\
        \beta_i & \gamma_{ij}
    \end{array}\right)\,,\\
    g^{\mu\nu} &= \left(\begin{array}{cc}
        -1/\alpha^2 & \beta^j/\alpha^2 \\
        \beta^i/\alpha^2 & \gamma^{ij}-\beta^i\beta^j/\alpha^2
    \end{array}\right)\,,
\end{align}
while an explicit expression for the extrinsic curvature
$K_{ij}$ in terms of the three-metric is given by
\begin{align}
\label{eq: extrinsic curvature}
    K_{ij} &= \frac{1}{2\alpha}\left(2\gamma_{ik}\partial_j\beta^k +
    \partial_k\gamma_{ij}\beta^k - \partial_t\gamma_{ij}\right)\,.
\end{align}
Additional relations that are useful when implementing the GRLBM are
given by the derivatives of the four-metric and its components, and can
be summarised as follows
\begin{align}
    \partial_\mu \alpha  &= \frac{\partial_\mu g^{00}}{2(-g^{00})^{3/2}}\,,\\
    \partial_\mu \beta_i &= \partial_\mu g_{0i}\,,\\
    \partial_\mu \beta^i &= \alpha^2\partial_\mu g^{0i} +
    2\alpha g^{0i}\partial_\mu\alpha\,,\\
    \partial_\mu \gamma_{ij} &= \partial_\mu g_{ij}\,,\\
    \partial_\mu \gamma^{ij} &= \partial_\mu g^{ij} +
    \frac{1}{\alpha^2}\left(\beta^i\partial_\mu\beta^j
    + \beta^j\partial_\mu\beta^i\right) -
    \frac{2}{\alpha^3}\beta^i\beta^j\partial_\mu\alpha\,.
\end{align}

\subsection{Geodesic Equation}
\label{subsec: Geodesic Equation}

Let $\tau$ be the proper time of the Eulerian observer and $\lambda$ be
an affine parameter along the photon null geodesic. The photon frequency
measured by the Eulerian observer is then defined as,
\begin{align}
    \nu :&= \frac{d\tau}{d\lambda}\,,
\end{align}
so that we can decompose the photon four-momentum in terms of the
four-velocity of the observer $n^\mu$ and the photon four-velocity this
observer measures $v^\mu$,
\begin{align}
    \label{eq: photon 4-momentum}
    p^\mu &= \frac{dx^\mu}{d\lambda} =
    \frac{d\tau}{d\lambda}\frac{dt}{d\tau} \frac{dx^\mu}{dt} = \nu
    \frac{1}{\alpha} \kappa^\mu = \nu(n^\mu + v^\mu)\,,
\end{align}
where ${dt}/{d\tau} = n^0 = {1}/{\alpha}$ and $\kappa^\mu :=
{dx^\mu}/{dt}$ is the tangent to the photon worldline.
The latter is related to the four-velocity of the Eulerian
observer and the three-velocity of the photon as observed
by such an observer by the relations
\begin{align}
    \frac{dx^\mu}{dt} &= \kappa^\mu = \alpha\left(n^\mu +
    v^\mu\right),\\ \frac{dx^0}{dt} &= \kappa^0 = 1 =
    \alpha\left(\frac{1}{\alpha} + v^0\right) \quad\Rightarrow\quad v^0 =
    0\,,\\
\label{eq: three vel connection}
    \frac{dx^i}{dt} &= \kappa^i = \alpha\left(n^i + v^i\right)
    =\alpha\left(-\frac{\beta^i}{\alpha} + v^i\right) = \alpha v^i - \beta^i\,,
\end{align}
where the normalisation of these four-vectors are
\begin{align}
  \kappa_\mu \kappa^\mu = 0, \qquad n_\mu n^\mu = -1, \qquad v_\mu v^\mu = v_i v^i = 1\,.
\end{align}
The coordinate-time dependent geodesic equations can then be written in
3+1-form as~\citep{Vincent2012}
\begin{align}
\label{eq: Geodesic Eq nu}
    \frac{d\nu}{dt} &= \nu(\alpha K_{ij}v^i v^j - v^i\partial_i\alpha)\,,\\
\label{eq: Geodesic Eq x}
    \frac{dx^i}{dt} &= \alpha v^i - \beta^i\,,\\
\label{eq: Geodesic Eq v}
    \frac{dv^i}{dt} &= \alpha v^j\left[v^i(\partial_j \ln\alpha-K_{jk}v^k)
    + 2K\indices{^i_j}-{}^3\Gamma\indices{^i_j_k}v^k\right]\nonumber\\
    &\quad-\gamma^{ij}\partial_j\alpha - v^j\partial_j\beta^i\,.
\end{align}

\section{Special-relativistic lattice-Boltzmann method for radiative transport}
\label{sec: SRLBMRT}

An obvious starting point to introduce our GRLBM is to briefly summarise
the special-relativistic approach proposed by~\citet{Weih2020c}, where it
is necessary to differentiate between the laboratory frame (LF) and the
fluid frame (FF), which we indicate using tilded variables.

Let $\mathcal{P}$ be a photon bundle at the spacetime position $x^\mu$,
three-velocity $v^i$, and four-momentum $p^\mu$. The evolution of its
Lorentz-invariant distribution function, $f_\nu=f(x^i, v^i, \nu,
t)=\tilde f_{\tilde\nu}$, is then governed by the radiative Boltzmann
equation
\begin{align}
  \label{eq: radiation Boltzmann equation}
    \frac{df_\nu}{d\lambda} & = \frac{\eta_\nu}{\nu^2} -
    \nu\kappa_{a\nu} f_\nu + \Gamma_\nu[f_\nu] =: \mathcal{C}_\nu[f_\nu]\,,
\end{align}
where $\nu := p^0 = {dt}/{d\lambda}$ is the frequency observed in the
laboratory frame, $\eta_\nu$ the fluid emissivity, $\kappa_{a\nu}$ the
fluid absorption coefficient, $\Gamma_\nu$ the scattering operator, and
$\mathcal{C}_\nu$ the total collision operator. The explicit form of the
scattering operator $\Gamma_\nu[f_\nu]$ depends on the underlying
scattering model. Following~\citet{Weih2020c}, we assume a homogeneous
iso-energetic scattering operator, which is simpler to express in the
comoving FF (see Appendix \ref{Appendix: Collision Operator} for full
derivation) as
\begin{align}
    \tilde\Gamma_{\tilde\nu}[\tilde f_{\tilde\nu}] & = \tilde \nu
    \left[\tilde\kappa_{0\tilde\nu}(\tilde E_{\tilde\nu} - \tilde
      f_{\tilde\nu}) + 3\tilde\kappa_{1\tilde\nu}\tilde v_i \tilde
      F^i_{\tilde\nu}\right]\,,
\end{align}
where $\tilde\kappa_{0\tilde\nu}$ and $\tilde\kappa_{1\tilde\nu}$ are the
zeroth and first-order scattering coefficients in the FF of an underlying
Legendre expansion (see Appendix \ref{Appendix: Collision Operator} for more
detail), accounting for the isotropic and forward scattering,
respectively.

Because in the FF we know both the explicit form of the scattering
operator, and obviously the fluid properties $\tilde\eta_{\tilde\nu},
\tilde\kappa_{a\tilde\nu}, \tilde\kappa_{0\tilde\nu},
\tilde\kappa_{1\tilde\nu}$, we can express the Boltzmann equation
\eqref{eq: radiation Boltzmann equation} in the FF as
\begin{align}
    \label{eq: fluid frame radiation Boltzmann equation}
    \frac{d\tilde f_{\tilde\nu}}{d\lambda} & =
    \frac{\tilde\eta_{\tilde\nu}}{\tilde\nu^2} -
    \tilde\nu\tilde\kappa_{a\tilde\nu} \tilde f_{\tilde\nu} + \tilde \nu
    \left(\tilde\kappa_{0\tilde\nu}(\tilde E - \tilde f_{\tilde\nu}) +
    3\tilde\kappa_{1\tilde\nu}\tilde v_i \tilde F^i_{\tilde\nu}\right)\,.
\end{align}
Next, we transform the affine parameter differential to the LF time
differential by using the chain rule and replace the distribution
function $\tilde f_{\tilde\nu}$ with the specific intensity $\tilde
I_{\tilde\nu} = \tilde\nu^3 \tilde f_{\tilde\nu}$. The Lorentz
transformation of the frequency can be derived by boosting the photon
four-momentum $p^\mu$ between inertial frames (see Appendix
\ref{Appendix: Lorentz Transformations})
\begin{align}
    \frac{d\tilde f_{\tilde\nu}}{d\lambda} & =
    \frac{dt}{d\lambda}\frac{d\tilde f_{\tilde\nu}}{dt} = \nu
    \frac{d\tilde f_{\tilde\nu}}{dt} = \frac{\tilde\nu}{A} \frac{d
      (\tilde I_{\tilde\nu} / \tilde\nu^3)}{dt}\,,
\end{align}
so that
\begin{align}
    \label{eq: fluid frame specific intensity equation}
    \frac{d\tilde I_{\tilde\nu}}{dt} & = A \tilde\nu^2 \frac{d\tilde
      f_{\tilde\nu}}{d\lambda} = A \tilde\nu^2 \mathcal{\tilde
      C}_{\tilde\nu}[\tilde f_{\tilde\nu}] \\
                                                 & = A
    \left[\tilde\eta_{\tilde\nu} - \tilde\kappa_{a\tilde\nu} \tilde
      I_{\tilde\nu} + \tilde\kappa_{0\tilde\nu}(\tilde \nu^3
      \tilde E_{\tilde\nu} - \tilde I_{\tilde\nu}) +
      3\tilde\kappa_{1\tilde\nu}\tilde v_i \tilde \nu^3 \tilde
        F^i_{\tilde\nu}\right]\,,
\end{align}
and where
\begin{align}
    A & := \gamma(1 - u_i v^i) = \frac{1 -
      u_i v^i}{\sqrt{1 - u_i u^i}}\,,
\end{align}
is the Doppler factor between the LF and FF, and $u^i$ the three-velocity
of the fluid measured in the LF. 

Since we are not interested in the monochromatic intensity $I_\nu$, but
rather in the total or ``grey'' (or frequency-integrated) intensity $I$,
we also define the total emissivity, opacities, and moments 
\begin{align}
    I :&= \int_0^\infty I_\nu\ d\nu,\\ \eta :&= \int_0^\infty
    \eta_\nu\,d\nu,\\ \kappa_\star :&= \frac{\int_0^\infty
      \kappa_{\star\nu}I_\nu \,d\nu}{\int_0^\infty I_\nu
      \,d\nu}, \quad \star\in\{a,0,1\}\\ E :&= \int_0^\infty \nu^3
    E_\nu\,d\nu\nonumber\\ &=
    \frac{1}{4\pi}\oint_{4\pi}\int_0^\infty \nu^3 f_\nu \ d\nu
    d\Omega = \frac{1}{4\pi} \oint_{4\pi}
    I\ d\Omega,\\ F^i :&= \int_0^\infty \nu^3
    F^i_\nu\,d\nu\nonumber\\ &=
    \frac{1}{4\pi}\oint_{4\pi}\int_0^\infty \nu^3 f_\nu v^i\ d\nu
    d\Omega = \frac{1}{4\pi} \oint_{4\pi} I v^i\ d\Omega.
\end{align}
Applying this 'grey' approximation to Eq.~\eqref{eq: fluid frame specific 
intensity equation}, we get,
\begin{align}
    \label{eq: fluid frame intensity equation}
    \frac{d\tilde I}{dt} & = A \left[\tilde\eta - (\tilde\kappa_a +
      \tilde\kappa_0) \tilde I + \tilde\kappa_0\tilde E +
      3\tilde\kappa_1\tilde v_i \tilde F^i\right]\\ & = A
    \left[\tilde\eta - (\tilde\kappa_a + \tilde\kappa_0) \tilde I +
      \tilde M\right]\,,
\end{align}
where we introduced the moment collision term
\begin{equation}
\tilde M :=
\tilde\kappa_0\tilde E + 3\tilde\kappa_1\tilde v_i \tilde F^i\,,
\end{equation}
in terms of the zero and first-order scattering coefficients $\kappa_0$
and $\kappa_1$. The next step is to transform the total intensities from
the FF to the LF by applying the respective Lorentz transformation law,
see Eq.~\eqref{eq: grey intensity transformation}. The mixed frame ODE
for the total LF intensity can then be expressed as,
\begin{align}
    \label{eq: mixed frame intensity equation}
    \frac{dI}{dt} & = \frac{\tilde\eta + \tilde M}{A^3} - A I
    (\tilde\kappa_a + \tilde\kappa_0)\,,
\end{align}
which is the same as Eq.~(B8) in~\citet{Weih2020c}. Specialisations to a
one-dimensional (two-dimensional) case is obtained with the following
change $A^3 \to A$ ($A^3 \to A^2$) in Eq.~\eqref{eq: mixed frame
  intensity equation}.

\subsection{Numerical discretization}
\label{subsec:num_imp}
When seeking a numerical solution we obviously need to discretize the
total intensity $I$ into the $\Ndir$ population intensities $I_\dir$,
\begin{align}
    \label{eq: discretized mixed frame intensity equation_1}
    I_\dir(x^i, t) & := I(x^i, v_\dir^i, t)\,, \\
    \label{eq: discretized mixed frame intensity equation_2}
    \frac{dI_\dir}{dt} & = \frac{\tilde\eta + \tilde M_\dir}{A_\dir^3} -
    A_\dir I_\dir (\tilde\kappa_a + \tilde\kappa_0)\,,
\end{align}
where, unlike the classical LBM, we did not include the weights in the
definition of the population intensities; a similar approach will be
necessary for the GRLBM, as we will see in Sec. \ref{sec:GRLBMRT}.

As a result, the weights $w_\dir$ of the numerical quadratures must be
included when calculating the moments
\begin{align}
    \label{eq: moment0}
    E & = \frac{1}{4\pi} \oint_{4\pi} I\,d\Omega \approx
    \sum_{\dir=0}^{\Ndir-1} w_\dir I_\dir\,, \\
    \label{eq: moment1}
    F^i & = \frac{1}{4\pi} \oint_{4\pi} v^i I\,d\Omega
    \approx \sum_{\dir=0}^{\Ndir-1} w_\dir I_\dir v^i_\dir\,, \\
    \label{eq: moment2}
    P^{ij} & = \frac{1}{4\pi} \oint_{4\pi} v^i v^j
    I\,d\Omega \approx \sum_{\dir=0}^{\Ndir-1} w_\dir I_\dir v^i_\dir
    v^j_\dir\,.
\end{align}
Note that the set of discretized velocities $v^i_\dir$ is restricted to
lie on the unit sphere due to the absolute speed of photons, so that
the Hermite-Gauss quadrature is not applicable anymore. Instead, we use
spherical quadratures in 2D and 3D, which aim to integrate Fourier- and
spherical-harmonics as well as possible. The number of directions in our
velocity stencil must be much higher than in the classical LBM to ensure
that the moments are integrated to an acceptable degree of
accuracy. Furthermore, besides the accuracy of the moment integrations, a
large number of populations is also essential for the free-streaming
scenario in optically-thin media, where a large number of homogeneously
distributed points must be used to achieve homogeneous propagation of
light beams.

\begin{figure}
    \centering \includegraphics[width=0.5\textwidth]{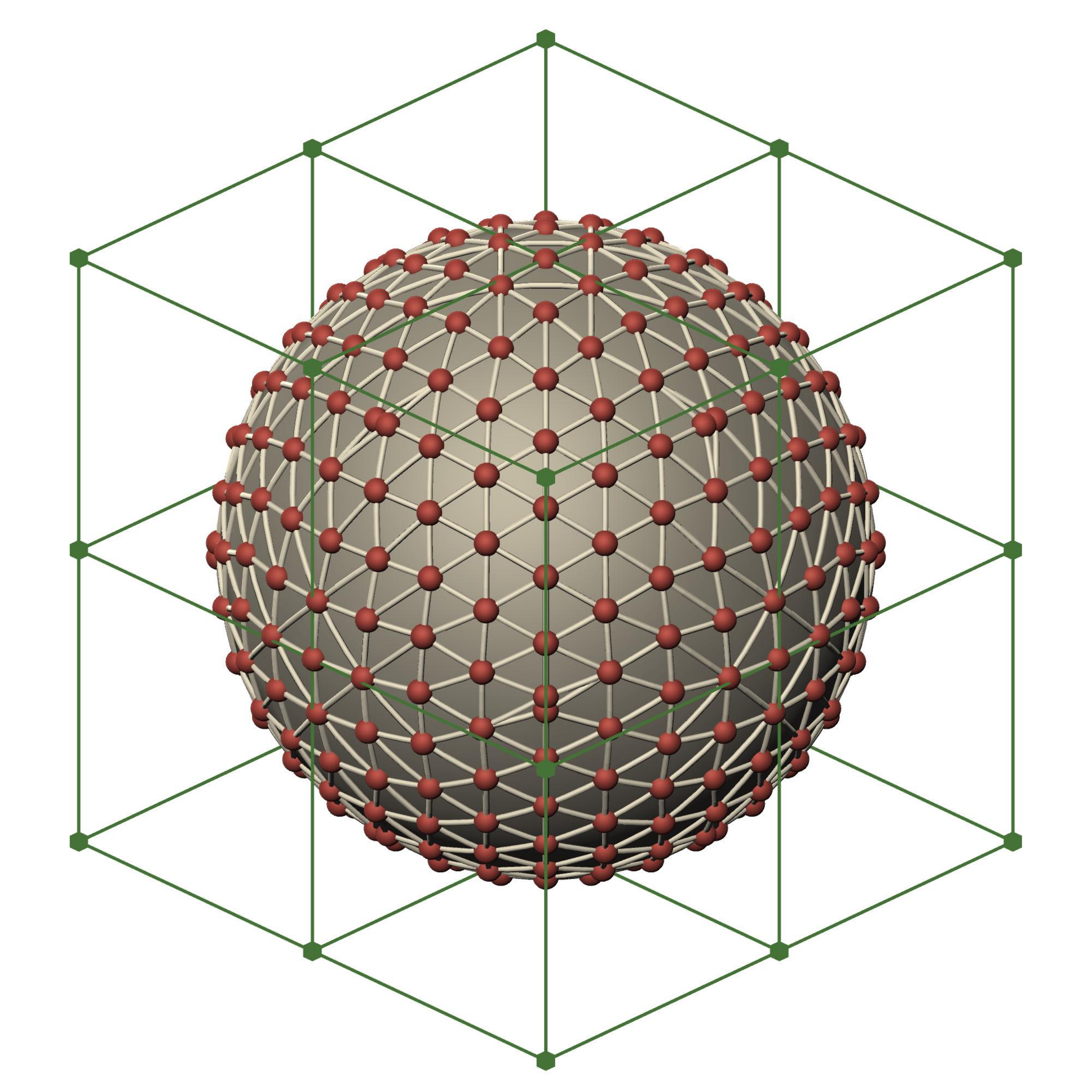}
    \caption{Lebedev stencil of order $p_{\rm Leb}=31$ with $351$
      discrete points (red dots) for the quadrature of the velocities on
      the two-sphere.}
    \label{fig: Lebedev velocity stencil}
\end{figure}

As a result, we use a Fourier quadrature with homogeneous velocity
distribution in 2D and a Lebedev quadrature of order $p_{\rm Leb}$ in 3D
as shown in Fig.~\ref{fig: Lebedev velocity stencil} for a unit two-sphere,
which is a 2D surface in a 3D space. Our tests have shown that in 2D, at
least $50$ directions and in 3D, at least $200$ directions are needed for
homogeneous propagation. Generally speaking, more directions are always
better, but no significant improvement can be seen in 2D beyond
$200$. This said, and as we will comment further in Sec. \ref{subsec:
  Adaptive stencil}, to reduce computational costs, the stencils can also
be suitably modified so as to use fewer directions while achieving the
desired accuracy.

Generally speaking, Eqs.~\eqref{eq: discretized mixed frame intensity
  equation_1} and~\eqref{eq: discretized mixed frame intensity
  equation_2} represent a stiff system of $\Ndir$ coupled ODEs. The
stiffness stems from the wide range of possible emissivity and opacity
values and mostly appears in optically-thick regimes. Therefore, a
forward Euler scheme will not be stable for all possible values of
$\tilde\eta$, $\tilde\kappa_a,\tilde\kappa_0,\tilde\kappa_1$. Instead, we
must employ an implicit integrator like the backward Euler
method. However, solving an $\Ndir\times \Ndir$ linear system every
timestep on every grid-point is not feasible given high number of
populations we need to use. A quicker alternative is the so-called
``Lambda-iteration'' scheme, that is, a fixed-point iteration method that
converges to the solution of the linear system~\citep{Hubeny2003}.

By expressing our system of ODEs in matrix form
\begin{equation}
  \label{eq:lambda_it}
  \frac{d\vec I}{dt} = \Lambda \vec I\,,
\end{equation}
we can split the matrix $\Lambda$ into two parts
\begin{equation}
\Lambda = \Phi + (\Lambda - \Phi) = \Phi + \Psi\,,
\end{equation}
where $\Phi$ should be easily invertible, \eg diagonal. Applying the
implicit Euler integrator, we can rearrange the system of ODEs to obtain
an expression for the new timestep at level $n$
\begin{align}
  \vec I_n &= \vec I_{n-1} + \Delta t[\Phi +\Psi] \vec I_n\,,
\end{align}
so that
\begin{align}
  \label{eq:In}
  \vec I_n &= [1 - \Delta t \Phi]^{-1} [\vec I_{n-1} + \Delta t \, \Psi\, \vec I_n]\,.
\end{align}
Since $\vec I_n$ appears also on the right-hand side of
Eq.~\eqref{eq:In}, this expression is not explicit. However, if we
replace $\vec I_n$ with a guess for it, $\vec I^{(0)}_n$, \eg the value
from the previous $n-1$ timestep $\vec I^{(0)}_n = \vec I_{n-1}$, the
equation above becomes an estimate for $\vec I_n$. Repeating this process
with the new estimate, we get an iterative scheme
\begin{align}
    \vec I^{(m+1)}_n &= [1 - \Delta t \Phi]^{-1} [\vec I_{n-1} + \Delta t
      \, \Psi \, \vec I^{(m)}_n]\,,
\end{align}
that converges to the solution of the initial system of ODEs,
\begin{align}
\vec I_n = \lim_{m\rightarrow\infty} \vec I^{(m)}_n\,.
\end{align}
In the case of Eq.~\eqref{eq: discretized mixed frame intensity
  equation_2}, the linear terms are given by $A_\dir I_\dir
(\tilde\kappa_a + \tilde\kappa_0)$, and the components in the FF moments,
$\tilde E$ and $\tilde F^i$, that are proportional to
$I_\dir$. Disentangling the linear terms from the FF moments is
algebraically and computationally very difficult and not necessary to
achieve good results with the Lambda-iteration method. Instead, we only
use the expression $A_\dir I_\dir (\tilde\kappa_a + \tilde\kappa_0)$ for
the operator split and apply the Lambda-iteration to our system of ODEs
in Eq.~\eqref{eq: discretized mixed frame intensity equation_1} and
in Eq.~\eqref{eq: discretized mixed frame intensity equation_2} to get
\begin{align}
    \label{eq: sr discretized Lambda-Iteration result}
    I_{\dir}^{(m+1)}(x^i_n, t_n) & =
    \frac{I_{\dir}^{(0)}(x^i_{n-1},t_{n-1}) + \Delta t \left({\tilde\eta +
        \tilde M_{\dir}^{(m)}}\right)/{A_\dir^3}}{1 + \Delta t A_\dir
      (\tilde\kappa_a + \tilde\kappa_0)}\,, 
\end{align}
where
\begin{align}
    \label{eq: sr discretized Lambda-Iteration result_2}
    \tilde M_{\dir}^{(m)} & := \tilde\kappa_0\tilde E^{(m)} +
    3\tilde\kappa_1\tilde v_{\dir,i} \tilde F^{i(m)}\,,
\end{align}
and with an upper limit on the number of iterations set to to $m \leq
100$.

As in the classical LBM case, before we start iterating Eq.~\eqref{eq: sr
  discretized Lambda-Iteration result}, we split it into two
steps\footnote{We recall that the $\star$ is commonly used throughout the
literature in the LBM for the post-collision populations $f^\star_\dir$,
to distinguish them from the usual populations $f_\dir$.}
\begin{align}
    \label{eq: SRLBMRT collision}
    I_{\dir}^{\star(m+1)}(y^i, t_{n-1}) & =
    \frac{I_{\dir}^{(0)}(y^i,t_{n-1}) + \Delta t \left({\tilde\eta + \tilde
        M_{\dir}^{(m)}}\right)/{A_\dir^3}}{1 + \Delta t A_\dir (\tilde\kappa_a +
      \tilde\kappa_0)}\,,
\end{align}
and
\begin{align}
    \label{eq: SRLBMRT streaming}
    I_\dir(x^i_n, t_n) & = I_\dir^{\star(\text{final})}(x^i_{n-1},
    t_{n-1})\,,
\end{align}
where Eq.~\eqref{eq: SRLBMRT collision} is the collision step, which now
consists of the Lambda-iteration scheme, locally at a grid-point
$y^i$. This equation should be iterated until all the post-collision
intensities $I^\star_\dir$ converge. In practice, we iterate until the
first three moments $E, F^i, P^{ij}$ converge to a desired precision.

\begin{figure}
    \centering
    \begin{tikzpicture}[scale=1.6]
      \draw[step=1cm,gray,very thin] (-1.3, -1.3) grid (1.3, 1.3);
      \draw[blue,thin,dashed] (0,0) circle (1cm);
      \draw[fill=cyan](-1/1.414213562, -1/1.414213562)circle(3pt);
      \node at (0.30, 0.15) {\(x^i_n\)};
      \node at (0.35 -1/1.414213562, -1/1.414213562) {\(x^i_{n-1}\)};
      \foreach \x in {-1,0,1} \foreach \y in {-1, 0, 1}
      { \draw[fill=black](\x,\y)circle(3pt); }
      \draw[red,thick,->] (0,0) -- ( 0,  1);
      \draw[red,thick,->] (0,0) -- ( 0, -1);
      \draw[red,thick,->] (0,0) -- ( 1,  0);
      \draw[red,thick,->] (0,0) -- (-1,  0);
      \draw[red,thick,->] (0,0) -- ( 1/1.414213562,  1/1.414213562);
      \draw[red,thick,->] (0,0) -- ( 1/1.414213562, -1/1.414213562);
      \draw[red,thick,->] (0,0) -- (-1/1.414213562,  1/1.414213562);
      \draw[red,thick,->] (0,0) -- (-1/1.414213562, -1/1.414213562);
    \end{tikzpicture}
    \begin{tikzpicture}[scale=3.2]
      \draw[step=1cm,gray,very thin] (-0.15,-0.15) grid
      (1.15,1.15); \draw[blue,thin,dashed] (0,1) arc (180:270:1);
      \draw[fill=cyan](1-1/1.414213562, 1-1/1.414213562)circle(3pt);
      \node at (1+0.1, 1-0.15) {\(x^i_n\)};
      \node at (1.25 -1/1.414213562,1-1/1.414213562) {\(x^i_{n-1}\)};
      \foreach \x in {0,1} \foreach \y in {0,1} {
        \draw[fill=black](\x,\y)circle(3pt); }
      \draw[green!50!black,thick,->] (0,1) --
      (1-1/1.414213562-0.07, 1-1/1.414213562+0.07);
      \draw[green!50!black,thick,->] (0,0) --
      (1-1/1.414213562-0.07, 1-1/1.414213562-0.07);
      \draw[green!50!black,thick,->] (1,0) --
      (1-1/1.414213562+0.07, 1-1/1.414213562-0.07);
      \draw[green!50!black,thick,->] (1,1) parabola
      (1-1/1.414213562+0.07, 1-1/1.414213562+0.07);
      \draw[red,thick,->] (1,1) -- (1-1/1.414213562, 1-1/1.414213562);
    \end{tikzpicture}
    \caption{Radiation streaming with spatial interpolation. \textit{Left
        panel:} Velocity vectors (red) do not reach the nearest neighbouring
      grid-points (black). \textit{Right panel:} Post-collision intensities
      must be interpolated (green arrows) to the off-grid grid-point (light
      blue dot) and streamed to the target grid-point (red
      arrow).}
    \label{fig: radiation streaming}
  \end{figure}
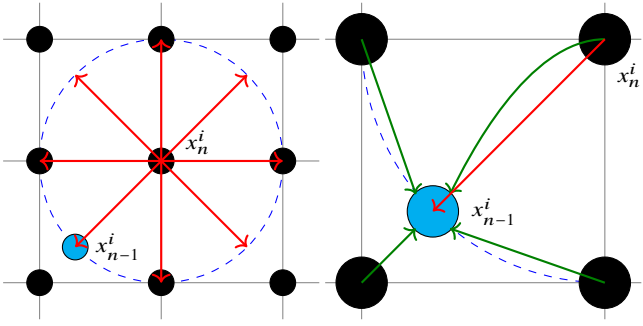

In the second step, \ie in the streaming step \eqref{eq: SRLBMRT
  streaming}, the velocities $v^i_\dir$ of the spherical stencils do not
reach the neighbouring grid-points (see left panel of Fig. \ref{fig:
  radiation streaming}). This requires a spatial interpolation of the
post-collision intensities $I_\dir^{\star(\text{final})}(y^i)$ to the
off-grid grid-point $x^i_{n-1}$. The streaming step then carries the
interpolated post-collision intensity from the off-grid source point
$x^i_{n-1}$ to the target point $x^i_n$ (see right panel of Fig.
\ref{fig: radiation streaming}). While our tests have shown
that a linear interpolation is sufficient in this step, it also
introduces numerical dispersion and breaks to a small extent the
perfectly conservative nature of the SRLBM.

Defining the relative mean-square error of the moments of as
\begin{align}
\epsilon :=  \left(\frac{\bar E_n - \bar E_{n-1}}{\bar
        E_n}\right)^2 + \sum_i \left(\frac{\bar F^i_n -
      \bar F^i_{n-1}}{\bar F^i_n}\right)^2 + \sum_{i,j}
    \left(\frac{\bar P^{ij}_n - \bar
        P^{ij}_{n-1}}{\bar P^{ij}_n}\right)^2\,.
\end{align}
We have tested the convergence rate of the Lambda-iteration scheme in an
optically thick regime with $\tilde\kappa_0 = 10^5$ with a threshold of
$\epsilon < 10^{-10}$. While the average iteration count remains at
roughly $4$, the maximum iteration count reaches the limit of $m=100$ at
the beginning of some simulations, suggesting that the initial intensity
distribution is not in equilibrium. This behavior is to be expected, as
we have no control over the initial pressure density (see Appendix
\ref{Appendix: Initial Data}), which is needed for adequate initial
data. Overall, our finding is that the first few timesteps allow the
radiation to equalise, drastically decreasing the needed iteration count
for the Lambda-iteration scheme.

\subsection{Coupling to background matter}
\label{subsec: Coupling to background Matter Fluid}

So far, we have only discussed how the radiation field is influenced by the
background matter fluid. Of course, we are equally interested in the
back-reaction that the radiation has on the fluid properties, that is,
energy and momentum. These contributions can be incorporated using the
following logical procedure.

\begin{enumerate}
\item For every timestep, compute the fluid frame emissivity $\tilde\eta$
  and opacity coefficients   $\tilde\kappa_{a/0/1},$ from the fluid rest-mass
  density $\rho$, temperature $T$, and three-velocity $u^i$.

\item As the conservative MHD variables are evolved in time, the $\Ndir$
  populations of the radiation intensities, $I_\dir$ are also evolved and
  the LF moments computed via the collision and streaming steps.

\item The LBM transforms the newly computed LF moments into the FF. 

\item From the FF moments, the radiative source terms $S_0$ and $S_k$
  given by Eqs.~\eqref{eq: source terms start}--\eqref{eq: source terms
    end} are computed as follows
\begin{align}
\label{eq: source terms start}
S_0 & = \gamma(\tilde\kappa_a \tilde E - \tilde\eta) +
\tilde\kappa H_0\,, \\
S_i & = \gamma(\tilde\kappa_a \tilde E -
\tilde\eta) u_i + \tilde\kappa H_i\,, 
\end{align}
where
\begin{align}
\tilde\kappa & = \tilde\kappa_a +
\tilde\kappa_s = \tilde\kappa_a + (\tilde\kappa_0 - \tilde\kappa_1/3)\,,
\\ H_i & = \gamma^3(F^m u_m - E) u_i + \gamma h_{im}F^m - \gamma h_{il}
u_m P^{lm}\,.
\label{eq: source terms end}
\end{align}
These quantities are then used for the subsequent MHD evolution.
\end{enumerate}

\section{General-relativistic lattice-Boltzmann method for radiative transport}
\label{sec:GRLBMRT}

\subsection{Mathematical Strategy}
\label{sec:Mathematics}

Within a general-relativistic formulation of the lattice-Boltzmann
equations for radiative transport, it is necessary to differentiate among
three frames: the Lab Frame (LF; for which we do not use any special
notation), the Eulerian (or inertial) frame (EF, for which we us barred
variables), and the Fluid Frame (FF; for which we employ tilded
variables). Transformations among all these frames need to be made using
Lorentz transformations.

Given the equivalence among these three frames, a decision should be
taken on the optimal frame where to express the evolution equation of the
total intensity. Inevitably, this decision also affects the frame in
which we discretize the velocity space. Because we want to preserve the
quadratures we employ in the SRLBM, the LF would not represent a useful
choice. Indeed, if we were to discretize the four-velocities $v^\mu$ in the
LF, neither the spatial components $v^k$, nor the three-velocity as seen
by the Eulerian observer $\bar{v}^k$ trace a spherical
shape, breaking the spherical quadrature. In addition, when deriving
quadrature rule in the LF we would need to choose weights that differ for
every grid-point and change over time in a dynamical spacetime (only a
local inertial frame preserves the isotropy in the quadrature weights). On the
other hand, using the FF introduces additional Lorentz-boosts when
comparing velocities, as the fluid velocity may vary substantially
between neighbouring grid-points. Finally, within the EF we can assume
that the four-velocities of neighbouring observers are similar enough,
allowing us direct comparison between directions without a
Lorentz-boosting. Doing so also implies that the local neighbourhood is
flat enough to ignore additional effects from parallel transport of
vectors between neighbouring grid-points. In view of these
considerations, we have concluded the EF to be the most convenient to
discretize the velocities and in the following we derive the total
intensity evolution equation only for the EF.

We start by expressing the Boltzmann equation in the FF and transform the
affine derivative into the LF time derivative via the chain rule,
\begin{align}
    \frac{d\tilde f_{\tilde\nu}}{d\lambda} & =
    \frac{dt}{d\lambda}\frac{\tilde f_{\tilde\nu}}{dt} =
    \frac{\nu}{\alpha} \frac{d\tilde f_{\tilde\nu}}{dt} =
    \frac{\tilde\nu}{\alpha A} \frac{d (\tilde I_{\tilde\nu} /
      \tilde\nu^3)}{dt}\,,
\end{align}
so that
\begin{align}
\frac{d(\tilde I_{\tilde\nu} / \tilde \nu^3)}{dt} & = \frac{\alpha
  A}{\tilde\nu} \frac{d\tilde f_{\tilde\nu}}{d\lambda} = \frac{\alpha
  A}{\tilde\nu} \mathcal{\tilde C}_{\tilde\nu}[\tilde f_{\tilde\nu}]\,,
\end{align}
and where we can use the same FF collision operator $\mathcal{\tilde
  C}_{\tilde\nu}[\tilde f_{\tilde\nu}]$ employed in Eq.~\eqref{eq: fluid
  frame specific intensity equation}.

Unlike the SRLBM, the LF is not a Lorentz frame, thus gaining additional
$\alpha$ (lapse) in the denominator, stemming from the time dilation
between the EF and the LF. In addition, the frequency $\tilde\nu$ is not
constant along the photon path due to gravitational redshift, so that we
cannot separate the frequency from the intensity as we did in
Eq.~\eqref{eq: fluid frame specific intensity equation}. As a result, we
need to perform the time integration before we can use the grey
approximation
\begin{align}
  \label{eq:time_int_Inu}
  & \int_{t_{n-1}}^{t_n}\frac{d(\tilde I_{\tilde\nu}/\tilde\nu^3)}{dt}
  dt = \frac{\tilde I_{\tilde\nu_n}}{\tilde\nu_n^3} - \frac{\tilde
    I_{\tilde\nu_{n-1}}}{\tilde\nu_{n-1}^3} = \int_{t_{n-1}}^{t_n} \alpha
  A \frac{\mathcal{\tilde C}_{\tilde\nu}[\tilde
      f_{\tilde\nu}]}{\tilde\nu}\ dt\,,
\end{align}
where we have introduced the short-hand notation $\tilde I_{\tilde\nu_n}
:= \tilde I_{\tilde\nu(t_n)}\left(x^i(t_n), \tilde v^i(t_n), t_n\right)$.

For the right-hand side of Eq.~\eqref{eq:time_int_Inu}, we use a forward
Euler integration and assume that the lapse and the fluid properties are
approximately constant to prevent strong coupling between the GRMHD and
radiation solvers. As a result, $\alpha = \alpha_{n-1} \approx \alpha_n$,
$\tilde\eta_{\tilde\nu} = \tilde\eta_{\tilde\nu_{n-1}} \approx
\tilde\eta_{\tilde\nu_n}$, and, $\tilde\kappa_{\star\tilde\nu} =
\tilde\kappa_{\star\tilde\nu_{n-1}} \approx
\tilde\kappa_{\star\tilde\nu_n}$, for $\star\in\{a,0,1\}$. Under these
assumptions, we can write
\begin{align}
     & \tilde I_{\tilde\nu_n} = \tilde s^3 \tilde I_{\tilde\nu_{n-1}} +
  \alpha\Delta t A_n \left[\tilde\eta_{\tilde\nu_n} -
    \tilde\kappa_{a\tilde\nu_n}\tilde I_{\tilde\nu_n}\right. \\ &
    \quad\qquad +\left.\tilde\kappa_{0\tilde\nu_n}(\tilde\nu_n^3 \tilde
    E_{\tilde\nu_n} - \tilde I_{\tilde\nu_n}) +
    3\tilde\kappa_{1\tilde\nu_n}\tilde\nu_n^3 \tilde v_{n,i} \tilde
    F^i_{\tilde\nu_n}\right]\nonumber\,,
\end{align}
so that the new variable $\tilde s$ measuring the ratio of the received
and emitted frequencies can be considered the ``redshift factor'' which
assumes different values in different frames, namely
\begin{align}
    \label{eq: redshift factor LF}
        s &= \frac{\nu_n}{\nu_{n-1}} \approx \frac{1}{1 - \Delta t
          (\alpha K_{ij} v^i v^j - v^i \partial_i\alpha)}\,, &({\rm LF})\,,\\
        \bar s &=
        \frac{\bar\nu_n}{\bar\nu_{n-1}} =
        \frac{\alpha_n\nu_n}{\alpha_{n-1}\nu_{n-1}} \approx
        \frac{\nu_n}{\nu_{n-1}} = s\,,  &({\rm EF})\,,\\
    \label{eq: redshift factor FF}
        \tilde s &= \frac{\tilde\nu_n}{\tilde\nu_{n-1}} =
        \frac{\nu_n}{\nu_{n-1}}\frac{A_n}{A_{n-1}} = \bar s
        \frac{A_n}{A_{n-1}} = s \frac{A_n}{A_{n-1}}\,,  &({\rm FF})\,.
\end{align}
Note that because the redshift factor is the same at all frequencies, we take it
out of the frequency integral in the grey approximation
$\int_0^{\infty}\dots d\tilde\nu_n$ and obtain
\begin{align}
    \tilde I_n &= \tilde s^4 \tilde I_{n-1} + \alpha\Delta t A_n
    \left[\tilde\eta - (\tilde\kappa_a + \tilde\kappa_0) \tilde I_n +
      \tilde M_n\right],\\ \tilde M_n &=\tilde\kappa_0\tilde E_n
    + 3\tilde\kappa_1\tilde v_{n,i}\tilde F^i_n\,.
\end{align}

As in the SRLBM, we can employ here the Lambda-Iteration scheme for the
evolution equation for the total intensity in the FF
\begin{align}
    \label{eq: gr Lambda-Iteration FF}
     \tilde I_n^{(m+1)} &= \frac{\tilde s^4 \tilde I_{n-1} +
       \alpha\Delta t A_n(\tilde\eta + \tilde M^{(m)})}{1 + \alpha\Delta
       t A_n(\tilde\kappa_a + \tilde\kappa_0)}\,,\\
     \tilde M^{(m)} &=
     \tilde\kappa_0 \tilde E^{(m)}_n + 3\tilde\kappa_1\tilde
     v_{n,i}\tilde F^{i(m)}_n\,, 
\end{align}
where $\tilde I_n^{(0)} = \tilde I_{n-1}$. On the other hand, for the
evolution equation in the EF, we have to transform the total intensities
$\bar I_n = \tilde I_n A_n^4$ and the redshift factor $\tilde s = s A_n /
A_{n-1}$ to obtain
\begin{align}
    \label{eq: gr Lambda-Iteration EF}
     & \bar I_{n}^{(m+1)} = \frac{s^4 \bar I_{n-1} +
      \alpha\Delta t \left(\tilde\eta + \tilde
        M^{(m)}\right) / A_n^3}{1 + \alpha\Delta t A_n(\tilde\kappa_a +
      \tilde\kappa_0)}\,\,, 
\end{align}
where, again, $\bar I_n^{(0)} = \bar I_{n-1}$. Next, we perform the
velocity discretisation
\begin{align}
    \bar I_\dir(x^i, t) &= s^4_\dir \bar I(x^i, \bar v^i_\dir, t),\\
    \bar I^\star_\dir(x^i, t) &= \frac{\bar I_\dir(x^i, t)}{s^4_\dir}\,,
\end{align}
and drop the timestep dependency on $A_n$ and replace it with a direction
dependency $A_\dir$.
By absorbing the redshift factor into the discretized intensities, we can
move the evaluation of the redshift factor to the streaming step.
\begin{align}
    \label{eq: GRLBMRT collision}
    \bar I^{\star(m+1)}_{\dir}(y^i, t_{n-1}) &= \frac{\bar
      I_{\dir}^{(0)}(y^i, t_{n-1}) + \alpha\Delta t
      \left(\tilde\eta + \tilde M_{\dir}^{(m)}\right) / A_\dir^3}{1 + \alpha\Delta t
      A_\dir(\tilde\kappa_a + \tilde\kappa_0)},\\ \tilde M_{\dir}^{(m)}
    &= \tilde\kappa_0 \tilde E^{(m)} + 3\tilde\kappa_1\tilde
    v_{\dir,i}\tilde F^{i(m)},\\
    \label{eq: GRLBMRT streaming}
    \bar I_\dir(x^i_n, \bar v^i_{\dir,n}, t_n) &= s_\dir^4 \bar
    I^{\star(\text{final})}(x^i_{\dir,n-1}, \bar v^i_{\dir,n-1},
    t_{n-1})\,.
\end{align}
Doing so, removes all the additional complexity from the collision step
\eqref{eq: GRLBMRT collision} and transfers it to the streaming step
\eqref{eq: GRLBMRT streaming}.

Comparing our newly derived collision operation~\eqref{eq: GRLBMRT
  collision} to that of the SRLBM \eqref{eq: SRLBMRT collision}, we can
see that they are identical except for the lapse $\alpha$. At first
glance, the streaming step might seem very similar as well. However, in
the special-relativistic scenario, the propagation direction of light
remains constant, \ie $v^i_n = v^i_{n-1}$, allowing the $\Ndir$ discrete
intensities to propagate independently from one another. In the
general-relativistic case, however, the direction of propagation of light
varies, \ie $\bar v^i_n \neq \bar v^i_{n-1}$, effectively introducing an
interpolation step in the velocity space and thus a direct
inter-dependency of the discretized intensities. For this reason, we did
not include the weights in the definition of the discretized intensities
in the previous section. Indeed, intensity interpolation in the velocity
space would not be possible when including the weights in the definition.

Note also that the streaming step is more complex in a
general-relativistic context (see Fig. \ref{fig: curved radiation
  streaming}).  First, we must determine the off-grid source point
$x^i_{\dir,n-1}$ (light blue dot), the source velocity $\bar
v^i_{\dir,n-1}$, and the redshift factor $s$ by integrating the geodesic
equations \eqref{eq: Geodesic Eq nu}--\eqref{eq: Geodesic Eq v} backward
in time. The initial data is the target point $x^i_n$ (black dot) and
target direction $\bar v^i_{\dir,n}$, and we solve the equations with a
fourth order adaptive Runge-Kutta-Fehlberg integrator.  We note that
repeating doing this operation for $\Ndir$ populations is computationally
expensive In Sec. \ref{subsec: Adaptive stencil}, we introduce an
interpolation scheme to reduce the number of ODEs drastically.

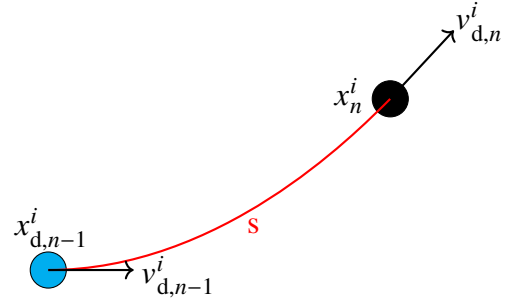
\begin{figure}
    \centering
    \begin{tikzpicture}[scale=4.5]
        \draw[fill=black](1,1)circle(1.5pt);
        \draw[fill=cyan](0, 0.5)circle(1.5pt);
        \draw[red,thick] (0, 0.5) parabola (1,1);
        \draw[black,thick,->] (0, 0.5) -- (0.25, 0.5);
        \draw[black,thick,->] (1, 1) -- (1.186, 1.2);
        \node[font=\Large] at (0.88, 1.01) {\(x^i_n\)};
        \node[font=\Large] at (0.01, 0.62) {\(x^i_{\dir,n-1}\)};
        \node[font=\Large] at (1.26, 1.22) {\(v^i_{\dir,n}\)};
        \node[font=\Large] at (0.38, 0.48) {\(v^i_{\dir,n-1}\)};
        \node[text=red, font=\Large] at (0.6, 0.63) {s};
    \end{tikzpicture}
    \caption{Curved radiation streaming of single direction. The origin
      of the radiation bundle (light blue dot) $x^i_{\dir,n-1}$ is
      computed by following the geodesic path (red line) backwards in
      time from the target point (black dot) $x^i_n$. Note that the
      velocities at the start and end do not match.}
    \label{fig: curved radiation streaming}
\end{figure}

Once the geodesic equations are solved, we would need to parallel
transport the source $\bar v^i_{\dir,n-1}$ velocity to the neighbouring
grid-points along some specified path. For simplicity, we assume that the
spatial discretization is fine enough and that the local curvature does
not vary significantly, so that the parallel transport can be replaced by
a simple interpolation. The testing we will discuss in Sec. \ref{sec:
  Implementation Tests} shows that this is a very reasonable
approximation, even in the vicinity of a black hole horizon. However,
special care must be taken if the radiation field is composed mainly of
photon bundles orbiting very close to the horizon.

Next, we perform velocity-space interpolation, for which we employ a
quadratic scheme in 2D and, due to the unstructured nature of Lebedev
stencils, a Voronoi interpolation scheme in 3D~\citep{Bobach2009}. As in
the SRLBM, linear interpolation is sufficient for the spatial
interpolation. However, due to the underlying curved spacetime, each grid
point experiences a different time dilation which affects the measured
intensities. Therefore, we transform all intensities from the neighbouring
gird points (B) to the receiving observer (A) at $x^i_n$ via a simple
algebraic expression
\begin{equation}
I_A =
\left(\frac{\alpha_A}{\alpha_B}\right)^4 I_B\,.
\end{equation}

\begin{figure*}
    \centering
    \includegraphics[height=0.33\textwidth]{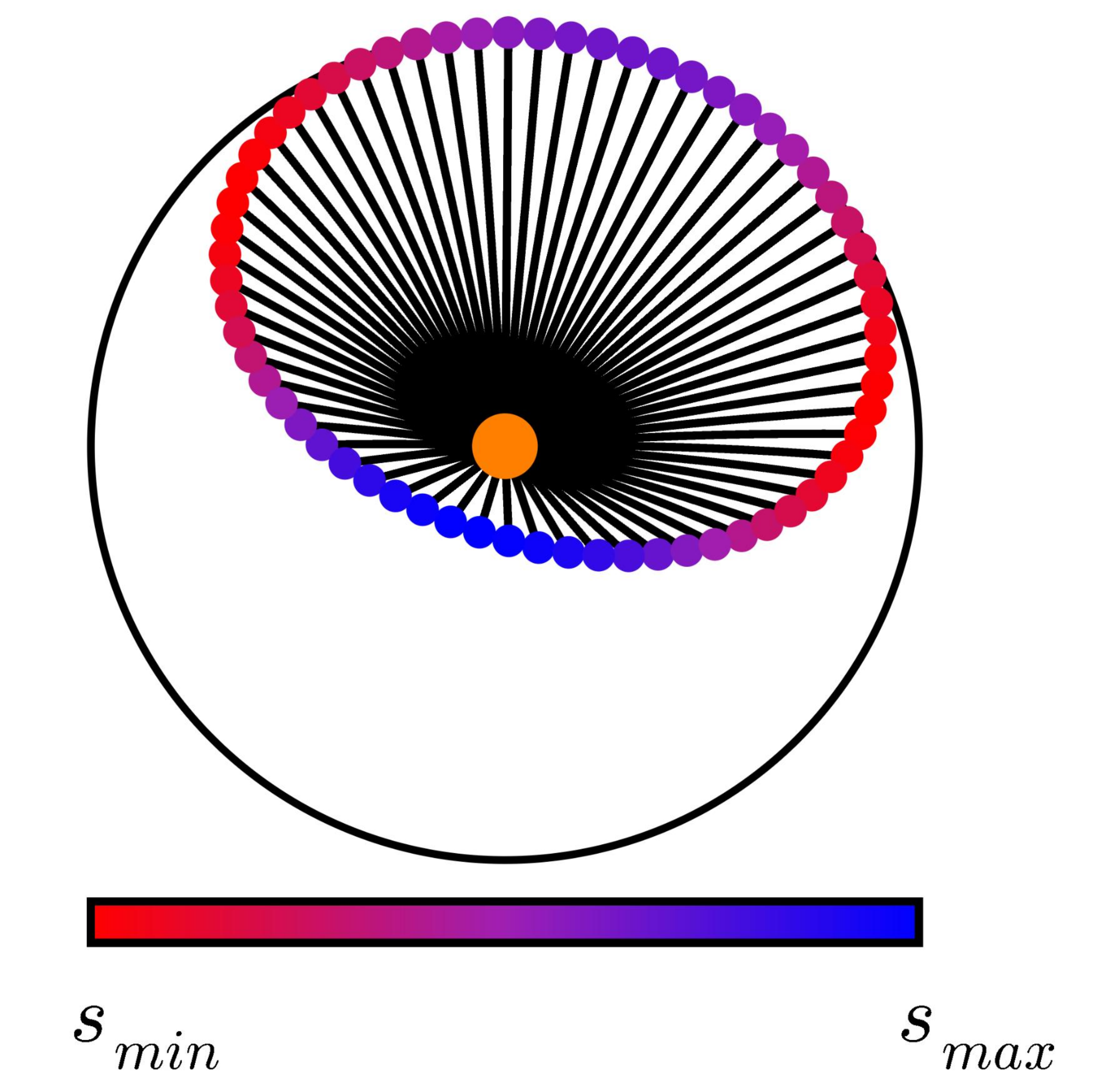}
    \hskip 1.0cm
    \includegraphics[height=0.33\textwidth]{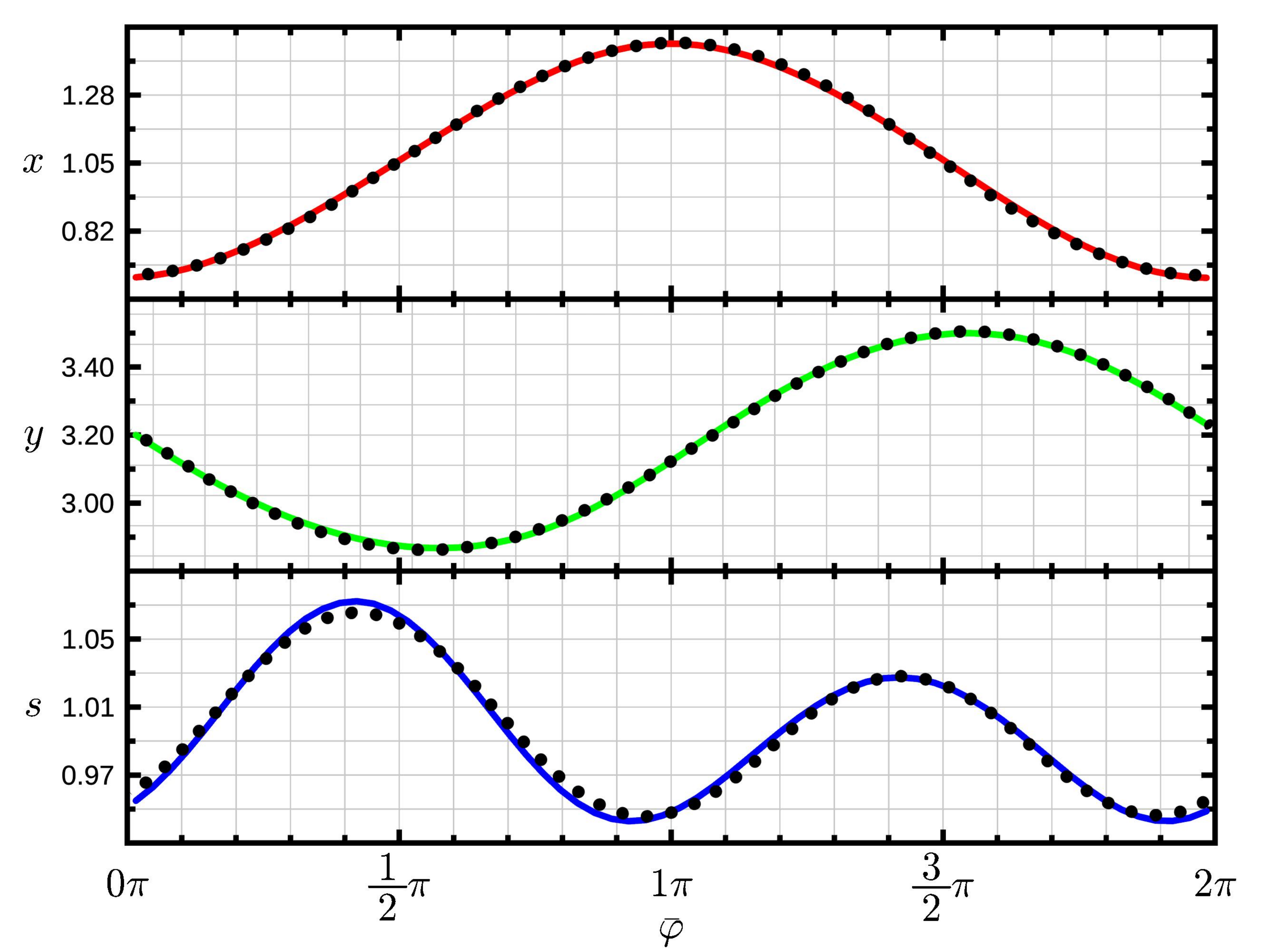}
    \caption{\textit{Left panel:} 64 photons emitted from $x^i=(1,3,0)$
      (orange dot) and integrated backwards in time for $\Delta t = 0.5$
      in a Kerr-Schild spacetime with $M=1$ and $a=0$. The dots around
      the orange dot show the final positions of the photons after the
      integration, which correspond to $x^i_{\dir,n-1}$. Their colouring
      shows their red- and blue-shift. The black circle mark an Euclidean
      two-sphere with radius $r = \Delta t = 0.5$.  \textit{Right panel:}
      The three plots show the final $x$, $y$, and $s$ components
      depending on the angle of emission $\bar\varphi$ in the EF. The
      black dots show instead the reconstruction of the functions from
      the Fourier transformation of only 5 photons. Note, that the
      discrepancy between the $s$ function and the Fourier reconstruction
      is due to the large timestep $\Delta t = 0.5$.}
    \label{fig: streaming geometry}
\end{figure*}

Finally, we are interested in the moments in the LF and to transform the
moments from the EF to the LF, we apply a tetrad transformation to the
relevant tensors. In particular, building the energy-momentum of the
radiation field in the Eulerian frame as
\begin{align}
    \bar T^{\mu\nu} & = \sum_\dir \left(\begin{array}{cc} 1 & \bar
      n^j_\dir \\ \bar n^i_\dir & \bar n^i_\dir \bar
      n^j_\dir\end{array}\right) w_\dir \bar I_\dir =
      \left(\begin{array}{cc} \bar E & \bar F^j \\ \bar F^i & \bar
        P^{ij}\end{array}\right)\,,
\end{align}
so that the energy-momentum of the radiation field in the LF will be
given by (see Appendix \ref{Apppendix: Tetrad Separation} for the
explicit expression of the tetrad)
\begin{align}
      T^{\mu\nu} &= e\indices{^\mu_\alpha} \, e\indices{^\nu_\beta} \bar T^{\mu\nu}\,.
\end{align}

Due to our definition of the discretized intensities, the moment
quadrature would also need to change
\begin{align}
  \label{eq: quadrature with s}
  \bar E = \sum_\dir \frac{w_\dir \bar I_\dir}{s^4_\dir} =
  \sum_\dir w_\dir \bar I^\star_\dir\,.
\end{align}
We should note that the inclusion of the frequency shifts $s_\dir$ in the
quadrature computation is very costly and increases the computational
time of factor of almost $15$. In the tests carried out here and
discussed below, we have evaluated the solution with and without the
frequency shift in Eq.~\eqref{eq: quadrature with s} finding only
negligible differences.

As a result, our implementation of the GRLBM is such that in the
Lambda-Iteration scheme the initial moments are computed with $\bar E
\approx \sum_\dir w_\dir \bar I_\dir$, leading to a minor error in the
initial moments. However, in all consecutive steps of the
Lambda-Iteration, we compute the moments from the iterated intensities
using the exact expression~\eqref{eq: quadrature with s}. Combined with
the iteration scheme's convergence criteria, this leads to a
self-correcting behavior, where any small mistakes in the initial moments
are subsequently corrected. At the same time, we note that the frequency
shift in the streaming step is essential for correctly propagating the
intensities.

As a concluding remark, we note that when entering optically-thick
regimes, the GRLBM is subject to the Courant-Friedrichs-Lewy (CFL)
condition like any solver in the diffusive limit. While it is advisable
to use the same CFL coefficient as that employed in the solution of the
GRMHD equations, usually around $0.2$, to ensure synchronisation, the
GRLBM can in principle handle much larger CFL numbers. Indeed, in the
tests presented below we employed a CFL coefficient of $0.9$ without
encountering problems even in diffusive regimes. In addition, in the
free-streaming case, the CFL coefficient can be further increased to be
up to $1.0$.

\subsection{Numerical Strategy}
\label{sec: Numerics}

\subsubsection{Harmonic Streaming}
\label{subsec: Harmonic Streaming}

In what follows, we discuss some of the most subtle issues when
developing a numerical infrastructure employing the GRLBM. To this scope,
we will restrict ourselves to a 2D scenario with only the position space
and frequency shift as this is simpler to explain, visualise, and
discuss.  However, the same strategy and all conclusions also apply for
the velocity space and can be extrapolated to 3D scenarios via the
spherical-harmonics decomposition.

As mentioned previously, it is numerically not feasible to solve
$N_{\text{grid}}\cdot \Ndir$ geodesic equations every time iteration for
the streaming step. However, we can drastically reduce the number of ODEs
that need to be solved by closely examining the geometric distribution of
the directions in which photons propagate from the emitters positions
$x^i_{\dir,n-1}$ and velocities $v^i_{\dir,n-1}$.

The left panel of Fig.~\ref{fig: streaming geometry} shows the
distribution of the photons for an emitter placed at $x^i = (1,3,0)$
outside the event horizon of a Kerr-Schild black hole with $M=1$ and $a=0$.
Note that in the EF, both the angular distribution and the
distribution of the redshift factor $s$ is far from being the isotropic
one expected in flat spacetime and shown with a circle. Rather, it
resembles an ellipse and the corresponding distributions are quantified
in the three plots in the right panel of Fig.~\ref{fig: streaming
  geometry}, which report the $x$, $y$ and $s$ distribution in terms of
the emission angle $\bar{\varphi}$ (in the IF). These distributions vary
from point to point and obviously become distorted as one approaches the
event horizon. However, given their smooth behaviour it is not difficult
to approximate them very accurately with a real Fourier harmonics
expansion of the type
\begin{align}
  \chi(\bar\varphi) &\approx \sum_{l=0}^{N_F-1} c_{\ell} B_{\ell}(\bar\varphi)\,,
\end{align}
where 
\begin{align}
    B_l(\bar\varphi) & := \left\{
    \begin{array}{ll}
        1, & l=0\, \\
        \sqrt{2}\cos\left(({{\ell}+1})\bar{\varphi}/2\right), & {\ell}=1,3,5,... \\
        \sqrt{2}\sin\left({{\ell}}\bar{\varphi}/2\right), & {\ell}=2,4,6,... 
    \end{array}\right. \\
\label{eq: evaluate Fourier transformation}
c_{\ell} & := \braket{\chi|B_{\ell}} = \frac{1}{2\pi}\int_0^{2\pi}
\chi(\bar\varphi)B_{\ell}(\bar\varphi)d\bar\varphi\,,
\end{align}
and, of course, the Fourier basis is orthonormal
\begin{align}
    \braket{B_i|B_j} & := \frac{1}{2\pi}\int_0^{2\pi}
    B_i(\bar\varphi)B_j(\bar\varphi)d\bar\varphi = \delta_{ij}\,.
\end{align}
 
In order to evaluate how many coefficients are necessary for an accurate
approximation we consider an extreme scenario in which the emitter is
outside but close to the event horizon of a Schwarzschild spacetime in
Kerr-Schild coordinates, \ie at the position $x^i_n/M =
(\sqrt2+0.1,\sqrt2+0.1)$, and the timestep is about two orders of
magnitude larger than that normally employed in GRMHD simulations, \ie
$\Delta t/M = 0.1$. The first nine coefficients for the Fourier harmonics
expansion of $x(\bar\varphi),y(\bar\varphi)$ and $s(\bar\varphi)$ for
this case are shown in Tab.~\ref{tab: Kerr Schild
  coefficients}.

\begin{table}
    \centering
    \setlength{\tabcolsep}{1pt}
    \begin{tabular}{|l|r|r|r||r|r|r|}
      & \multicolumn{3}{c}{${\rm Kerr-Schild~coordinates}$}   
      & \multicolumn{3}{c}{${\rm pseudo-Cartesian~coordinates}$}  \\
      \hline
      $c_k$ & $x(\bar\varphi)$ & $y(\bar\varphi)$ & $s(\bar\varphi)$
      & $x(\bar\varphi)$ & $y(\bar\varphi)$ & $s(\bar\varphi)$ \\
      \hline
        $c_0$ & $ 2.189220$ & $ 2.189220$ & $ 1.410883$ & $ 2.141385$ & $ 2.141385$ & $ 1.414574$ \\
        $c_1$ & $-0.063087$ & $ 0.020200$ & $ 0.003369$ & $-0.018760$ & $ 0.016436$ & $ 0.005429$ \\
        $c_2$ & $ 0.000124$ & $-0.059765$ & $ 0.004685$ & $ 0.000000$ & $-0.009044$ & $ 0.021124$ \\
        $c_3$ & $ 0.000021$ & $-0.000253$ & $-0.006217$ & $-0.000016$ & $-0.000207$ & $-0.000044$ \\
        $c_4$ & $ 0.000392$ & $ 0.000300$ & $ 0.018515$ & $ 0.000235$ & $-0.000112$ & $ 0.000024$ \\
        $c_5$ & $-0.000101$ & $-0.000249$ & $-0.000356$ & $ 0.000003$ & $-0.000000$ & $ 0.000000$ \\
        $c_6$ & $ 0.000293$ & $-0.000185$ & $ 0.000109$ & $ 0.000000$ & $ 0.000003$ & $ 0.000000$ \\
        $c_7$ & $-0.000002$ & $ 0.000005$ & $ 0.000071$ &             &             &             \\
        $c_8$ & $ 0.000004$ & $ 0.000001$ & $ 0.000058$ &             &             &             \\
          \hline
    \end{tabular}
    \caption{First nine Fourier harmonic-expansion coefficients in a
      Schwarzschild spacetime in Kerr-Schild coordinates for an emitter
      at $x^i_n = (\sqrt2+0.1,\sqrt2+0.1)$, with $\Delta t = 0.1$, using
      an adaptive RK45 solver.}
    \label{tab: Kerr Schild coefficients}
\end{table}

As would be expected from the general behaviour shown in the right panel
of Fig.~\ref{fig: streaming geometry}, we find that the first three
coefficients are sufficient to obtain a very good approximation and
indeed the coefficients $c_3$ and higher-order are at least two orders of
magnitude smaller in size. Similar estimates apply also for the frequency
shift $s$, but the more complex dependence in this case requires at least
the first five coefficients. A more quantitative measure of the error
made can be obtained when comparing the exact values for $x,y,$ and $s$
for 200 photons with those obtained with different numbers of the Fourier
expansion order $N_F$, which is reported in Tab.~\ref{tab: Fourier
  expansion error}
\begin{table}
    \centering
    \begin{tabular}{|l|r|r|r|}\hline
        $N_F$ & $\epsilon_{\text{rel}}(x)[\%]$ & $\epsilon_{\text{rel}}(y)[\%]$ & $\epsilon_{\text{rel}}(s)[\%]$ \\\hline
        3 & $0.062347$ & $0.064227$ & $3.707103$\\ 
        5 & $0.038384$ & $0.041219$ & $0.086021$\\ 
        7 & $0.000557$ & $0.000547$ & $0.017651$\\ 
        9 & $0.000395$ & $0.000368$ & $0.001043$\\\hline
    \end{tabular}
    \caption{Maximal relative error in emitted $x^i_e,s_e$ values of
      Fourier expansion in $\%$ after streaming for $\Delta t = 0.1$.
      The origin point is $x^i_r=(\sqrt{2}+0.1,\sqrt{2}+0.1)$ in a
      Kerr-Schild spacetime with $M=1$ and $a=0$.}
    \label{tab: Fourier expansion error}
\end{table}
Clearly, the error on the frequency shift is the largest for $N_F=3$ and
drops significantly when including the next higher-order
coefficients. These errors should be considered as upper values, as the
match further improves with smaller timesteps $\Delta t$ and further away
from the event horizon.

We should also note that these considerations are also dependent on the
spacetime considered and the coordinates employed to describe it, which
may or may not be better suited to describe photon motion. For instance,
in the case of a Schwarzschild spacetime described in pseudo-Cartesian
coordinates~\citep{Muller2009}, the first three Fourier harmonics
coefficients would also be sufficient also to accurately capture the
frequency shift (see Tab.~\ref{tab: Kerr Schild coefficients}). Overall,
we have found that using the first five coefficients $c_0$-$c_4$ is
sufficient and robust for most cases of interest.

\begin{figure}
    \centering
    \includegraphics[width=0.42\textwidth]{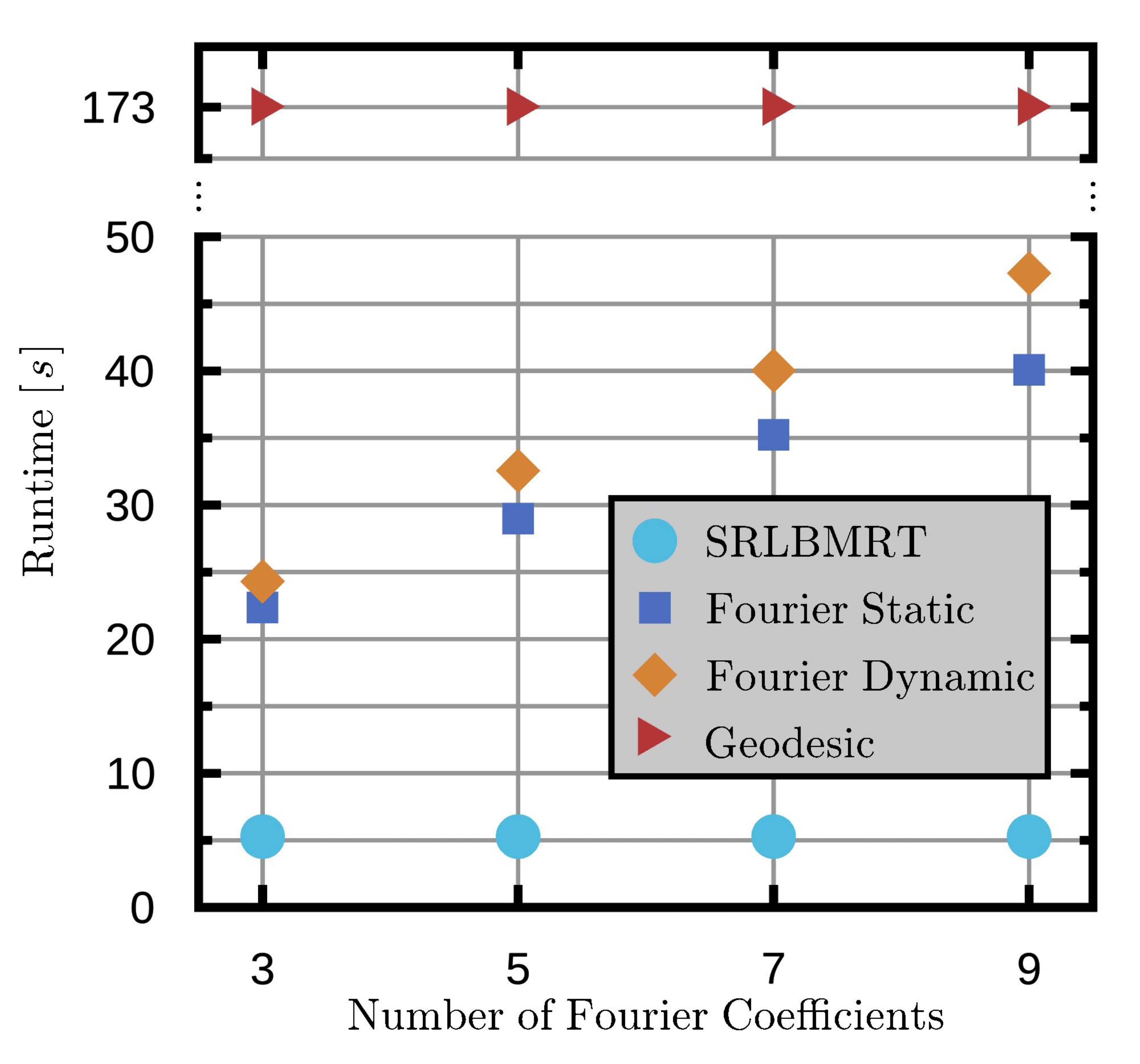}
    \caption{Total runtime in seconds as a function of the Fourier
      coefficients $N_F$ using a stencil with $\Ndir=200$ populations on
      a $200\times 250$ Cartesian grid. The background spacetime is that
      of a Schwarzschild black hole in Kerr-Schild coordinates and the
      evolution was carried out till $t=5.1\,M$ with a timestep of
      $\Delta t = 0.015$ using an adaptive RK45 solver for the solution of
      the geodesic equations \eqref{eq: Geodesic Eq nu}-\eqref{eq:
        Geodesic Eq v}.}
    \label{fig: Fourier Expansion Runtime}
\end{figure}

All in all, the Fourier expansion described above has the advantage that
instead of solving $\Ndir$ ODEs at every grid-point, we only need to
solve $N_F$ ODEs, which we use to calculate the Fourier coefficients
$c_{\ell}$ and hence approximate the emitter positions and velocities for
all $\Ndir$ photons. In the case of a static spacetime, this procedure
only needs to be done once, while the Fourier coefficients must be
recalculated at the beginning of every timestep for a dynamical
spacetime. In 2D, a Fourier stencil with $5$ directions and, therefore
$5$ Fourier coefficients is sufficient. In 3D, the same reasoning holds
regarding spherical-harmonic coefficients, $c_{\ell m}$. Our experience
suggests to use a Lebedev-5 stencil (\ie with $p_{\rm Leb}=5$) with $14$
directions, which leads to $((5+1)/2)^2 = 9$ spherical-harmonic
coefficients (${\ell}=0,1,2,\ldots$, $m\in\{-{\ell},{\ell}\}$)
\begin{align}
    \chi(\bar\theta,\bar\varphi) &\approx
    \sum_{{\ell}=0}^{N_{\ell}}\sum_{m=-{\ell}}^{{\ell}} c_{\ell m} Y_{\ell
      m}(\bar\theta,\bar\varphi)\,,
\end{align}
Of course, using a different number of ODEs directly impacts the code
performance and Fig.~\ref{fig: Fourier Expansion Runtime} is meant to
offer a runtime comparison of the four streaming algorithms, namely,
\begin{itemize}
    \item a flat streaming, without any velocity space interpolation (this
      is essentially the SRLBM).
    \item a geodesic streaming, where we solve the geodesic equations for
      every photon, every timestep.
    \item a Fourier streaming with static coefficients.
    \item a Fourier streaming with updating the coefficients every timestep.
\end{itemize}

A quick look at Fig.~\ref{fig: Fourier Expansion Runtime} is then
sufficient to realise that for the reference case of five Fourier
coefficients, the harmonic streaming leads to a speedup of a factor of
five. Furthermore, it is easy to appreciate that recalculating the
Fourier coefficients has only a minor impact on the runtime as most of
the extra runtime is spent in evaluating the Fourier transformation
\eqref{eq: evaluate Fourier transformation} and the velocity
interpolation. All things considered, the GRLBMRT code using harmonic
streaming, is about six times slower than the special-relativistic
equivalent. Further profiling tests have shown that roughly $20\%$ of the
extra time is spent in the velocity interpolation, while the bulk of the
extra run time comes from evaluating the Fourier coefficients. This is
also the reason why it is essential to use the lowest number of Fourier
coefficients.

\subsubsection{Adaptive stencil}
\label{subsec: Adaptive stencil}

\begin{figure*}
  \centering
  \includegraphics[width=0.65\textwidth]{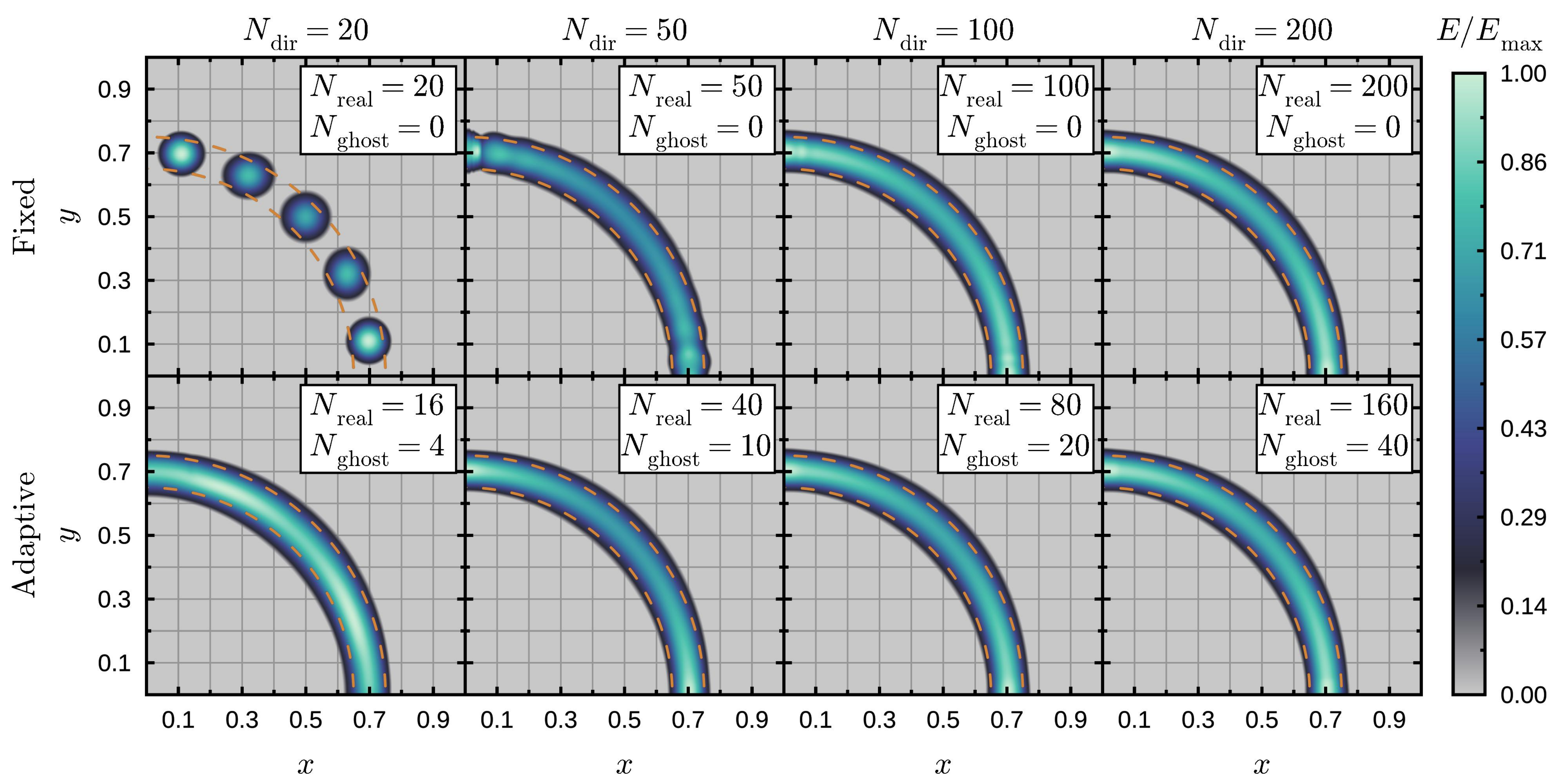}
  \vskip 0.25cm
  \includegraphics[width=0.65\textwidth]{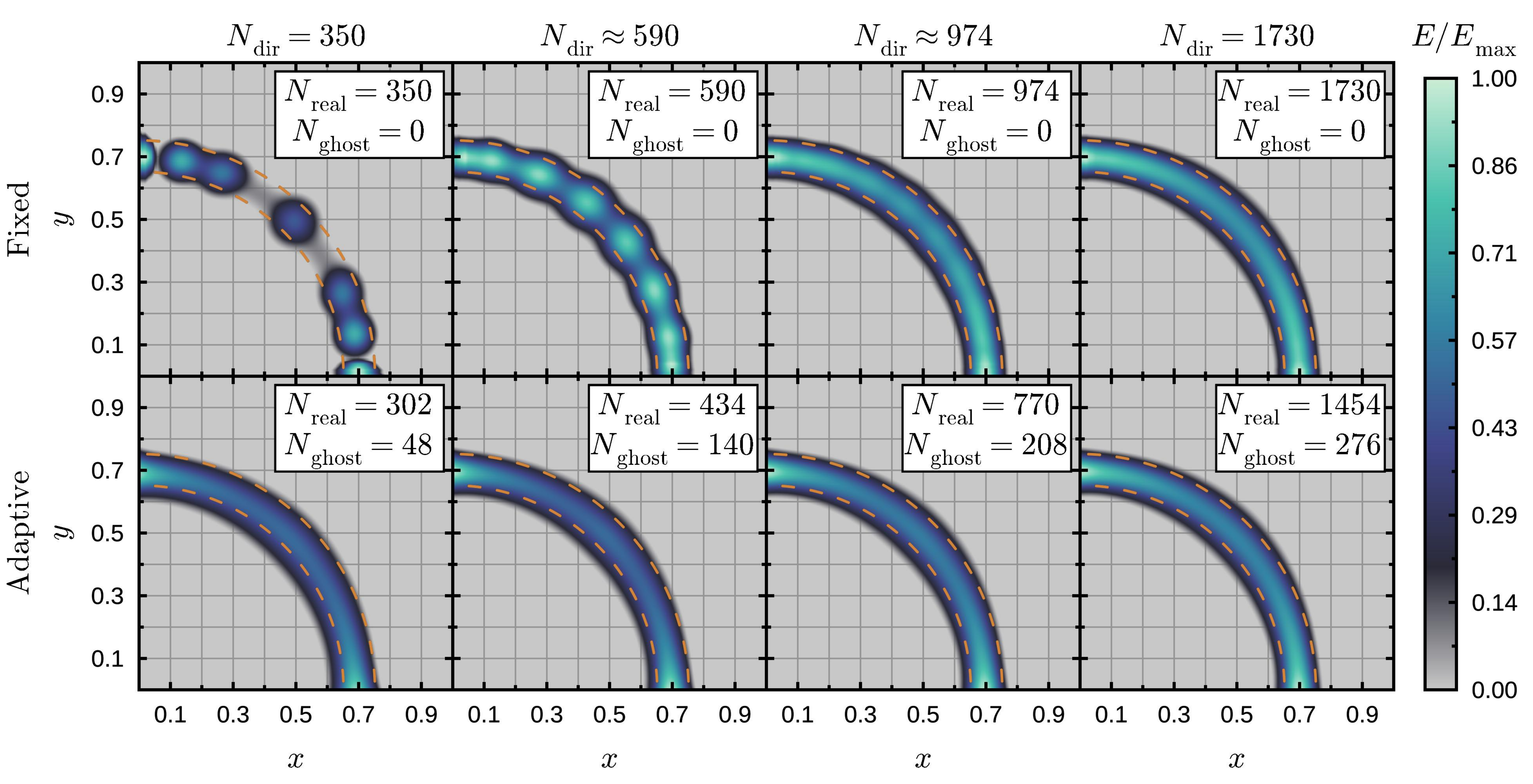}
  \caption{\textit{Top panel:} Energy distribution for the 2D
    sphere-wave test $t=0.7$. The top panels report the results for
    fixed-streaming stencils $(N_{\rm ghost}=0)$ for increasing number of
    directions $(N_{\rm real})$ from left to right. The bottom panels,
    on the other hand, show the results when employing adaptive-streaming
    stencils with the same number of total directions $(N_{\rm dir})$
    matching the top panels and highlighting that fewer directions are
    sufficient to obtain comparable solutions. The grid has extent
    $x,y\in[-1,1]$ with $dx = dy = 0.006$ ($300^2$ grid-points) and a
    timestep $dt=0.006$. Adaptive streaming uses $20\%$ ghost
    directions. \textit{Bottom panel:} The same as top panel but at
    $z=0$ of a 3D simulation with, $x,y,z\in[-0.9,0.9]$, $dx = dy = dz =
    0.01$ ($180^3$ grid-points), and $dt=0.009$. In both panels, The
    orange dashed lines indicate the analytical position of the sphere
    wave.}
    \label{fig: Sphere Wave}
\end{figure*}

Quite generically, the optically thin regime is the most challenging for
LBMs. This is true in special relativity and is made worse in curved
spacetimes. This is because, without sufficient fluid scattering, the
discretized intensities are not redistributed enough, which leads to beam
separation (see, for instance, the top left panel in left part of
Fig.~\ref{fig: Sphere Wave}). Obviously, it is possible to increase the
directional resolution to improve this behaviour, but given long enough
free-streaming paths, beam separation will tend to occur. To resolve some
of these issues, we have introduced an adaptive stencil whereby at each
grid-point we properly orient the stencils to capture the directions in
which most of the photon bundles propagate. This improves significantly
the directional resolution and hence the handling of the optically thin
regime with the GRLBM.

\begin{figure}
  \centering
    \includegraphics[width=0.45\columnwidth]{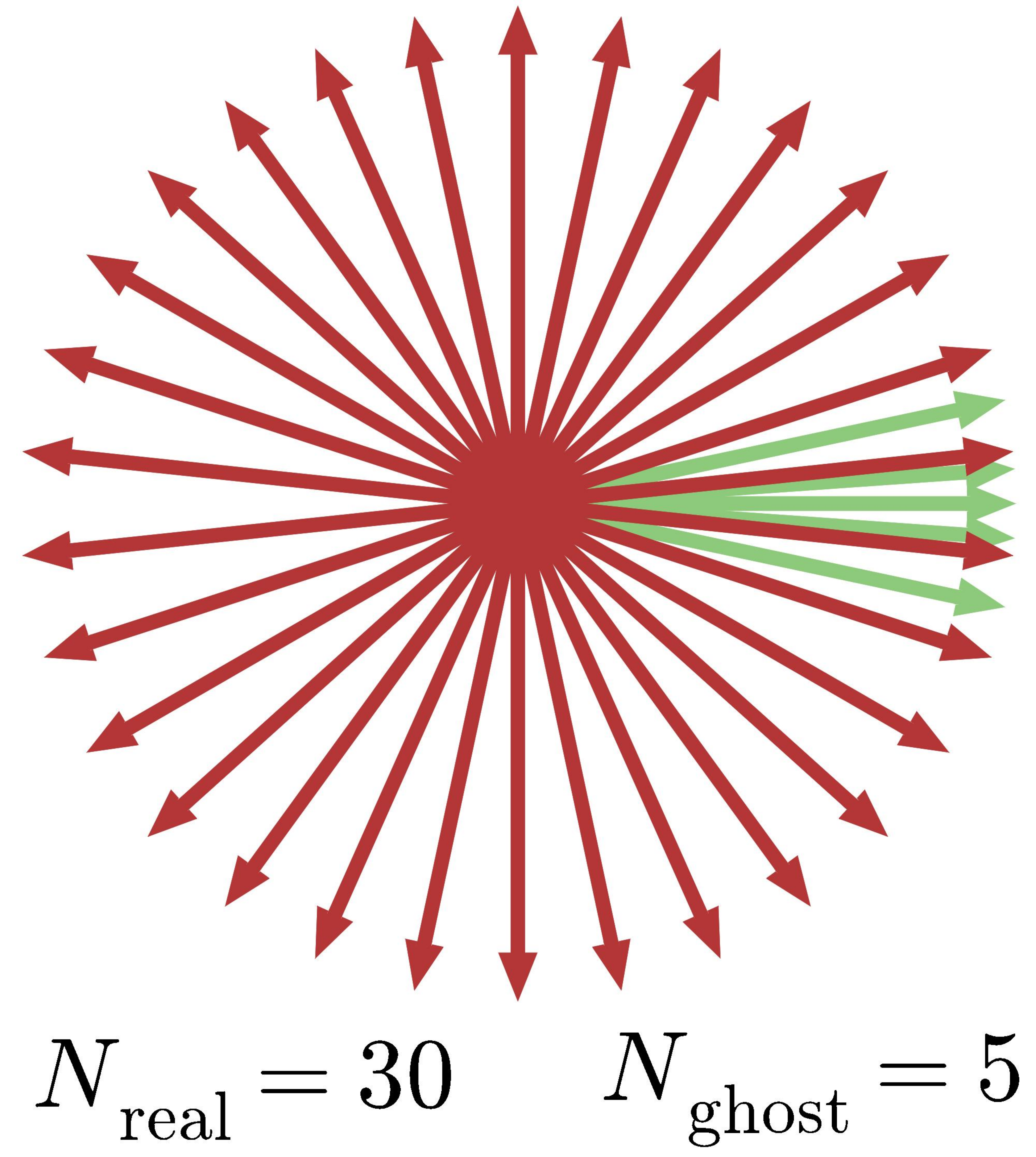}
    \hfill
    \includegraphics[width=0.45\columnwidth]{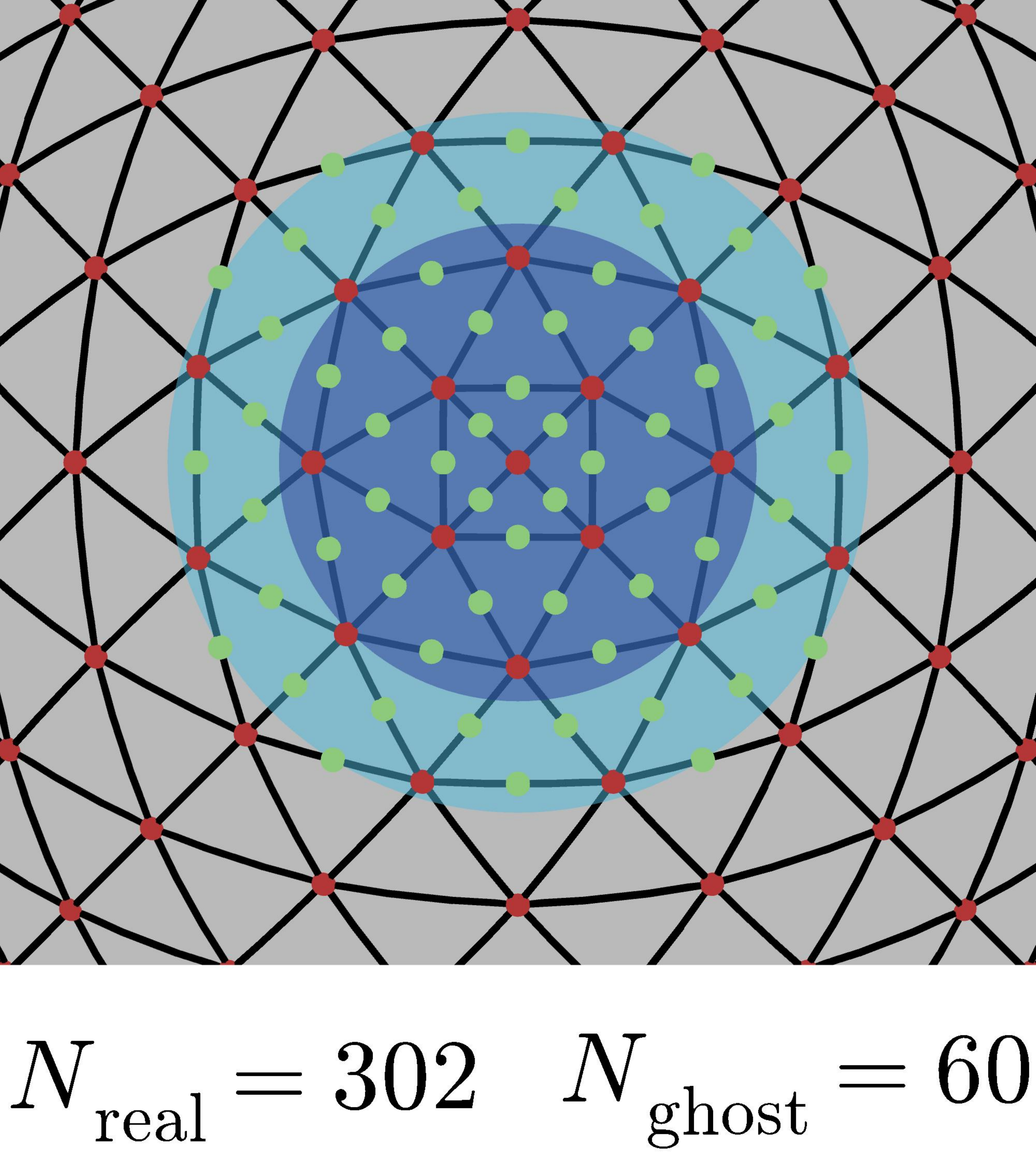}
    \caption{\textit{Left Panel:} Two-dimensional adaptive stencil with
      30 real (red arrows) and five ghost (green arrows)
      directions. \textit{Right Panel:} North pole of a three-dimensional
      adaptive stencil with two refinement levels (blue shadings). Shown
      in red are the real directions, while green is used for the
      ghost directions.}
    \label{fig: adaptive stencil}
\end{figure}

More specifically, in our adaptive stencil we start from considering the
basic Fourier and Lebedev stencil we have already discussed and we then
introduce additional ghost intensities in the ``forward'' direction of the
stencil (see Fig.~\ref{fig: adaptive stencil} for a diagrammatic
representation).

In 2D, the forward direction is defined by a vanishing polar angle, $\phi
= 0$ and ghost directions are added following to the distribution
$D(\phi) = \arccos(1-2\phi) / \pi$ in the interval $\pm{\pi}/{8}$ around
the forward direction. The function $D(\phi)$ essentially provides a
distribution of directions that is not isotropic, concentrating the
stencils around the forward direction. Using this distribution, we
determine how many ghost directions need to be added between two existing
directions and we then arrange all ghost directions between two
directions uniformly for the quadratic velocity- interpolation scheme
(see left panel of Fig.~\ref{fig: adaptive stencil} for a schematic
representation).

In 3D, the logic is very similar and we define the forward direction by a
vanishing azimuthal angle, $\theta = 0$ and introduce two refinement
levels (these are shown with light-blue and dark-blue shaded regions in
the right panel of Fig.~\ref{fig: adaptive stencil}). We then add a ghost
direction at the center of every triangle whose center lies within the
light-blue refinement level, while in the dark-blue refinement level we
add the center of each triangle edge, thus adding three ghost
directions. The two refinement levels can be used exclusively or layered
inside each other (see the right panel of Fig.~\ref{fig: adaptive
  stencil}).

Referring to the original directions as to the ``real'' directions, the
counting of direction is then simply given by $N_\text{pop} =
N_{\text{real}} + N_{\text{ghost}}$, where the ghost directions have
quadrature weights of $w_\dir = 0$ and do not contribute to the numerical
integrals. However, they contribute to the streaming step by improving
the accuracy of the velocity interpolation in the forward direction and
the corresponding flux. In addition, we allow the stencil to rotate and
align its refined forward direction with the flux (see Appendix
\ref{Appendix: Initial Data} for details). We also define a proper naming
convention to express the various options of the adaptive stencils. More
specifically, a 2D adaptive stencil is named after the number of real
directions (same as the order of stencil) and the number of ghost
directions, so that, for instance, the stencil in the left panel of
Fig.~\ref{fig: adaptive stencil} is the Fourier-30-5. Similarly, the
naming notation of the adaptive 3D stencils is given by listing the
refinement levels from low to high in terms of $\pi$, so that, for
example, the adaptive stencil shown in the right panel of Fig.~\ref{fig:
  adaptive stencil} is referred to as Lebedev-29-0.15-0.1, with the first
refinement level at $0.15\pi$, and the second level at $0.1\pi$.

While the advantages in terms of direction resolution offered by the
adaptive stencil are obvious, a final remark should be made on the
related additional costs. Within a GRLBM scheme, where the velocity
interpolation is already intrinsically present, the use of adaptive
stencils does not lead to any significant additional computational
cost. This is not the case for the SRLBM scheme, where we would need to
introduce a velocity interpolation, drastically impacting the
computational complexity. A more detailed cost analysis can be found in
Section \ref{subsec: Performance Analysis}.

\begin{figure*}
    \centering
    \includegraphics[width=0.65\textwidth]{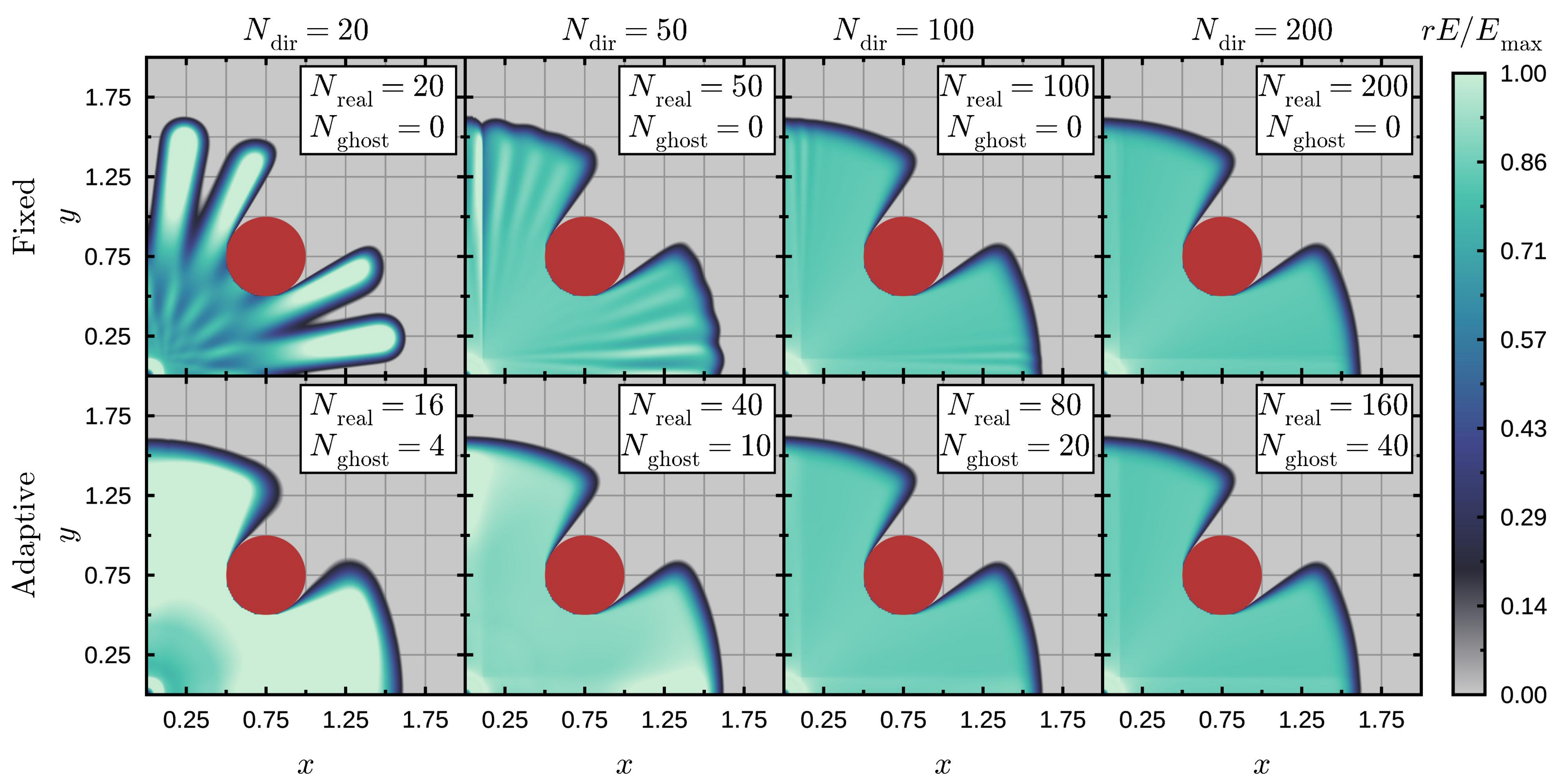}
    \vskip 0.25cm
    \includegraphics[width=0.65\textwidth]{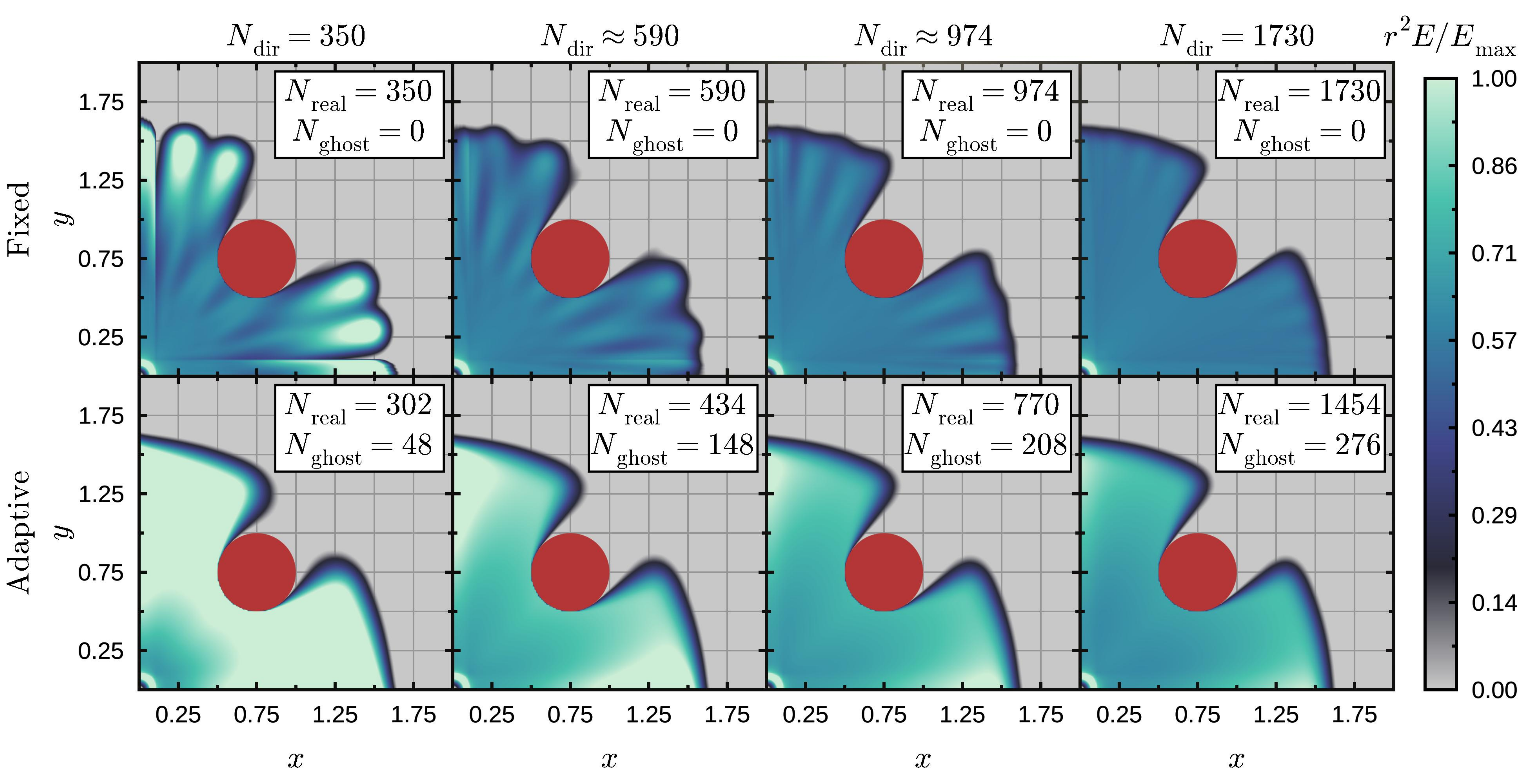}
    \caption[Shadow casting 2D and 3D.]{The same as in Fig.~\ref{fig:
        Sphere Wave} but for the shadow-casting problem at $t=1.5$ and
      where the colormap reports the rescaled energy density, \ie $r^{}E$
      ($r^{2}E$) so that the solution should be constant in 2D (3D). The
      2D data (top panel) refers to a grid with $x,y\in[-0.2,1.7]$, $dx
      = dy = 0.01$ ($190^2$ grid-points), and $dt=0.009$. The 3D data
      (bottom panel), on the other hand, refers to a grid with
      $x,y,z\in[-0.2,1.7]$, $dx = dy = dz = 0.01$ ($190^3$ grid-points),
      and $dt=0.009$. In all cases, the red-filled circle indicates the
      position of the sphere casting the shadow.}
     \label{fig: Shadow}
\end{figure*}

\section{Numerical Tests}
\label{sec: Implementation Tests}

This section is dedicated to the presentation of a series of standard and
non-standard tests in 2D and 3D aimed at validating the correctness of
the GRLBM described above and to measure its performance and a variety of
conditions and choices for the stencil order. We will start with
considering the flat-spacetime tests by~\citet{Weih2020c} and then move
on to curved (but fixed) spacetimes. For consistency, we set the CFL
factor to $0.9$ for all tests presented here.

\subsection{Flat Spacetime Tests}
\label{subsec: Flat Spacetime Tests}

\subsubsection{Sphere Wave}
\label{subsubsec: Sphere Wave}

This test consists of a spherical light pulse propagating radially
outwards in flat spacetime with initial data given by $E = 1$ and $F^i =
0$ inside a sphere of radius $r\leq0.1$. The top panel of Fig.~\ref{fig:
  Sphere Wave} reports the results of 2D calculations at $t=0.7$, with
the different sub-panels reporting the results for different combination
of the real and ghost directions $N_{\rm real}$ and $N_{\rm ghost}$ (we
report a single quadrant but the calculations do not enforce any
symmetry). When concentrating on the top portion with fixed (or uniform)
stencil ($N_{\rm ghost}=0$), it is possible to note that beam separation
occurs when using $20$ or $50$ real directions. Increasing them to 100
directions leads to an improvement, but artefacts in the energy
distribution can still be found right next to the coordinate axes. Such
artefacts, which are due to the spatial interpolation being trivial along
the principal directions, can be further moderated when using 200
directions, which leads to a propagating shell that is mostly uniform in
the polar direction. When using the adaptive stencil, as shown in the
bottom portion of the top panel of Fig.~\ref{fig: Sphere Wave}, we
obtain a uniform propagating shell already with only $N_{\rm real}=16$
and $N_{\rm ghost}=4$ the quality of the result
increases as $N_{\rm real}$ and $N_{\rm ghost}$ are increased. The
improvements tend to saturate between 100 and 200 directions and going
beyond 200 directions does not improve the results significantly.

The bottom panel of Fig.~\ref{fig: Sphere Wave} reports the equivalent
results of 3D calculations restricted to the plane with $z=0$, where it
is possible to appreciate that the beam separation with the fixed stencil
(top sub-panels) is more pronounced and still clearly visible with $590$
directions. Overall, also in 3D, the adaptive stencil and the addition of
ghost directions (bottom sub-panels) produces a much smoother and
homogeneous distribution of the energy density, while retaining the error
along the principal directions. Since the interpolation is intrinsically
more diffusive in 3D while it remains very accurate near the coordinate
axes, it is not surprising that the accumulation of energy along
these axes is even more pronounced in the 3D simulations. All in
all, this test in the free-streaming regime shows that the adaptive
stencil provides a significant improvement over the fixed stencil: it
yields the needed isotropy and reduces the computational costs.

\subsubsection{Shadow Casting}
\label{subsubsec: Shadow Casting}

For the following test, we use the same initial data but keep the
intensities in the sphere of radius $r_s<0.1$ fixed to the initial
values, giving a constant emitting light source similar to a star. We
also introduce an optically thick sphere of radius $r_p=0.25$ at position
$(x_p,y_p,z_p) = (0.75, 0.75, 0.00)$, with an absorption coefficient
$\kappa_a=10^{10}$, effectively absorbing all radiation.

\begin{figure*}
  \centering
  \includegraphics[width=0.65\textwidth]{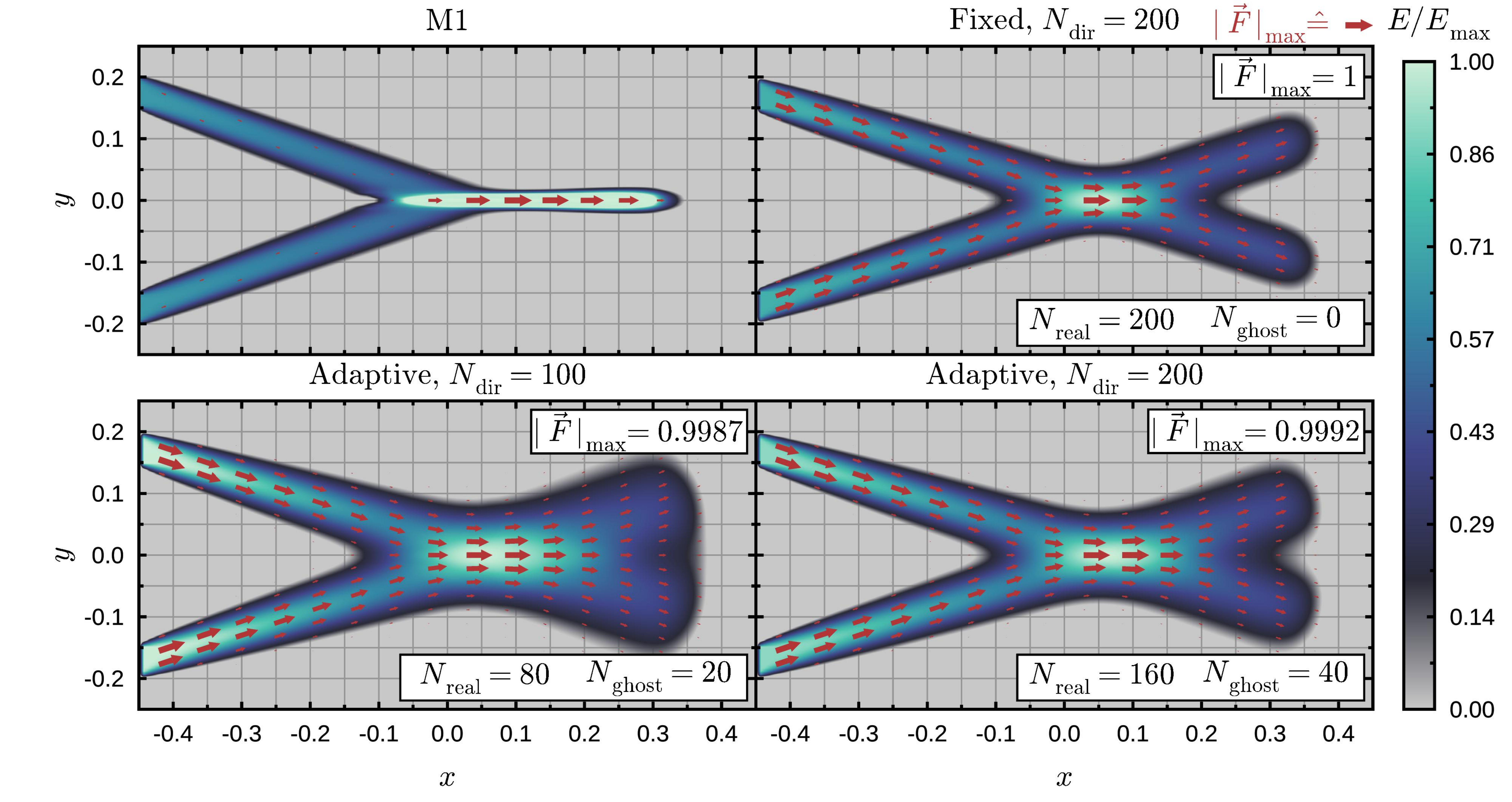}
  \vskip 0.25cm
  \includegraphics[width=0.65\textwidth]{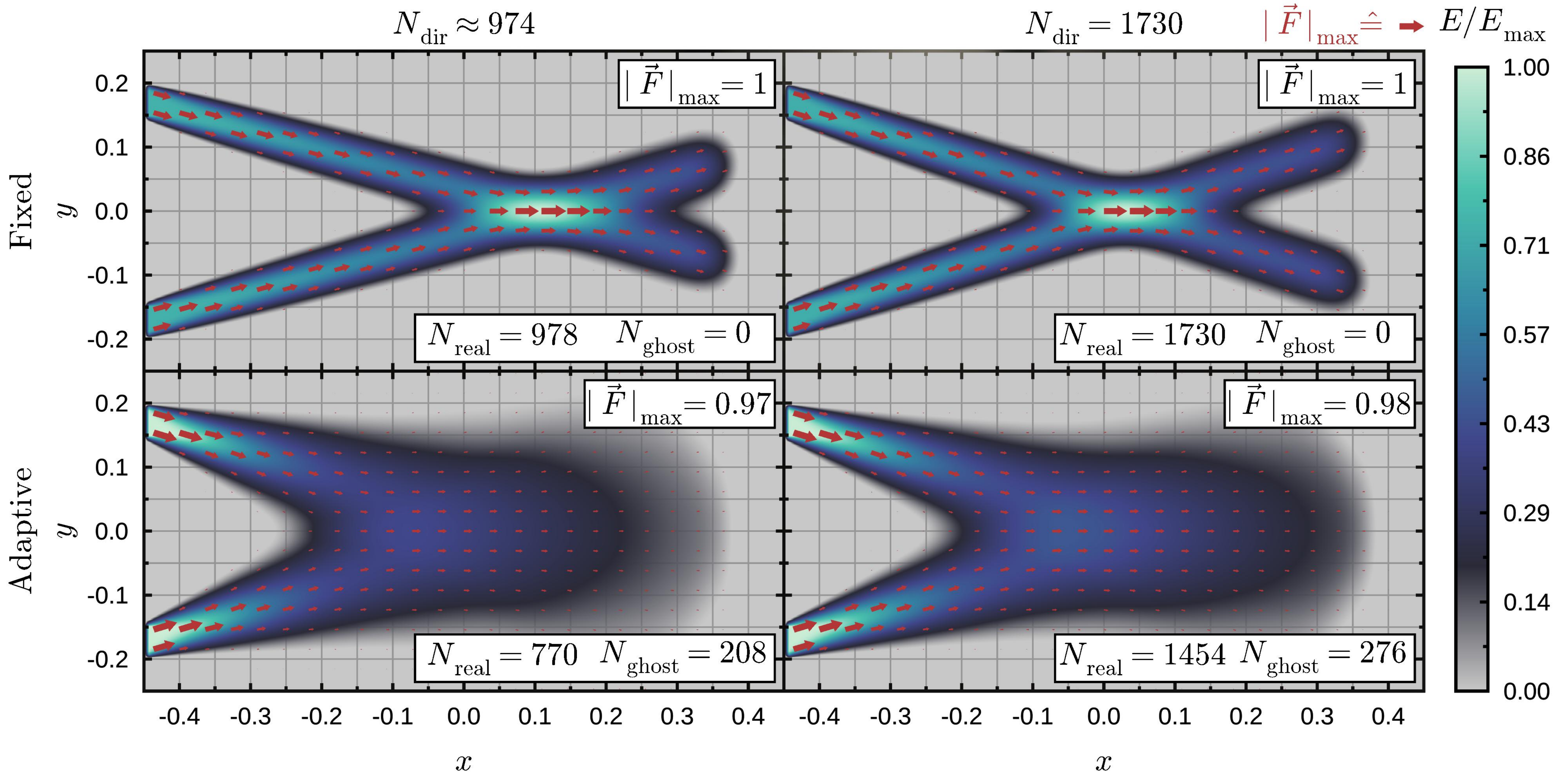}
  \caption[Beam crossing 2D and 3D.]{The same as in Fig.~\ref{fig: Sphere
      Wave} but for the beam-crossing problem at $t=0.85$. The top-left
    panel reports the solution with the M1 approach, which obviously
    fails in this test. The 2D data (top panel) refers to a grid with
    $x\in[-0.5, 0.5]$, $y\in[-0.25, 0.25]$, $dx = dy = 0.005$
    ($200\times100$ grid-points), and $dt = 0.0045$. The 3D data (bottom
    panel), on the other hand, refers to a grid with $x\in[-0.5, 0.5]$,
    $y\in[-0.25,0.25]$, $z\in[-0.125,0.125]$, $dx = dy = dz = 0.005$
    ($200\times100\times50$ grid-points), and $dt=0.0045$.}
      \label{fig: Beam Crossing}
\end{figure*}

Figure \ref{fig: Shadow} shows the results for the 2D and 3D simulations,
respectively. Note how in the 2D case (top panels), a shadow is
cast downstream of the sphere, while elsewhere the energy density
falls-off like $1/r$, as expected. Furthermore, when using a fixed
stencil (top portion of the top panel), it is possible to notice evident
beam-separation artefacts when using only 50 directions, but also that
these are considerably suppressed when increasing to 100 directions,
leaving some inaccuracies only along the principal directions when using
200 directions. At the same time, the shadow cast when employing the
adaptive streaming (bottom portion of the top panel) is already very
sharp with only $N_{\rm real}=40$ and $N_{\rm ghost}=10$ directions and
the solution further improves with 100 directions, being comparable to
that with twice as many fixed directions and saturating after that.

When considering the 3D simulations (bottom panels), this test shows all
of its complexity and challenges when using a fixed stencil. While a
shadow is cast already with few directions, the quality of the
energy-density solution is poor and does not improve significantly when
doubling the number of directions. With adaptive streaming, the shadow
casting is much better, but the inverse square law seems broken due to
the more substantial bias toward the principal directions. In
Sec.~\ref{subsubsec: Radiating Star}, we will look closer at the inverse
square law and see that the discrepancy is not as severe as it might
seem.

\subsubsection{Beam Crossing}
\label{subsubsec: Beam Crossing}

While the two previous tests provide ideal testbeds to highlight the
advantages of the adaptive stencil, due to most of the light at one grid
cell moving in the same direction, this is not the case for another
standard test, namely, the beam-crossing test. We recall that this test
amounts to evolving two beams of radiations along directions that
eventually cross and interact leading to a local increase in the energy
density. After the crossing, the two beams should continue their motion
along the initial direction of propagation, but this is not always
reproduced by radiative-transfer approaches, for which this represents a
very challenging test. A classical example of this failure is given by
the M1-moment scheme, where the photon momenta are actually linearly
combined and after the crossing a single beam is produced propagating in
the combined direction. Indeed, the ability of successfully perform this
test has been reported only for more advanced approaches, such as the
SRLBM~\citep{Weih2020c} or MonteCarlo approaches~\citep{Foucart2017}.

We set up two mono-directional constantly emitting beams with the same
energy density $E=1$ and flux density norm $|F^i|=1$. The beams are
emitted at the boundary cells where $x=-0.5$, with a flux direction of
components $(0.3, 0.1)$ and $(0.3, -0.1)$ between $y\in[-0.2,-0.15]$ and
$y\in[0.15, 0.2]$ respectively. Note that the initial data is not
identical for all the stencil configurations.  When using the adaptive
stencil, a delta-like intensity distribution where only a single
direction holds any intensity while all others are zero leads to high
interpolation errors in velocity space. Depending on the resolution, each
stencil has its own maximal intensity distribution, and therefore a
maximum value of $|F^i|$ it can resolve without interpolation errors
becoming too big (see Appendix \ref{Appendix: Initial Data} for a
detailed discussion of this issue). As a result, higher-order stencils,
especially stencils with more ghost directions and/or higher levels of
refinement, have a higher initial flux density. This artificial maximum
flux boundary does not limit the fixed streaming method. Instead, we find
the direction vector closest to the desired direction in our stencil and
set the corresponding intensity to one and all others to zero.

Figure~\ref{fig: Beam Crossing} shows the results of the beam-crossing
test, again reporting the outcome of the 2D simulations in the top panels
and those of the 3D simulations in the bottom panel. Concentrating on the
former first, we show in the top-left sub-panel the solution obtained
with the M1-moment scheme as computed by~\citet{Weih2020c}. Note how
after the crossing the two beams merge into one, therefore failing the
test. The remaining sub-panels on the top of Fig.~\ref{fig: Beam
  Crossing} show the results of our novel GRLBM, with the top-right one
being with a fixed stencil and the bottom sub-panels showing the results
with adaptive stencils. Clearly, in all cases the test is passed already
with $N_{\rm real}=80, N_{\rm ghost}=20$ directions and the solution
improves as the number of directions is increased.

\begin{figure*}
    \centering
    \includegraphics[width=0.45\textwidth]{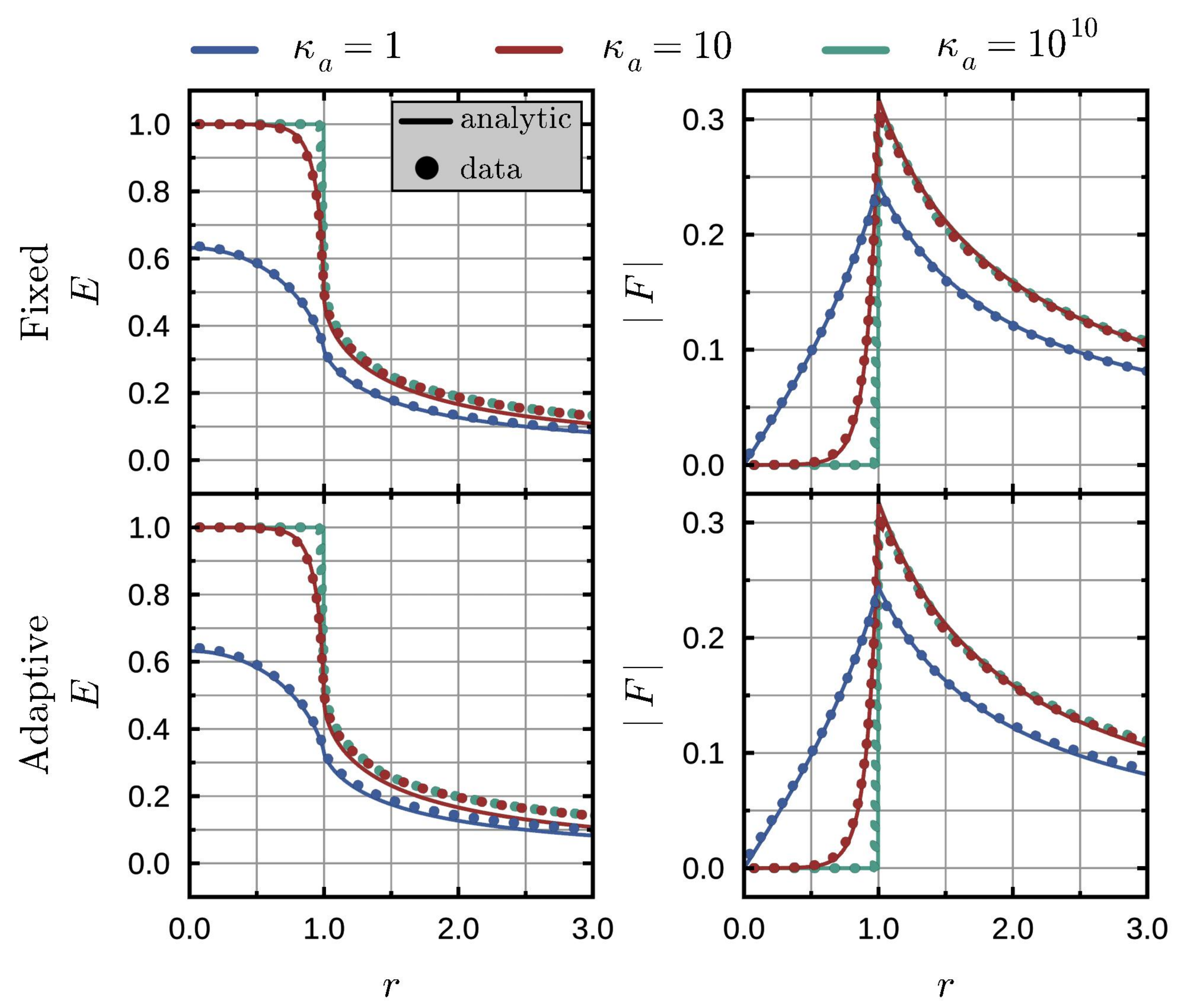}
    \hskip 1.0cm
    \includegraphics[width=0.45\textwidth]{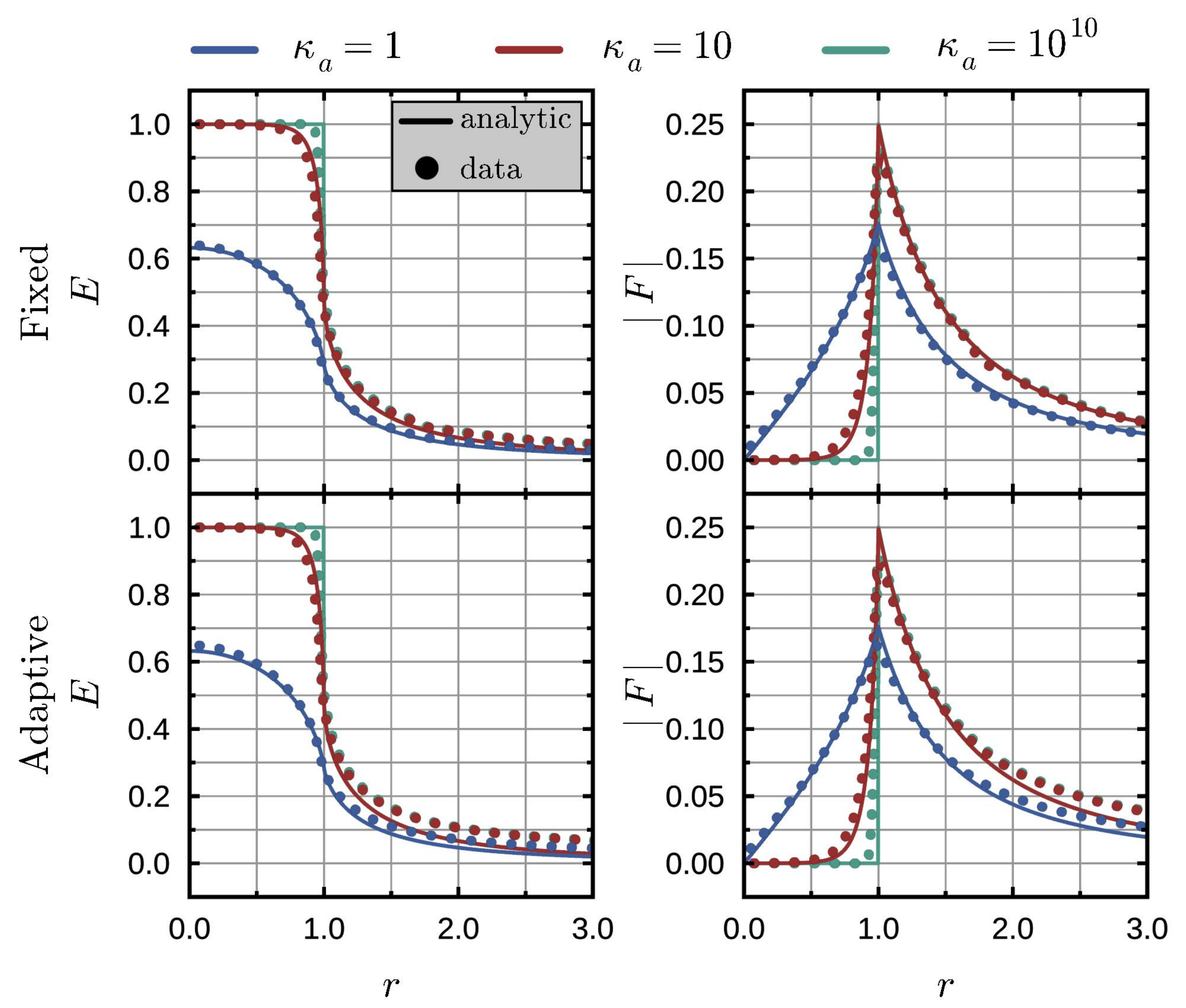}
    \caption[Radiating star 2D and 3D.]{Results of the radiating-sphere
      test reporting with filled circles of different colours the
      numerical 1D profiles for the energy (left part) and the and flux
      density (right part) for different values of $\kappa_a = \eta$;
      shown with solid lines are the corresponding analytic
      solutions. Both in the left and in the right panels, the top
      sub-plots refer to the fixed streaming, while the bottom ones to
      the adaptive streaming. The 2D data (left panel) refers to a grid
      with $x,y\in[-4, 4]$, $dx = dy = 0.02$ ($400^2$ grid points), and
      $dt=0.018$. The 3D data (right panel), on the other hand, refers to
      a grid with $x,y,z\in[-4, 4]$, $dx = dy = dz = 0.05$ ($160^3$ grid
      points), and $dt=0.045$.}
    \label{fig: Star}
\end{figure*}

The bottom panel of Fig.~\ref{fig: Beam Crossing} shows instead the
results in 3D which are much poorer and that the velocity-space
resolution is too low to resolve sharp, distinct beams with the adaptive
approach. Indeed, even with an adaptive Lebedev stencil of order $p_{\rm
  Leb}=65$ with $1454$ real and $276$ ghost directions, the diffusion is
so pronounced that the beams hardly cross.  We believe this behaviour is
mostly the result of the ``mono-directional'' prescription in which our
adaptive stencil is implemented. More precisely, in the present approach
the adaptive stencils aligned naturally along the direction of
propagation of the radiation, which is assumed to be \text{only
  one}. Since before crossing there is a single direction of propagation
for each beam, the adaptive approach works very well and leads to two
distinct and sharp beams. At the crossing, however, the radiation field
will have two distinct directions of propagation and the mono-directional
adaptive stencil will adapt to their average, leading to a wrong
alignment. As a result the stencil has less resolution in the beam
directions, which leads to additional numerical diffusion in the velocity
space interpolation.

While this phenomenology indicates that the use of a mono-directional
adaptive stencil is not satisfactory in conditions where the radiation is
not propagating along a main direction, it also provides a useful hint on
how to improve it. In particular, much of the diffusion can be removed by
making the adaptive stencil even more adaptive in at least two different
ways. First, a new implementation can be made in which the adaptive
stencil is either ``switched'' on or off depending on the local
conditions of the radiation field. Second, a different approach could
consist in decomposing the angular distribution intensity more finely and
set the adaptive stencil not to follow a single direction but the two
directions in which the angular distribution of the intensity is
peaked. This would essentially transform the
mono-directional adaptive stencil developed here into a multi-directional
one. Since these algorithmic modifications affect a considerable part of
the numerical infrastructure, we have decided to explore this possibility
in future work.

\subsubsection{Radiating Sphere}
\label{subsubsec: Radiating Star}

The next test corresponds to a dense sphere with a sharp boundary to
vacuum, radiating constantly and homogeneously from its
surface~\citet{Smit:97}. Hence, as initial data we set the emissivity and
absorption opacity to be constant and equal $\eta = \kappa_a$ inside a
sphere of radius $r < 1$. Eventually, this system will reach a steady
state for which the analytic solution for the distribution function is
\begin{align}
&f(r,\mu) = 1 -
    \exp\left[-\kappa_a\left(r\mu + \sqrt{1 - r^2(1 -
        \mu^2)}\right)\right]\,,
\end{align}
and
\begin{align}
&    f(r,\mu) = 1 -
    \exp\left[-2\kappa_a\sqrt{1 - r^2(1 - \mu^2)}\right]\,,
\end{align}
for $r < 1, \mu\in[-1,1]$ and for $r \geq 1, \mu\in[\sqrt{1 -1/r^2},1]$,
respectively, and where $\mu := \cos\theta$. The moments can then be
computed by integrating the distribution functions and are given by
\begin{align}
    E(r) &=
    \frac{1}{\pi}\int_{\mu_{\text{min}}}^{\mu_{\text{max}}}\frac{f(r,\mu)}{\sqrt{1
        - \mu^2}} d\mu, & F(r) &=
    \frac{1}{\pi}\int_{\mu_{\text{min}}}^{\mu_{\text{max}}}\frac{\mu
      f(r,\mu)}{\sqrt{1 - \mu^2}} d\mu\,,
\end{align}
and
\begin{align}
    E(r) &= \frac{1}{2}\int_{\mu_{\text{min}}}^{\mu_{\text{max}}}f(r,\mu)
    d\mu, & F(r) &=
    \frac{1}{2}\int_{\mu_{\text{min}}}^{\mu_{\text{max}}}\mu f(r,\mu)
    d\mu\,,
\end{align}
in 2D and 3D, respectively.

Figure~\ref{fig: Star} shows the results for the 2D (filled circles in
the left panels) and 3D simulations (filled circles in the right panels)
for different values of the absorption opacity. In both cases, the
numerical results closely follow the analytical solution (solid lines of
different colour) inside the radiating sphere where $\kappa_a = \eta \neq
0$, with relative errors in the energy density below $0.0062\%~(0.023\%)$ and
flux density below $0.234\%~(0.233\%)$ with the fixed (adaptive)
stencil. Outside the sphere, in the free-streaming region, both methods
tend to have an energy density that is slightly larger than the analytical one.
The error is smaller for the fixed stencil than for the adaptive stencil
$1.06\%$ and $1.81\%$ respectively. A similar behaviour is shown also by the flux
density, that has a smaller error with the fixed stencil, while the
adaptive approach overshoots slightly, although the relative error is
$0.043\%$ and $0.046\%$ respectively. Very similar behaviours are shown
also in the case of full 3D simulations, where the relative errors
are slightly larger but mostly because of the smaller spatial resolution,
\ie $160^3$ grid-points to be contrasted with the $400^2$ grid-points of
the 2D simulations.

Overall, this test shows that both the fixed and the adaptive stencil
produce correct and accurate results in optically intermediate to thick
regimes. It also demonstrates that the fixed stencil is more accurate in
the free-streaming limit than the adaptive counterpart both in 2D and in
3D at least in scenarios with very high symmetry, as the one considered
here. Under more general conditions, however, we expect the adaptive
stencil to provide comparable if not better accuracy.

\begin{figure*}
  \centering
    \includegraphics[width=0.49\textwidth]{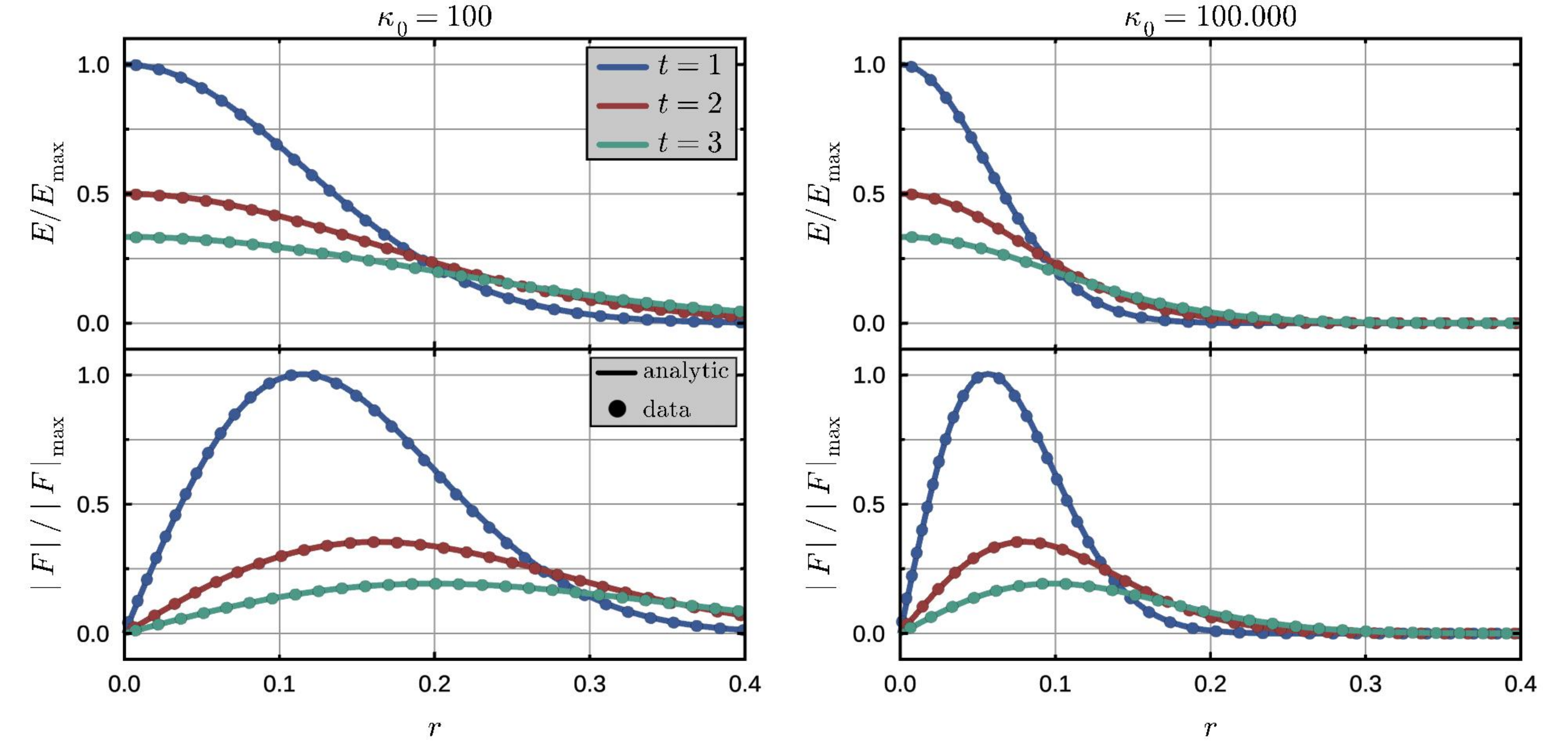}
    \hskip 0.25cm
    \includegraphics[width=0.49\textwidth]{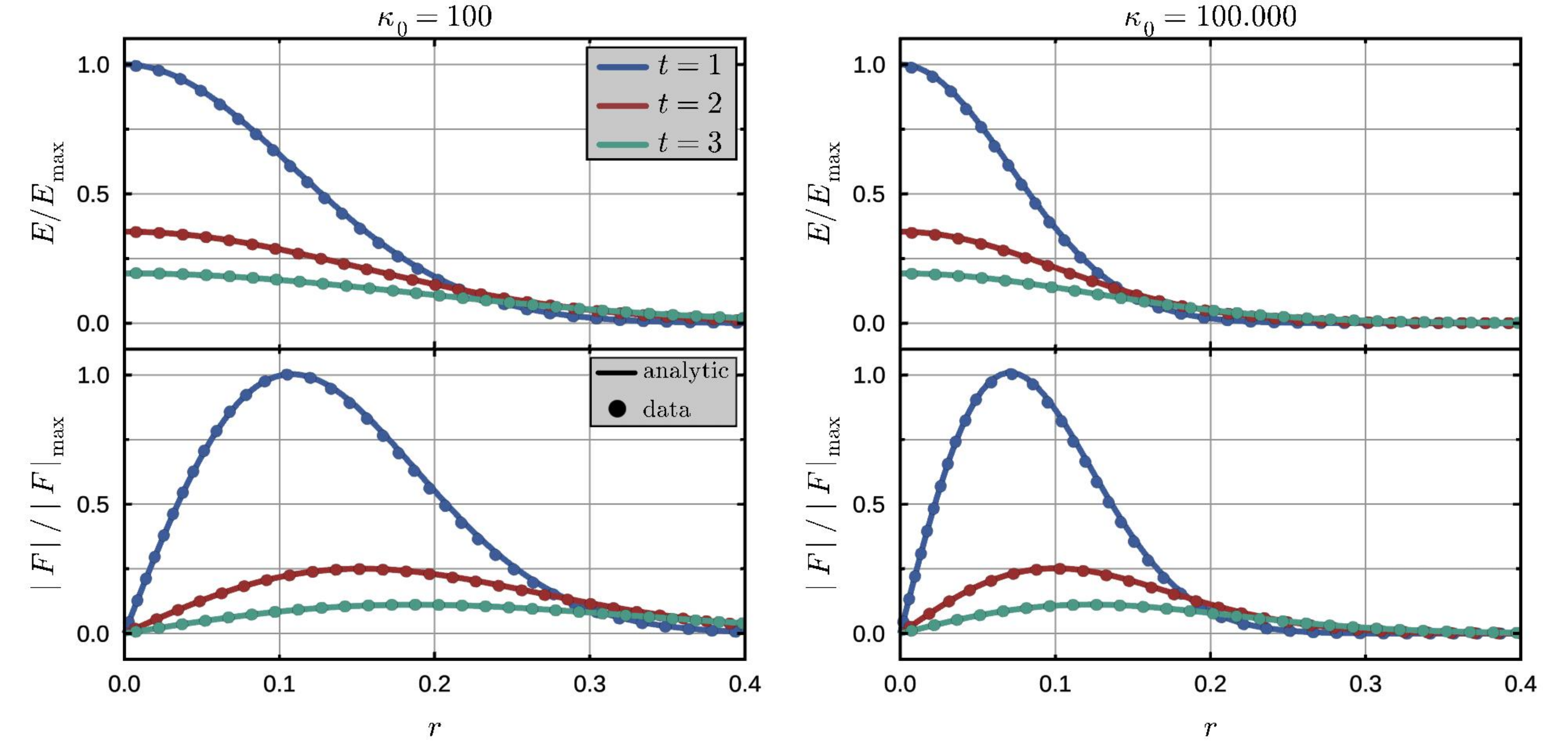}
    \caption[Diffusion 2D and 3D.]{Shown in the left and right panels are
      the energy density (top rows) and flux density (bottom rows) of the
      radiation-diffusion test at times $t=1,2,3$; numerical data is
      shown with filled circles, while solid lines are used for the
      analytic solution. For each panel, the left (right) column refers
      to an opacity $\kappa_0=10^2$ ($\kappa_0=10^5$); absorption,
      emission, and forward scattering are set to zero $\kappa_a = \eta =
      \kappa_1 = 0$. The left panel refers to 2D simulations on a grid
      with $x,y\in[-0.5, 0.5]$, $dx = dy = 0.005$ ($200^2$ grid-points),
      and $dt = 0.0045$, while the right panel reports the results of the
      3D simulations on a grid with $x,y,z\in[-0.5, 0.5]$, $dx = dy = dz
      = 0.01$ ($100^3$ grid-points), and $dt = 0.0009$. Because of the
      different resolutions, the 2D and 3D results refer to
      $\mathscr{Pe}=1$ (left column), $\mathscr{Pe}=1000$ (right column)
      and $\mathscr{Pe}=0.5$ (left column), $\mathscr{Pe}=500$ (right
      column), respectively.}
    \label{fig: Diffusion}
\end{figure*}

\subsubsection{Static and boosted diffusion}
\label{subsubsec: Diffusion}

So far, all our tests have only tested pure streaming, absorption, and
emission. To properly test if the collision step still behaves correctly
even with the addition of ghost directions, we test the scattering regime
in the diffusive limit of the radiative-transfer equation. Following the
standard approach for a static diffusion test~\citep{Pons2000,
  Kuroda2016, Weih2020b, Radice2022}, we set our initial data for the
energy density to a Gaussian according to the analytical solution at
$t_0=1$
\begin{align}
\label{eq:diffusion_E}
E(r,t) &= {t^{-d/2}} \exp\left(-\frac{r^2}{4\mathscr{D}t}\right)\,, \\
\label{eq:diffusion_F}
F(r,t) &= \frac{r E}{2 t (1 + a\, \mathscr{Pe})}\,, \\
\mathscr{D} &= \frac{1}{d\kappa_0} (1 + a\, \mathscr{Pe})\,,
\end{align}
where $\mathscr{D}$ is the diffusion coefficient and includes corrections
for additional numerical diffusion proportional to the P\'eclet number
$\mathscr{Pe}$, where $a=0.64~(a=0.75)$ in 2D (3D) simulations, and $d\in\{2,3\}$
the number of spatial dimensions. 

The diffusion test is also the first instance in which the
Lambda-Iteration in the collision step [see Eq.~\eqref{eq:lambda_it} and
  related discussion] has to be employed and iterated for multiple
steps. In the most extreme case, \ie of a 3D simulation with $\kappa_0 =
10^5$, we measured a maximum number of $100$ and an average of $3.69$
iteration steps per timestep; for all other simulations, we measure
smaller values both the maximum and the average number of iteration
steps. It is also worth remarking that our initial-data approach only
sets a value for the energy and flux density, with no control over the
pressure tensor $P^{ij}$, which is calculated after the first iteration
(see Appendix \ref{Appendix: Initial Data} for details). As a result, the
initial value of the pressure are ``inconsistent'' with the prescriptions
of $E$ and $F$ and the first couple of timesteps are needed to drive the
pressure tensor to consistent values. As a result, the initial steps are
also those where the number of steps in the Lambda-iteration is the
largest, dropping drastically once the pressure is consistently computed.

Figure~\ref{fig: Diffusion} shows the evolution of the energy and flux
density of the static-diffusion test either in 2D (left panel) or in 3D
(right panel), reporting with filled circles the numerical solutions and
with solid lines of the same colour the corresponding analytic
solutions. All results refer to the adaptive-streaming approach and show
that numerical values match the analytical solution very well, even for
the extreme case of $\kappa_0 = 10^5$, with a maximum relative error of
$0.016\%$. Not shown in Figure~\ref{fig: Diffusion} are
the results when employing the fixed-streaming approach, as the results
look indistinguishable from the adaptive streaming and the relative error
is of the same order.

\begin{figure}
    \centering
    \includegraphics[width=\columnwidth]{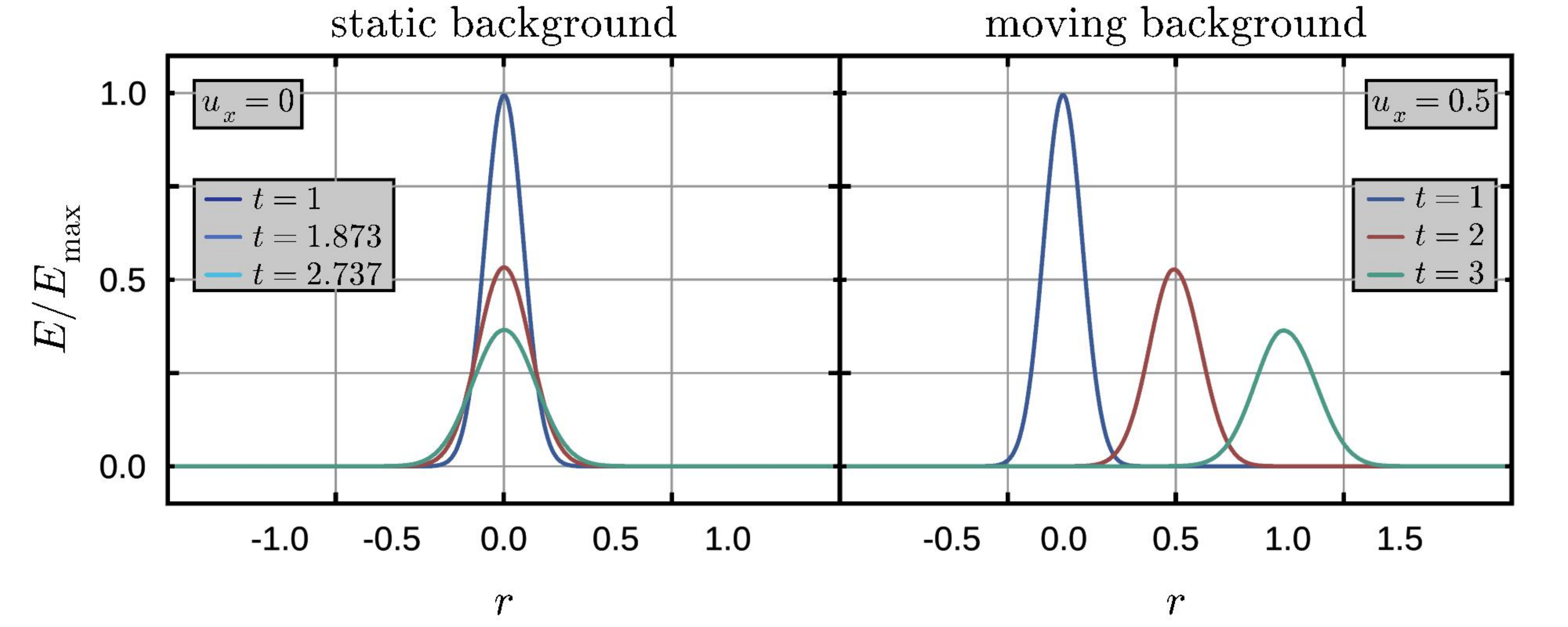}
    \includegraphics[width=\columnwidth]{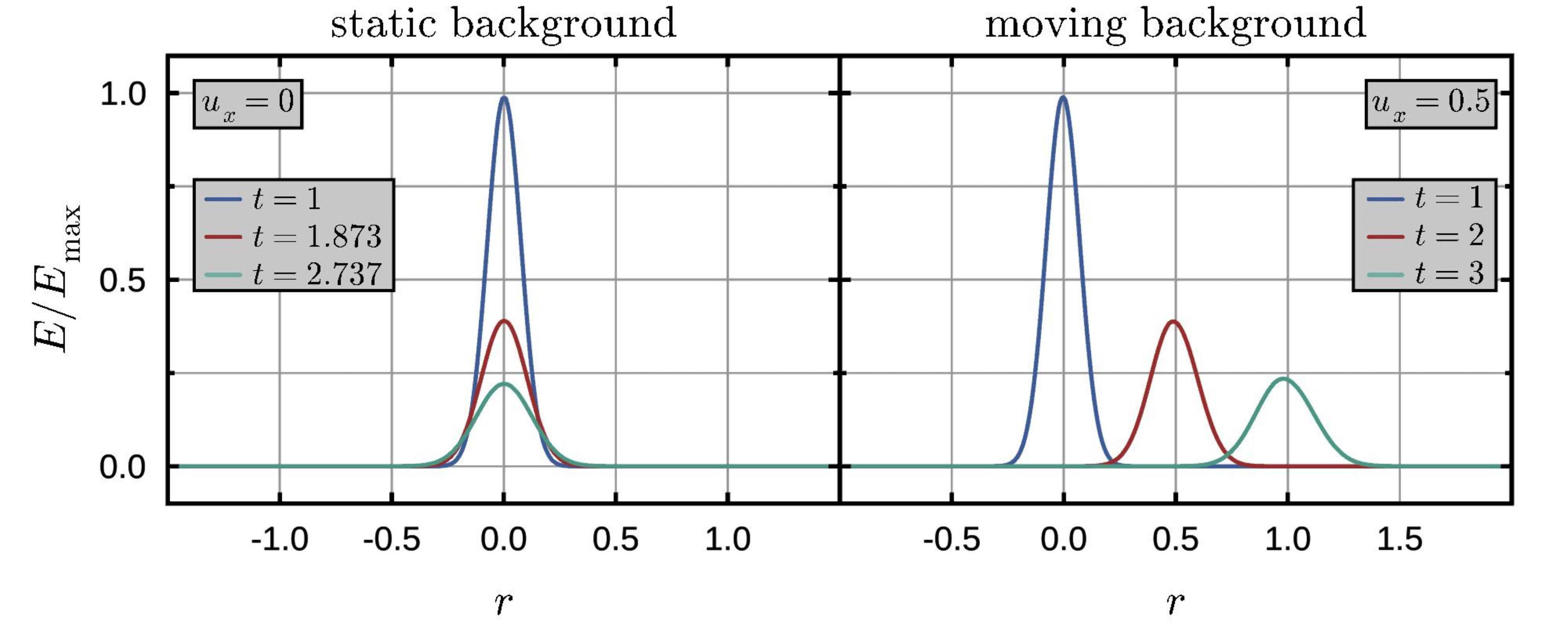}
    \caption[Moving Diffusion 2D and 3D.]{Moving radiation-diffusion test
      at three different times and $\kappa_0 = 1000$; the left columns
      report the values of the normalised energy density in the FF (where
      the fluid is at rest $u_x=0$) while the right columns those in the
      LF (where the fluid moves with $u_x=0.5$). The top panel refers to
      the 2D simulations on a grid with $x\in[-1,2]$, $y\in[-0.5, 0.5]$,
      $dx = dy = 0.001$ ($300\times100$ grid-points), and $dt = 0.0009$,
      while the bottom panel to the 3D data on a grid with $x\in[-1,2]$,
      $y,z\in[-0.5, 0.5]$, $dx = dy = 0.001$ ($300\times100^2$
      grid-points), and $dt = 0.0009$.}
      \label{fig: Moving Diffusion}
\end{figure}

A more challenging diffusion test can be made when considering the
diffusion in a moving medium. In this case, we follow~\citet{Radice2022}
and \citet{Musolino2023}, and repeat the diffusion test in a moving
background fluid with $u_x = 0.5$. For the initial data at $t=1$, we use
the same energy and flux density as in the previous diffusion test and to
account for the moving fluid, we need to Lorentz-boost the initial
energy-momentum tensor $\tilde T^{\mu\nu}$ from the FF to the LF with
$T^{\mu\nu}$. Since we need the entire energy momentum tensor, we also
require the (isotropic) pressure in the FF, which takes the diagonal form
$\tilde P^{ij}/\tilde E = \delta^{ij}/\mathscr{D}$ in the diffusive
limit. Considering the additional numerical diffusion, this leads to the
following pressure tensor
\begin{align}
    \tilde P^{ij} &= \frac{\tilde E}{\mathscr{D}(1 + a\,
      \mathscr{Pe})} \delta^{ij}\,.
\end{align}
Note that time dilation in terms of the inverse Lorentz factor
$\gamma^{-1} = \sqrt{1-u_x^2} = 0.866$ needs to be taken into account
when comparing the results in the LF and in the FF.

\begin{figure*}
    \centering
    \includegraphics[width=0.9\textwidth]{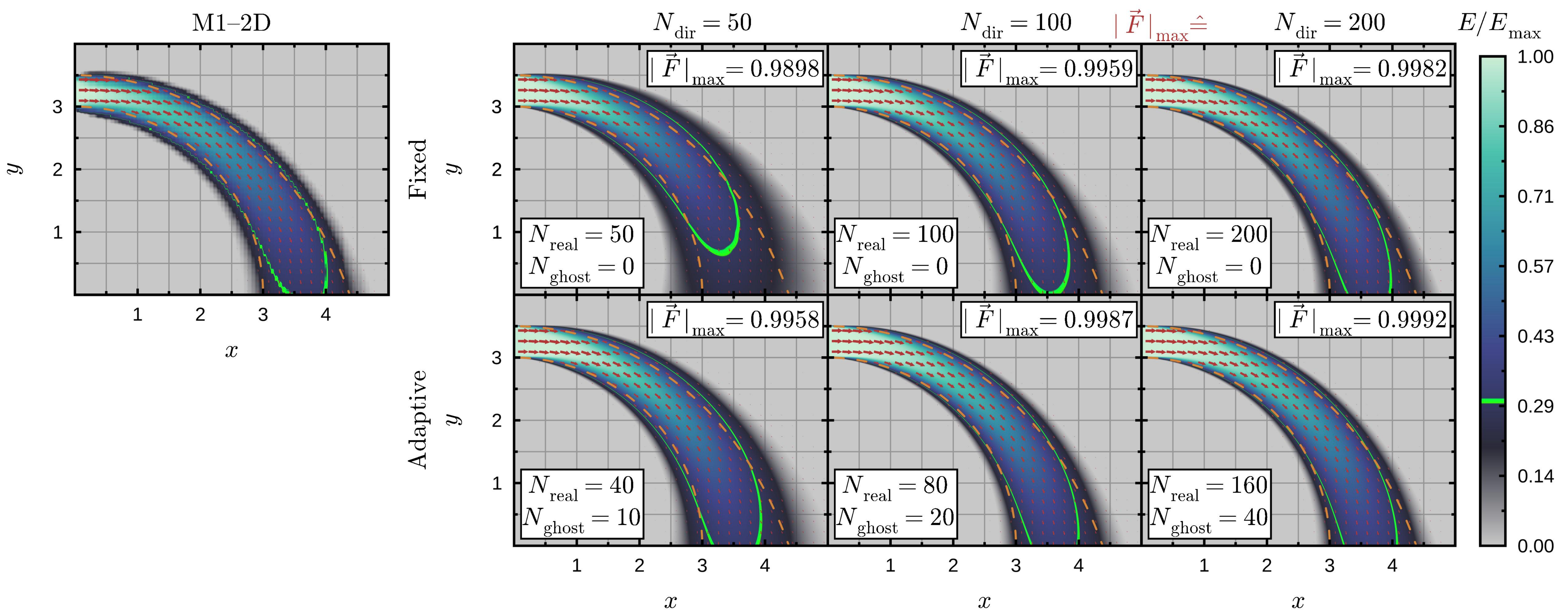}
    \vskip 0.25cm
    \includegraphics[width=0.9\textwidth]{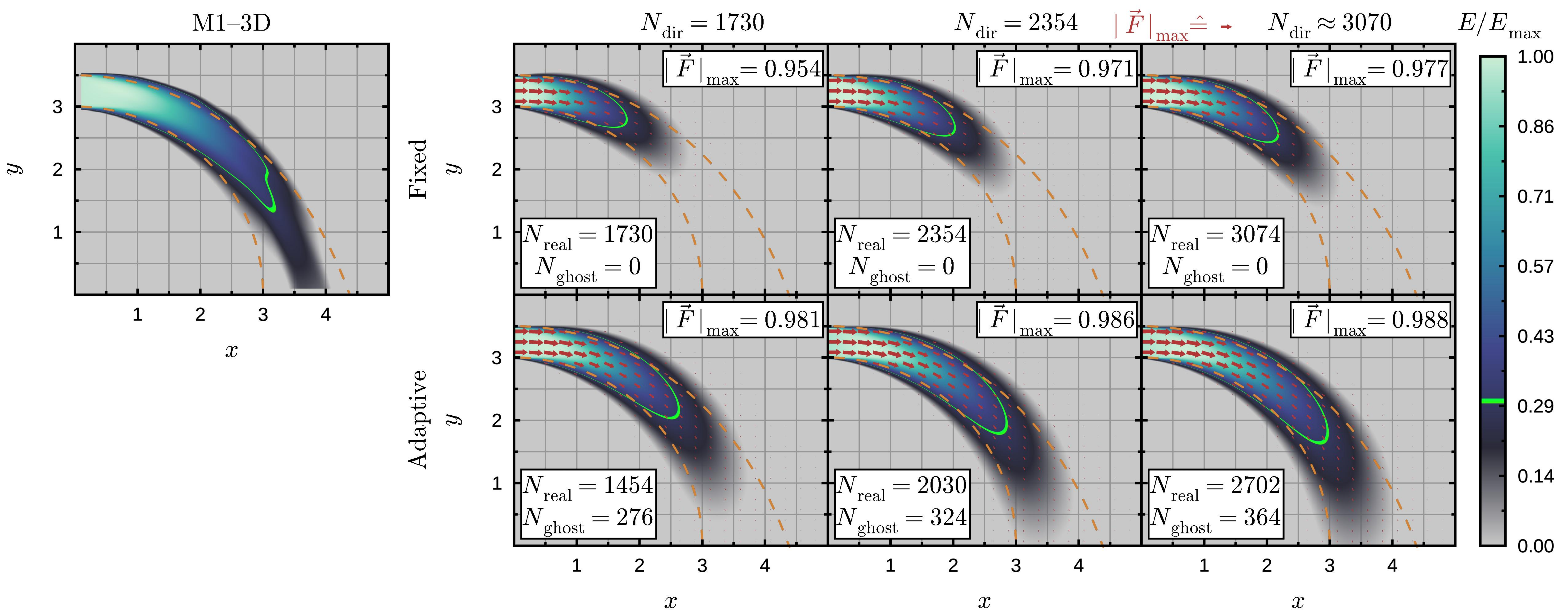}
    \caption[Curved beam 2D and 3D.]{Normalised energy density in a
      vacuum curved-beam test at $t=10\,M$ in a Schwarzschild spacetime
      expressed Cartesian Kerr-Schild coordinates. The orange dashed
      lines show the trajectories of the bounding geodesics, while the
      bright-green region in the colormaps marks the range $0.29-0.30$ to
      highlight the progression of the beam. The top panel reports the 2D
      data on a grid with $x\in[0,5]$, $y\in[0,4]$, $dx = dy = 0.02$
      ($250\times200$ grid-points), and $dt=0.018$. The bottom panel, on
      the other hand, refers to the 3D data computed on a grid
      $x\in[0,5]$, $y\in[0,4]$, $z\in[-0.25,0.25]$, $dx = dy = dz =
      0.025$ ($200\times160\times20$ grid-points), and $dt=0.0225$.  In
      both panels, the top rows report the data with either a fixed or
      adaptive stencils, respectively, while the top-left sub-panels
      offer as comparison with the solution computed with the M1 method,
      either in 2D~\citep[top panel,][]{Weih2020}, or in 3D
      by~\citep[bottom panel,~][]{Musolino2023}.}
    \label{fig: Curved Beam}
\end{figure*}

Figure~\ref{fig: Moving Diffusion} shows that the results of the boosted
diffusion test (right panel) with adaptive streaming and compares it with
the results of the static case. Obviously, in the former case, the energy
density is both diffused and advected at the expected rate, \ie by $0.5$
space units per time units. When comparing the height of the Gaussian
peaks, it is possible to see they are very similar, with the energy being
slightly smaller in the case of the boosted diffusion most likely as a
result of the errors introduced by the Lambda-Iteration scheme. At any
rate, the relative difference between the two solutions is very small and
below $0.8\%$; a similar behaviour has been observed
also with the fixed stencils and is not reported in Fig.~\ref{fig: Moving
  Diffusion}.

\subsection{Curved Spacetime Tests}
\label{subsec: Curved Spacetime Tests}

\subsubsection{Curved Beam}
\label{subsubsec: Curved Beam}

We now turn our attention to tests in curved spacetimes and thus
validating the code in the solution of the newly derived equations
\eqref{eq: GRLBMRT collision} and \eqref{eq: GRLBMRT streaming}. The
first of these tests is the well-known ``curved beam'' test, consisting
in the emission of a beam of radiation close but outside of the unstable
radial photon orbit of a Schwarzschild black hole (we recall that this
orbit is at $r=3\,M$ for a black hole of mass $M$). The behaviour
expected in this test is that the beam of radiation will not propagate
unchanged on a straight line, but will instead be ``bent'', suffer
diffusion, and redshift.

For this test, we use a Cartesian Kerr-Schild Metric with $M=1$ and $a=0$
with a surrounding vacuum, \ie $\kappa_0 = \kappa_1 = \kappa_a = \eta =
0$. For the initial data, both in the 2D and 3D simulations, we set the
energy density to $1$ and the flux density towards the positive
$x$-direction in the LF to have norm $1$. In 2D, this is done at
$x/M=0$ and at $y/M\in[3,3.5]$, while in 3D, the beam has a square
section initially placed at $y/M\in[3,3.5]$ and $z/M\in[3,3.5]$.

Figure~\ref{fig: Curved Beam} shows the results with the 2D (top series
of panels) and with the 3D simulations (bottom series of panels);
furthermore, in each case the upper part reports the results with the
fixed stencils, while the lower part refers to the adaptive-stencil
approach, and the top-left sub-panels show a comparison to the M1 method,
either in 2D~\citep[top panel,][]{Weih2020}, or in 3D by~\citep[bottom
  panel,~][]{Musolino2023}. Starting with the 2D results Fig.~\ref{fig:
  Curved Beam} highlights that when using $200$ directions and a fixed
stencil, the results of the GRLBM are comparable or better than those
obtained with the M1 method.  This can be best appreciated when
concentrating of the width of the region in bright green, which marks a
special position in the colormap and that we use to mark the propagation
of the beam. Interestingly, this test shows more than others the
advantage of the adaptive stencil, which produces comparable results
already with $100$ directions and provides remarkably good results
already with only $50$ directions.

When considering the simulations in 3D, the results shown in the bottom
panel of Fig.~\ref{fig: Curved Beam} that using a sufficient number of
direction leads to a beam propagation that suffers only mildly of
diffusion at the edges and that the shape of the beam is closer to the
expected one than with the M1 method, where the tendency to focusing,
already encountered in the beam-crossing test, is present. Furthermore,
these considerations apply both in the case of fixed stencils (top part)
and of the adaptive one, although the latter shows an overall better
performance. In summary, also the results of the curved-beam tests
clearly indicate the ability of the GRLBM to handle the propagation of
radiation in the free-streaming regime accurately and with minor
diffusion also in the presence on spacetimes in strong curvature.

\subsubsection{Lensed thin disc around a black hole}
\label{subsubsec: Thin disc}

As our final test, we consider a somewhat different setup and present a
novel test that could be employed when considering novel methods handling
radiative-transfer problems in curved spacetimes. In particular,
following the recent work of the Event Horizon Telescope collaboration on
the imaging of supermassive black holes~\citep{Akiyama2019_L1_etal,
  EHT_SgrA_PaperI_etal} and their physical
interpretation~\cite{Akiyama2019_L5_etal, Kocherlakota2021,
  EHT_SgrA_PaperVI_etal}, we test the ability of the GRLBM in modelling
the image of a black hole surrounded by a thin disc of matter emitting
radiation. More specifically, we consider a Schwarzschild of mass $M=1$
and an infinitesimally thin disc with inner and outer edges at $r_{\rm
  in}=6\,M$ and $r_{\rm out}=12\,M$, respectively. For simplicity, we
assume the fluid to have zero velocity and set the energy density to
unity everywhere in the disc and the flux density to zero apart from a
thin layer above and below the disc, and keep them constant throughout
the simulation. We then put an orthographic camera at $(0,0,-16)$ with a
$10$ degree tilt towards the disc, spanning the complete width and height
of our numerical domain.

\begin{figure}
    \centering
    \includegraphics[width=0.48\textwidth]{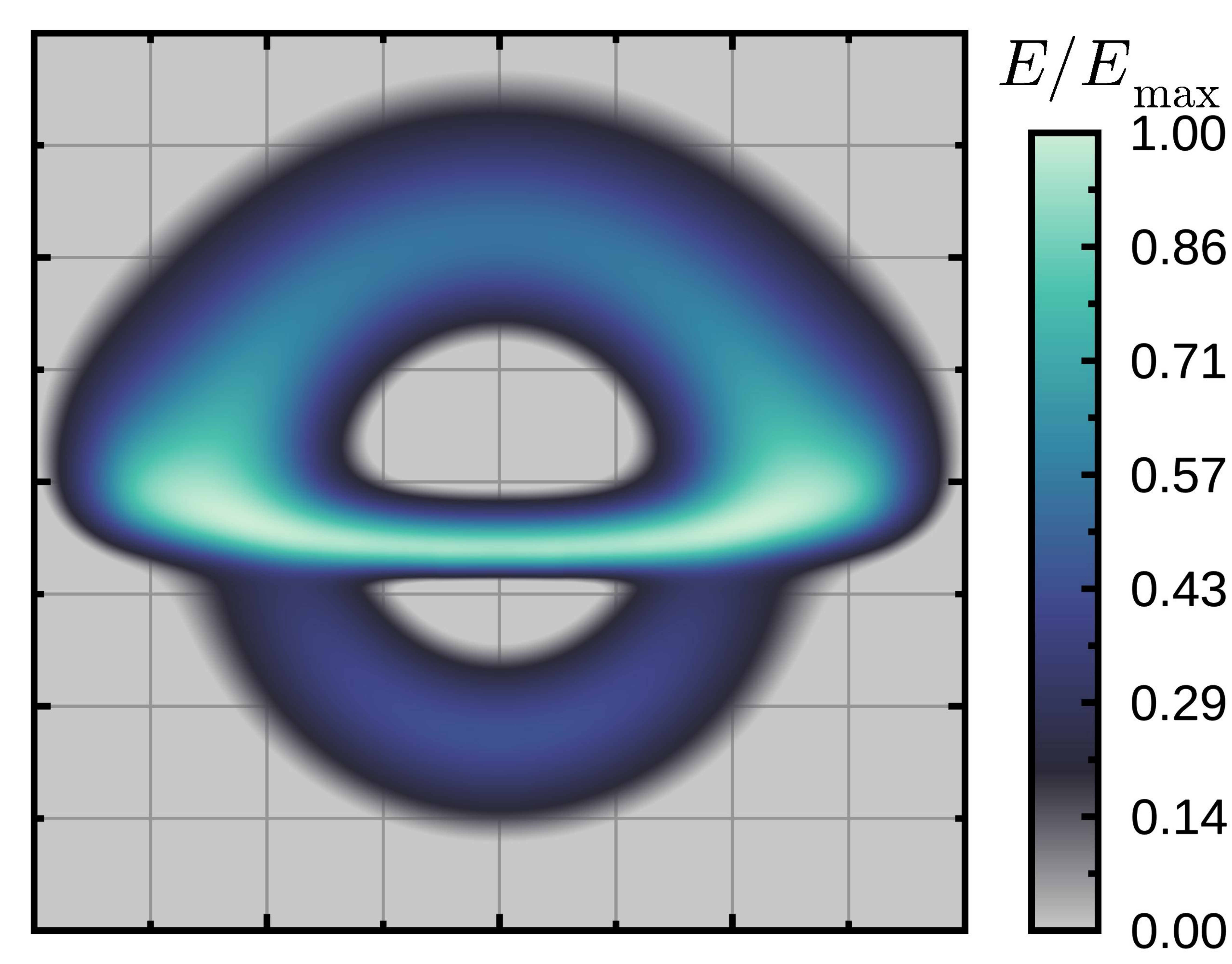}
    \caption{Normalised energy density from a thin disc around a
      Schwarzschild black hole in Cartesian Kerr-Schild coordinates on a
      grid with $x\in[-14,14]$, $y\in[-10,15]$, $z\in[-18,14]$, $dx = dy
      = dz = 0.166667$ ($171\times153\times195$ grid-points), and
      $dt=0.15$. The data refers to a fixed stencil with a Lebedev order
      of $p_{\rm Leb}=65$ ($1454$ directions).}
    \label{fig: Thin Disc}
\end{figure}

Figure~\ref{fig: Thin Disc} shows the lensed image of the thin disc when
using $1454$ directions and a fixed Lebedev stencil of order $p_{\rm Leb}=65$.
While the image is obviously rather diffused since the orthographic camera only
detects radiation passing orthogonally through the camera plane. While
this test is not as efficient as the standard ray-tracing approach
adopted to do black-hole imaging, and where the radiative-transfer
equation is solved along the photon geodesics~\citep[see,
  \eg][]{Gold_EHT:2020_etal}, it is quite remarkable that the GRLBM is
able to reproduce the basic features of this lensed image, namely the
intensity enhancement of the forward part of the disc (no Doppler
boosting is possible because the fluid is assumed to have zero velocity),
the lensing of the backward part of the disc, and even the lensed image
of the lower sheet of the disc. Clearly, comparatively sharper images are
possible when increasing the number of directions and the background grid
resolution. In summary, the lensed thin-disc image represents a rather
inexpensive test that the GRLBM passes successfully and that could be
employed also in future implementations of general-relativistic
treatments of the radiative-transfer problem.

\begin{figure*}
    \centering
    \includegraphics[width=0.95\columnwidth]{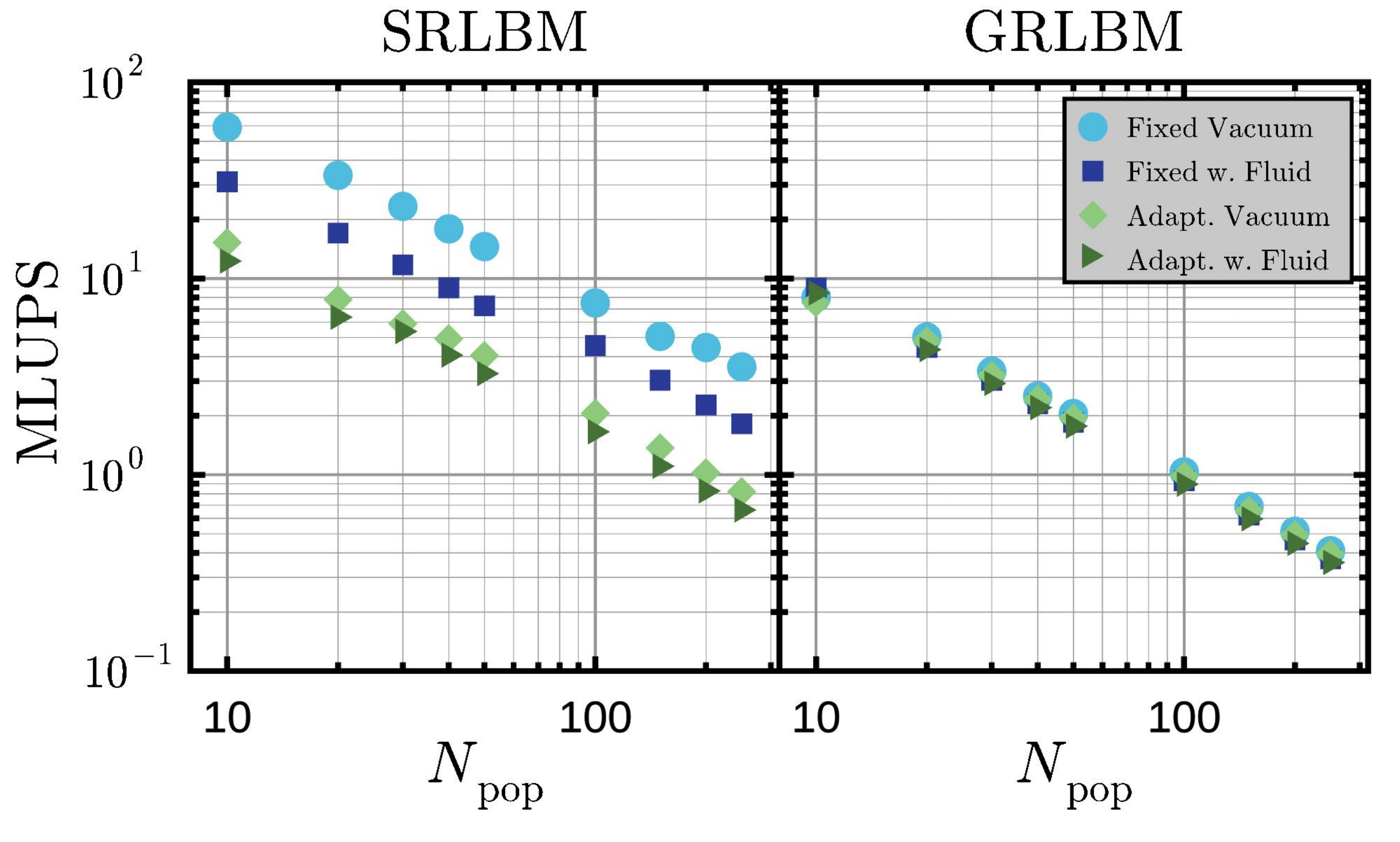}
    \hspace{0.25cm}
    \includegraphics[width=0.95\columnwidth]{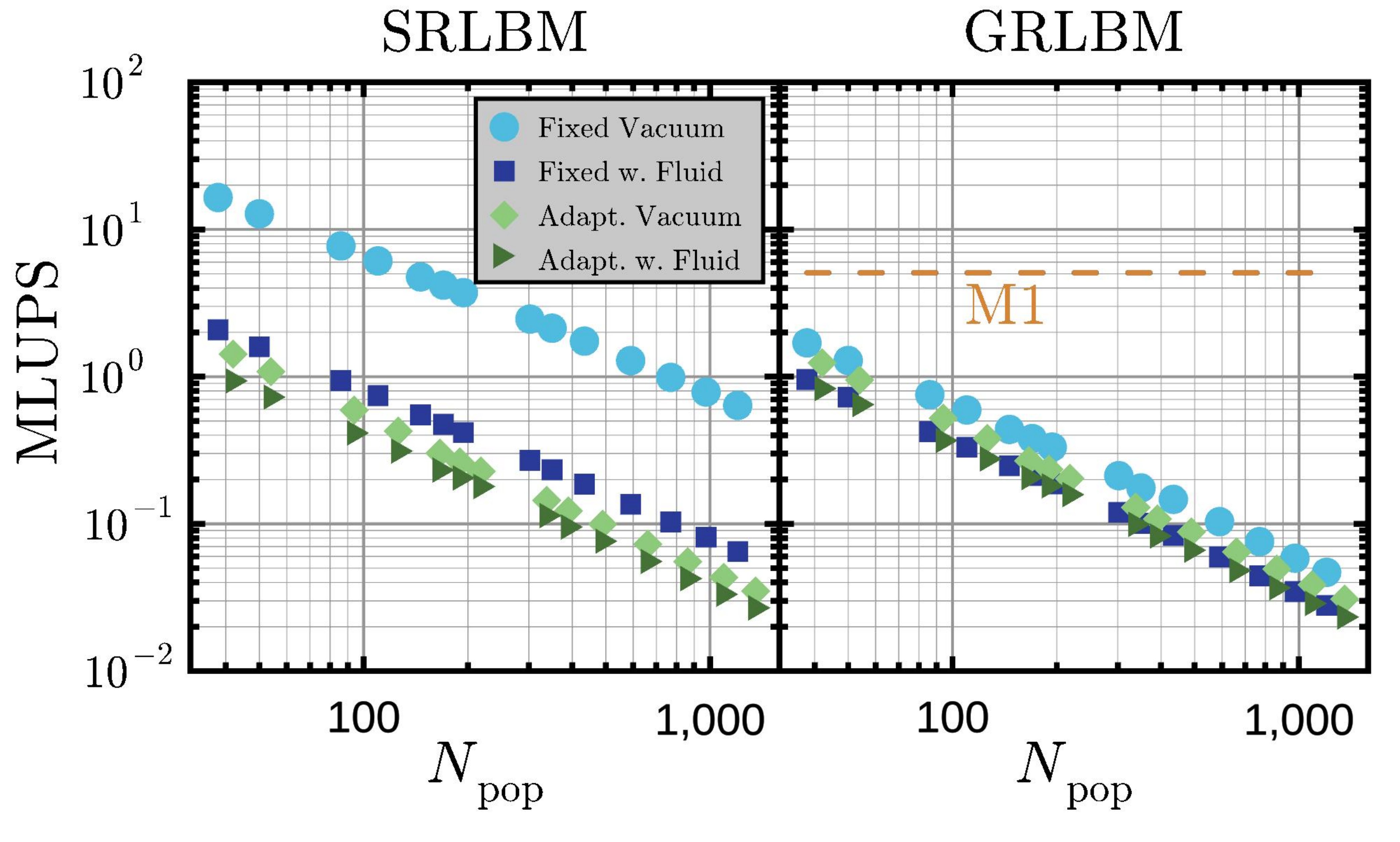}
    \caption[Performance analysis 2D and 3D.]{Performance analysis in
      terms of million lattice updates per second (MLUPS) and for four
      different scenarios: vacuum with $\kappa_0 = \kappa_1 = \kappa_a =
      \eta = u^i = 0$ and random initial moments $E, F^i$, and a
      homogeneous fluid with $\kappa_0 = 100, \kappa_1 = 20, \kappa_a =
      \eta = 10, u^x = 0.5$. The left and right panels report the 2D and
      3D data with $400^2$ and $100^3$ grid-points, respectively. In each
      case, the left and right columns report the results of the SRLBM
      and of the GRLBM. Each test is run ten times with $n_t = 50$ and we
      report the best result for each configuration.}
    \label{fig: Benchmark}
\end{figure*}

\subsection{Performance analysis}
\label{subsec: Performance Analysis}

While in the previous sections we have clearly demonstrate the ability of
the newly developed GRLBM to provide an accurate solution of the
radiative-transfer equation in flat and curved spacetimes, we have not
yet discussed the computational costs associated with the method and, in
particular, how these costs vary when considering either a fixed stencil
or an adaptive one. We recall that the tests have shown that the
adaptive-streaming algorithm captures well the optically-thick regime and
gives better results than the fixed-streaming method in the
free-streaming regime, with the exception of the beam-crossing test where
it is more diffusive. Given the higher computational costs of the
adaptive approach, it is useful to measure how larger such costs are,
both in flat and curved spacetimes.

Hence, we ran the code on a dual socket system with two Intel Xeon Silver
4314 CPUs with $32$ cores running at $2.4\,\text{GHz}$ each and measured
the performance $P$ in million lattice updates per second (MLUPS)
\begin{align}
    P &= \frac{n_x n_y n_z n_t}{10^6 \tau_{\text{tot}}}~\text{MLUPS}\,.
\end{align}
The two scenarios considered are those of a vacuum with $\kappa_0 =
\kappa_1 = \kappa_a = \eta = u^i = 0$ and random initial moments $E,
F^i$, together with that of a homogeneous fluid with $\kappa_0 = 100,
\kappa_1 = 20, \kappa_a = \eta = 10, u^x = 0.5$, thus leading to four
different benchmarks. For the GRLBM we use a Schwarzschild spacetime.

Figure~\ref{fig: Benchmark} reports the results of these measurements
showing, as in previous cases, the values for the 2D and 3D simulations
on the left and right panels, respectively. When considering the
performance of the SRLBM (left portions of each panel), the data both in
2D and 3D clearly indicates that the adaptive-stencil approach has a very
high impact in the pure-streaming scenario in vacuum and leads to a
significant speed-down. This is not the case when when considering also the
collisions, the difference between fixed and adaptive approaches being
less pronounced because the collision step takes up a significant portion
of the total runtime.

On the other hand, when looking at the GRLBM (right portions of each
panel), all four benchmarks give similar results, thus indicating that
the adaptive streaming has only a minor impact on the performance. This
is because in this case also the fixed-streaming approach requires
interpolations and these represent a good portion of the computational
cost. Indeed, the addition of the collision step increases the total
runtime only slightly. From these results, we conclude that
adaptive-streaming is not particularly advantageous in generic scenarios
investigated with the SRLBM, and that the additional computational costs
are compensated only in regimes that are close to the free streaming. By
contrast, the adaptive-streaming approach, which was specially designed
for curved spacetimes, provides the optimal approach in all scenarios
that could be of interest for the use of the GRLBM.

We conclude these considerations on the performance of the GRLBM by
comparing its efficiency with that of the M1 scheme in the 3D
scenarios. This is shown with a dashed horizontal line in the right
portion of the right panel in Fig.~\ref{fig: Benchmark}, which highlights
that the M1 approach always performs better than the GRLBM and that the
additional costs can be even of two orders of magnitude larger in the
case of large number of directions. While our GRLBM code is far from
being optimised and its vectorisation with SIMD instructions could easily
lead to a speed-up of a factor 10, it is unlikely that it will be less
expensive than the M1 approach. Hence, and not differently from
MonteCarlo approaches, the GRLBM should be seen as an appealing approach
for a more accurate but also more expensive solution of the
general-relativistic radiative-transfer equation.

\section{Conclusion and Outlook}
\label{sec: Conclusion and Outlook}

We have extended the special-relativistic lattice-Boltzmann method
(SRLBM) for radiation transport by~\citet{Weih2020c} to curved
spacetimes, thus allowing for the solution of the radiative-transfer
equation in curved spacetimes as those explored by GRMHD simulations of
high-energy astrophysical phenomena. We recall that the lack of a closure
relation is a significant advantage of the SRLBM for radiation transport
over the M1 scheme~\citet{Weih2020c}. In particular, to compute the
pressure tensor in the intermediate regime between optically thin and
optically thick plasmas, moment-based approaches such as the M1 method,
rely on the interpolation between closure relations in the optically thin
and thick limits. The SRLBM, on the other hand, does not rely on any
closure relation, as it allows for direct computation of any moment of
the radiation field via the stencil quadrature.

The novel general-relativistic lattice-Boltzmann method (GRLBM) approach
presented here is based on three main strategies:
\begin{itemize}
\item[] \textit{(i)} the streaming of carefully selected photons along
  null geodesics and interpolating their final positions, velocities, and
  frequency shifts to all photons in a given velocity
  stencil. Furthermore, in order to make the streaming along null
  geodesics numerically feasible, we introduce a spherical-harmonics
  extrapolation scheme, reducing the number of geodesic equations to
  solve drastically.

\item[]\textit{(ii)} the use of transformation laws between the
  laboratory frame, the Eulerian frame, and the fluid frame, enabling us
  to perform the collision step in the fluid frame, thus retaining the
  collision operator of the special-relativistic case with only minor
  modifications. As a result, we are able to model the evolution of the
  frequency-independent (``grey'') radiation field as it interacts with a
  background fluid via absorption, emission, and scattering in a curved
  background spacetime.

\item[]\textit{(iii)} the introduction of an adaptive stencil, which is
  suitably distorted in the direction of propagation of the photon
  bundle, reduces the computational costs of the method while
  improving its performance in the optically-thin regime.
\end{itemize}

To verify the validity of the adaptive streaming scheme, we performed a
series of tests in flat spacetime and compared them to fixed streaming,
M1, and analytical solutions. We found that adaptive streaming gives the
same results as fixed streaming in optically thick and intermediate
regimes. In the free-streaming limit, adaptive streaming fixes the beam
separation problem of fixed streaming but introduces more diffusion in
beam crossing. The curved beam and the thin-disc test demonstrate the
validity of the newly derived curved spacetime lattice-Boltzmann
equations. The additional computational cost of adaptive streaming does
not justify its usage in flat spacetime simulations unless one works on a
memory-bound system. However, it has a negligible performance impact on
the curved spacetime code, and we highly recommend using it in this
context. 

The discussion of the mathematical and numerical strategy developed for
the GRLBM, and proposed here for the first time, is followed by the
presentation of a series of standard and non-standard tests in 2D and 3D
aimed at validating the correctness of the GRLBM under a variety of
physical and numerical conditions. In all cases, we have demonstrated the
ability of the GRLBM to provide an accurate solution of the
general-relativistic radiative-transfer equation, thus opening the way to
the use of the GRLBM in direct numerical simulations of astrophysical
plasmas.

Overall, the results of the tests indicate that the adaptive-streaming
approach is not particularly advantageous in generic scenarios
investigated with the SRLBM, where the additional computational costs are
not compensated but in those regimes that are close to the free
streaming. By contrast, because it was specially designed for curved
spacetimes, the adaptive-streaming provides the optimal method in all
scenarios that could be of interest for the use of the GRLBM.

The results presented here are meant mostly as a proof-of-principle of
the feasibility and robustness of the GRLBM approach. Hence, much more
can be done in future work to further optimise the techniques employed
and increase the efficiency of the solution, \eg by reducing the
floating-point accuracy of the populations from $64\,\text{bit}$ to
$32\,\text{bit}$ (which would also benefit GPU implementations), the use
of the Voronoi interpolation in 3D for the velocity, or other less
expensive quadratures on the two-sphere. However, given the intrinsically
higher complexity of the approach, we regard the GRLBM as an appealing
but also more expensive approach to the solution of the
general-relativistic radiative-transfer equation. In this respect, the
GRLBM is not dissimilar to equivalent Monte Carlo approaches, that are
intrinsically more computationally intensive, although more accurate than
the simpler moment-based approaches.

As a final remark, we note that the GRLBM could be used in combination
with the M1 scheme in a way that is logically similar to the
Guided-Moments approach suggested by~\cite{Izquierdo2024b}. More
specifically, the new scheme would use the M1 method in its standard form
but replace the closure relation for the pressure tensor with a
simplified GRLBM scheme that only computes the pressure tensor. We will
explore this approach in future work.

\section*{Acknowledgements}

We thank C. Musolino, L. R. Weih, and S. Succi for helpful discussions
and V. Lindenstruth for financial support. Support comes from the State
of Hesse within the Research Cluster ELEMENTS (Project ID 500/10.006) and
through the European Research Council Advanced Grant ``JETSET: Launching,
propagation and emission of relativistic jets from binary mergers and
across mass scales'' (grant No. 884631). LR acknowledges the Walter
Greiner Gesellschaft zur F\"orderung der physikalischen
Grundlagenforschung e.V. through the Carl W. Fueck Laureatus Chair.

\section*{Data Availability}
The corresponding author will share this paper's data at a reasonable
request.

\section*{Code Availability}

The 2D and 3D GRLBM codes developed for this work are freely accessible
on \texttt{github} and can be found
at~\url{https://github.com/Tom-Olsen/2dRadiation}
and~\url{https://github.com/Tom-Olsen/3dRadiation}, respectively.


\bibliographystyle{mnras}
\bibliography{aeireferences, bibtex}

\newpage
\appendix
\section{Lorentz Transformations}
\label{Appendix: Lorentz Transformations}

Let $\mathcal{A}$ and $\mathcal{B}$ be two inertial frames and $u^i$ the
three-velocity of $\mathcal{B}$ as measured by $\mathcal{A}$. The
Lorentz factor and the Lorentz boost matrix are defined as,
\begin{align}
    \gamma &= \frac{1}{\sqrt{1-u_i u^i}},\\
    \Lambda\indices{^\mu_\nu} &= \left(\begin{array}{cc}
        \gamma & -\gamma u_j \\
        -\gamma u^i & \delta\indices{^i_j} + (\gamma-1){u^i u_j}/{u_k u^k}
    \end{array}\right).
\end{align}
Let $P^\mu = \nu N^\mu$ be a photon's four-momentum, $\nu$ its frequency,
$N^\mu = (1, n^i)$ its four-velocity and $n^i$ its three-velocity. By
Lorentz boosting the time component of the four-momentum, we can derive
the transformation law of the frequency,
\begin{align}
    \tilde P^0 &= \tilde\nu \tilde N^0 = \Lambda\indices{^0_\mu} P^\mu = \nu\Lambda\indices{^0_\mu} N^\mu = \nu \left(\Lambda\indices{^0_0} N^0 + \Lambda\indices{^0_i} N^i\right),\\
    \Rightarrow \tilde\nu &= \nu \gamma(1 - u_i n^i),
\end{align}
and by Lorentz boosting the spatial components of the four-momentum, we
can derive the transformation law of the three-velocity,
\begin{align}
    \tilde P^i &= \tilde\nu \tilde N^i = \Lambda\indices{^i_\mu} P^\mu = \nu \left(\Lambda\indices{^i_0} N^0 + \Lambda\indices{^i_j} N^j\right),\\
    \Rightarrow \tilde n^i & = \frac{\nu}{\tilde\nu} \left[n^i -\left(1 - \frac{\gamma u_j n^j}{\gamma + 1}\right)\gamma u^i \right].
\end{align}
Through the principles of symmetry, we can immediately derive the inverse
transformation laws, yielding the following results,
\begin{align}
    A &= \gamma(1 - u_i n^i) = \frac{1}{\gamma(1 + u_i \tilde n^i)},\\
    \nu & = \frac{\tilde\nu}{A}, \qquad \tilde\nu = \frac{\nu}{A},\\
    n^i & = A \left[\tilde n^i + \left(1 + \frac{\gamma u_j\tilde n^j}{\gamma + 1}\right) \gamma u^i\right],\\
    \tilde n^i & = \frac{1}{A} \left[n^i -\left(1 - \frac{\gamma u_j n^j}{\gamma + 1}\right)\gamma u^i \right],
\end{align}
where $A$ is called the Doppler factor.

The specific intensity $I_\nu$ is the energy $dE_\nu$ per time $dt$, area
$d^{D-1}x$, frequency $d\nu$, and solid angle $d\Omega$, or simply the
'energy per everything',
\begin{align}
    I_\nu &= \frac{dE_\nu}{dt\ d^{D-1}x\ d\nu\ d\Omega},\\
    dE_\nu &= \nu\ dN,\\
    dN &= f\ d^Dr\ d^Dp,\\
    d^Dp &= p^{D-1}\ dp\ d\Omega = \nu^{D-1}\ d\nu\ d\Omega,\\
    d^Dr &= dt\ d^{D-1}x.
\end{align}
The letter $D$ denotes the dimensionality of space, $N$ is the number of
photons we are looking at, and $f$ is the Lorentz invariant phase space
density of photons. Putting everything together we get,
\begin{align}
    I_\nu &= \nu^D f.
\end{align}
We already know how the frequency transforms and that the phase space
density is Lorentz invariant. From this, we can directly derive the
transformation law of the specific intensity, which is
dimension-dependent,
\begin{align}
\label{eq: intensity transformation}
    I_\nu &= \frac{\tilde I_{\tilde\nu}}{A^D}, \qquad \tilde I_{\tilde\nu} = I_\nu A^D.
\end{align}
The transformation law of the 'grey' or total intensity directly follows as,
\begin{align}
\label{eq: grey intensity transformation}
    I &= \int_0^\infty I_\nu\ d\nu \quad \Rightarrow \quad I = \frac{\tilde I}{A^{D+1}}, \qquad \tilde I = I A^{D+1}.
\end{align}
The last transformation law we need is that of the solid angle $d\Omega$.
For this, we take a close look at the following invariant Lorentz scalar,
\begin{align}
    e &= u_\mu u_\nu T^{\mu\nu} = \oint_{S} \underbrace{u_\mu u_\nu p^\mu p^\nu}_a  \underbrace{\frac{I\ d\Omega}{\nu^2}}_b.
\end{align}
The factor $a$ is invariant because it is a full contraction and,
therefore, a Lorentz scalar. This means that the factor $b$ must also be
invariant. However, we already know how the total intensity $I$ and the
frequency $\nu$ transform. Therefore, the solid angle $d\Omega$ must
transform as,
\begin{align}
    d\Omega &= d\tilde\Omega A^{D-1}, \qquad d\tilde\Omega = \frac{d\Omega}{A^{D-1}}.
\end{align}

\section{Collision Operator}
\label{Appendix: Collision Operator}

The most general representation of the Lorentz-invariant collision
operator in Eq.~\eqref{eq: radiation Boltzmann equation} is given
by,
\begin{align}
\label{eq: full radiation collision operator}
    \Gamma[f_\nu] &= \oint_{4\pi}\int_0^\infty \nu'^3[f'_{\nu'} (1 -
      f_\nu) R^{I} - f_\nu(1 - f'_{\nu'}) R^{O}] d\nu' d\Omega'\,.
\end{align}
The incoming and outgoing scattering kernels, $R^{I} = R^{I}(\nu, \nu',
n^i, n'^i)$ and $R^{O} = R^{O}(\nu, \nu', n^i, n'^i)$, depend on the
underlying scattering model, which is usually expressed in the fluid
frame, where the scattering center rests. By assuming iso-energetic
scattering, we can neglect the frequency dependence of the scattering
kernels and expand them in a Legendre series,
\begin{align}
\label{eq: iso-energetic scattering kernel}
    \tilde R(\tilde \nu, \tilde \nu', \tilde n^i, \tilde n'^i) &= \tilde
    R^{I}(\tilde \nu, \tilde \nu', \tilde n^i, \tilde n'^i) = \tilde
    R^{O}(\tilde \nu, \tilde \nu', \tilde n^i, \tilde n'^i)\\ &\approx
    (\frac{1}{2}\tilde \Phi_0 + \frac{3}{2}\tilde \Phi_1 \tilde n_i
    \tilde n'^i) \delta(\tilde \nu - \tilde \nu')\,.
\end{align}
We then define the scattering opacities in terms of the Legendre
coefficients, $\tilde\kappa_{l,\tilde\nu} = 2\pi \tilde\nu^2
\tilde\Phi_l$. The collision operator in the fluid frame then becomes,
\begin{align}
    \tilde\Gamma[f_{\tilde\nu}] &= \oint_{4\pi}\int_0^\infty
    \tilde\nu'^3(f'_{\tilde\nu'} - f_{\tilde\nu}) \tilde R\ d\tilde\nu'
    d\tilde\Omega'\\ &= \oint_{4\pi} \tilde\nu^3(f'_{\tilde\nu} -
    f_{\tilde\nu}) (\frac{1}{2}\tilde \Phi_0 + \frac{3}{2}\tilde \Phi_1
    \tilde n_i \tilde n'^i)\ d\tilde\Omega'\\ &= \frac{\tilde\nu}{4\pi}
    \oint_{4\pi} (f'_{\tilde\nu} - f_{\tilde\nu})
    (\tilde\kappa_{0,\tilde\nu} + 3\tilde\kappa_{1,\tilde\nu} \tilde n_i
    \tilde n'^i)\ d\tilde\Omega'\\
\label{eq: radiation collision operator FF}
    &= \tilde\nu \left[\tilde\kappa_{0,\tilde\nu}(\tilde
    E_{\tilde\nu} - f_{\tilde\nu}) + 3\tilde\kappa_{1,\tilde\nu} \tilde
  n_i\tilde F^i_{\tilde\nu}\right]\,,
\end{align}
where we have used the monochromatic radiation moments,
\begin{align}
    E_\nu &= \frac{1}{4\pi} \oint_{4\pi} f_\nu d\Omega,\\
    F^i_\nu &= \frac{1}{4\pi} \oint_{4\pi} f_\nu n^i d\Omega\,.
\end{align}
Note that the final form of our collision operator in Eq.~\eqref{eq:
  radiation collision operator FF} is not Lorentz-invariant anymore and
thus only valid in the fluid frame.

\section{Tetrad Separation}
\label{Apppendix: Tetrad Separation}

For our curved spacetime LBM scheme, we need a tetrad to transform any
tensors between the LF and the EF. The tetrad
$\boldsymbol e$ must then obey,
\begin{align}
\label{eq: eta = e e g}
    \eta_{\bar\mu\bar\nu} &= g_{\mu\nu} e\indices{^\mu_{\bar\mu}} e\indices{^\nu_{\bar\nu}}
    &\Leftrightarrow &&
    \boldsymbol{\eta} &= \boldsymbol{e}^T \boldsymbol{g} \boldsymbol{e}\,.
\end{align}
Due to the symmetry of $g_{\mu\nu}$ and $\eta_{\bar\mu\bar\nu}$ Equation
\eqref{eq: eta = e e g} contains only 10 unique equations, but 16
unknowns. The missing six constraints correspond to the six degrees of
freedom of the Lorentz group, consisting of three spatial rotations and
three velocity boosts. This can be shown by boosting Eq.~\eqref{eq:
  eta = e e g} with two boost matrices,
\begin{align}
\label{eq: boosted tetrad equation}
    \eta_{\bar\mu\bar\nu}\Lambda\indices{^{\bar\mu}_{\bar\alpha}} \Lambda
    \indices{^{\bar\nu}_{\bar\beta}} &= g_{\mu\nu}
    e\indices{^\mu_{\bar\mu}}\Lambda\indices{^{\bar\mu}_{\bar\alpha}}
    e\indices{^\nu_{\bar\nu}}\Lambda\indices{^{\bar\nu}_{\bar\beta}},\\ \Rightarrow
    \eta_{\bar\alpha\bar\beta} &= g_{\mu\nu}
    \underbrace{e\indices{^\mu_{\bar\mu}}
      \Lambda\indices{^{\bar\mu}_{\bar\alpha}}}_{b\indices{^\mu_{\bar\alpha}}}
    \underbrace{e\indices{^\nu_{\bar\nu}}
      \Lambda\indices{^{\bar\nu}_{\bar\beta}}}_{b\indices{^\nu_{\bar\beta}}}\,.
\end{align}
The invariance of the Minkowski metric leaves it unchanged under
Lorentz-transformations $\Lambda\indices{^{\bar\mu}_{\bar\alpha}}$.
Thus, the boosted tensor ${b\indices{^\mu_{\bar\alpha}} =
  e\indices{^\mu_{\bar\mu}}\Lambda\indices{^{\bar\mu}_{\bar\alpha}}}$
also obeys Eq.~\eqref{eq: eta = e e g} and is, therefore, a valid
tetrad.

While in most cases, we are not interested in the exact spatial
orientation of the IF, the boost of the IF is important. We
can read the boost of the IF directly from the tetrad by comparing
the first column of the tetrad to the four-velocity of the IF. Let
$u^\mu$ be the four-velocity of the IF as seen in the LF. The
IF sees itself being at rest in its frame of reference, $\bar u^\mu
= (1,\bar0^k)$. Transforming the four-velocity between these two
reference frames then yields,
\begin{align}
    u^\mu &= e\indices{^\mu_\nu} \bar u^\nu = e\indices{^\mu_0} \bar u^0
    + e\indices{^\mu_i} \bar u^i = e\indices{^\mu_0} 1 +
    e\indices{^\mu_i} 0^i = e\indices{^\mu_0}\,.
\end{align}

The fundamental ansatz of the tetrad separation is to constraint the
tetrad to a lower triangular matrix and then split it into four separate
tensors
\begin{align}
    \boldsymbol{e} &= (e\indices{^\mu_\nu}) = \left(\begin{array}{cccc}
        e\indices{^0_0} & 0 & 0 & 0 \\
        e\indices{^1_0} & e\indices{^1_1} & 0 & 0 \\
        e\indices{^2_0} & e\indices{^2_1} & e\indices{^2_2} & 0 \\
        e\indices{^3_0} & e\indices{^3_1} & e\indices{^3_2} & e\indices{^3_3}
    \end{array}\right) = \boldsymbol{ABCD},\\
    \boldsymbol{A} &= \left(\begin{array}{cccc}
        e\indices{^0_0} & 0 & 0 & 0 \\
        e\indices{^1_0} & 1 & 0 & 0 \\
        e\indices{^2_0} & 0 & 1 & 0 \\
        e\indices{^3_0} & 0 & 0 & 1
    \end{array}\right),
    \boldsymbol{B} = \left(\begin{array}{cccc}
        1 & 0 & 0 & 0 \\
        0 & e\indices{^1_1} & 0 & 0 \\
        0 & e\indices{^2_1} & 1 & 0 \\
        0 & e\indices{^3_1} & 0 & 1
    \end{array}\right),\nonumber\\
    \boldsymbol{C} &= \left(\begin{array}{cccc}
        1 & 0 & 0 & 0 \\
        0 & 1 & 0 & 0 \\
        0 & 0 & e\indices{^2_2} & 0 \\
        0 & 0 & e\indices{^3_2} & 1
    \end{array}\right),
    \boldsymbol{D} = \left(\begin{array}{cccc}
        1 & 0 & 0 & 0 \\
        0 & 1 & 0 & 0 \\
        0 & 0 & 1 & 0 \\
        0 & 0 & 0 & e\indices{^3_3}
    \end{array}\right).\nonumber
\end{align}
The six zeroes fix the six degrees of freedom of the Lorentz group,
meaning that our final tetrad spatial orientation and boost are already
determined.

Inserting the separation ansatz into the matrix representation of
Eq.~\eqref{eq: eta = e e g} allows us to write it in an iterative
manner
\begin{align}
    \boldsymbol{\eta} = \boldsymbol{e}^T \boldsymbol{g} \boldsymbol{e}
    &= \boldsymbol{D}^T \boldsymbol{C}^T \boldsymbol{B}^T \boldsymbol{A}^T \boldsymbol{g}^0
       \boldsymbol{A}   \boldsymbol{B}   \boldsymbol{C}   \boldsymbol{D}\\
    &= \boldsymbol{D}^T \boldsymbol{C}^T \boldsymbol{B}^T \boldsymbol{g}^1
       \boldsymbol{B}   \boldsymbol{C}   \boldsymbol{D}\\
    &= \boldsymbol{D}^T \boldsymbol{C}^T \boldsymbol{g}^2
       \boldsymbol{C}   \boldsymbol{D}\\
    &= \boldsymbol{D}^T \boldsymbol{g}^3
       \boldsymbol{D} = \boldsymbol{g}^4.
\end{align}
Going back to index notation and taking a closer look at the first step,
$\boldsymbol{g}^1 = \boldsymbol{A}^T \boldsymbol{g}^0 \boldsymbol{A}$, we
can see that it only affects the first column and row of the resulting
intermediate matrix,
\begin{align}
\label{eq: g1 master}
    g^1_{\mu\nu} &= g^0_{\alpha\beta} A\indices{^\alpha_\mu}
    A\indices{^\beta_\nu},\\ g^1_{ij} &= g^0_{\alpha\beta}
    A\indices{^\alpha_i} A\indices{^\beta_j} = g^0_{00} A\indices{^0_i}
    A\indices{^0_j} + 2g^0_{0k} A\indices{^0_i} A\indices{^k_j} +
    g^0_{ab} A\indices{^a_i} A\indices{^b_j}\nonumber\\ &=
    g^0_{00}\cdot0\cdot0 + 2g^0_{0k}\cdot0\cdot A\indices{^k_j} +
    g^0_{ab} \delta\indices{^a_i} \delta\indices{^b_j} = g^0_{ij},\\
\label{eq: 0 part}
    g^1_{0i} &= g^0_{\alpha\beta} A\indices{^\alpha_0}
    A\indices{^\beta_i} = A\indices{_\beta_0} A\indices{^\beta_i} =
    A_{00} A\indices{^0_i} + A_{j0} A\indices{^j_i} \nonumber\\ &=
    A_{00}\cdot0 + A_{j0} \delta\indices{^j_i} = A_{i0} \overset{!}{=}
    0_i,\\
\label{eq: -1 part}
    g^1_{00} &= g^0_{\alpha\beta} A\indices{^\alpha_0}
    A\indices{^\beta_0} = A\indices{_\beta_0} A\indices{^\beta_0}
    \overset{!}{=} -1\,.
\end{align}
Furthermore, we can conclude from Eq.~\eqref{eq: -1 part} that the
column four-vector $A\indices{^\mu_0}$ in $\boldsymbol{A}$ must be
timelike. Eq.~\eqref{eq: 0 part} tells us that the covariant
spatial components of $A\indices{_\mu_0}$ must vanish, meaning that it is
orthogonal to a three-dimensional hypersurface of constant coordinate time.
For any metric, there only is one such future-directed four-vector, the
four-velocity of the Eulerian observer,
\begin{align}
    A\indices{^\mu_0} = n^\mu = \frac{1}{\alpha} \left(1,
    -\beta^i\right), &&& n_\mu = (-\alpha, 0_i)\,.
\end{align}
The resulting intermediate tensor $\boldsymbol{g}^1$ now takes the form,
\begin{align}
    \boldsymbol{g}^1 = \left(\begin{array}{cc}
         -1 & 0_j \\
        0_i & g_{ij}
    \end{array}\right)\,,
\end{align}
meaning we successfully ``diagonalised'' the first row and column.

Due to the nature of our split, the following iterations will not alter
the diagonalised part from the previous iteration. To continue we repeat
the calculations in \eqref{eq: g1 master} to \eqref{eq: -1 part} but for
$\boldsymbol{g}^2 = \boldsymbol{B}^T \boldsymbol{g}^1 \boldsymbol{B}$.
In the following, we introduce capitalised indices, which run from two to
three, $I,J,K\in\{2,3\}$.
\begin{align}
    g^2_{ij} &= g^1_{ab} B\indices{^a_i} B\indices{^b_j},\\ g^2_{IJ} &=
    g^1_{ab} B\indices{^a_I} B\indices{^b_J} = g^1_{11} B\indices{^1_I}
    B\indices{^1_J} + 2g^1_{1K} B\indices{^1_I} B\indices{^K_J} +
    g^1_{AB} B\indices{^A_I} B\indices{^B_J}\nonumber\\ &=
    g^1_{11} \times 0 \times 0 + 2g^1_{1K} \times 0 \times  B\indices{^K_J} +
    g^1_{AB} \delta\indices{^A_I} \delta\indices{^B_J} =
    g^1_{IJ},\\ g^2_{1I} &= g^1_{ab} B\indices{^a_1} B\indices{^b_I} =
    B\indices{_b_1} B\indices{^b_I} = B_{11} B\indices{^1_I} + B_{J1}
    B\indices{^J_I} \nonumber\\ &= B_{11} \times 0 + B_{J1}
    \delta\indices{^J_I} = B_{I1} \overset{!}{=} 0_I,\\ g^2_{11} &=
    g^1_{ab} B\indices{^a_1} B\indices{^b_1} = B\indices{_b_1}
    B\indices{^b_1} \overset{!}{=} 1\,.
\end{align}
In contrast to the previous iteration the three-vector $B\indices{^i_1}$
in $\boldsymbol{B}$ must now be spacelike instead of timelike, while
still having a vanishing contravariant part in the last two components.
We introduce an Euclidean $2+1$-split inspired by the $3+1$-formalism to
construct a vector with precisely these properties,
\begin{align}
\label{eq:2+1 split start}
    \tilde{g}_{ij} &= \left(\begin{array}{cc} \tilde{\alpha}^2 +
      \tilde{\beta}_K\tilde{\beta}^K & -\tilde{\beta}_J
      \\ -\tilde{\beta}_I & \tilde{\gamma}_{IJ}
    \end{array}\right),\\
    \tilde{g}^{ij} &= \left(\begin{array}{cc} 1/\tilde{\alpha}^2 &
      \tilde{\beta}^J/\tilde{\alpha}^2
      \\ \tilde{\beta}^I/\tilde{\alpha}^2 &
      \tilde{\gamma}^{IJ}+\tilde{\beta}^I\tilde{\beta}^J/\tilde{\alpha}^2
    \end{array}\right),\\
    \tilde{\gamma}_{IJ} &= \tilde{g}_{IJ},\qquad \tilde{\gamma}^{IJ} =
    (\tilde{\gamma}_{IJ})^{-1} = \tilde{g}^{IJ} -
    \tilde{\beta}^I\tilde{\beta}^J/\tilde{\alpha}^2,\\ \tilde{\beta}_I &=
    -\tilde{g}_{1I},\qquad \tilde{\beta}^I =
    \tilde{\gamma}^{IJ}\tilde{\beta}_J = \tilde{\alpha}^2
    \tilde{g}^{1I},\\ \tilde{\alpha} &= \frac{1}{\sqrt{\tilde{g}^{11}}} =
    \sqrt{\tilde{g}_{11} - \tilde{\beta}_I\tilde{\beta}^I},\\ \tilde{n}_i
    &= (\tilde{\alpha}, 0_I),\qquad \tilde{n}^i =
    \frac{1}{\tilde{\alpha}}(1,\tilde{\beta}^I)^T, \qquad
    \tilde{n}_i\tilde{n}^i = 1\,.
\label{eq:2+1 split end}
\end{align}
We mark all $2+1$-split quantities with a tilde to differentiate between
the two splits. By construction, the three-vector $\tilde{n}^i$ has
precisely the properties we are looking for, giving us the solution for
the tensor $\boldsymbol{B}$,
\begin{align}
    B\indices{^i_1} = \tilde{n}^i = \frac{1}{\tilde{\alpha}} \left(1,
    \tilde{\beta}^I\right), &&& \tilde{n}_i = (\tilde{\alpha}, 0_I)\,.
\end{align}
For the tensor $\boldsymbol{C}$, we repeat the procedure with an even
simpler $1+1$-split, and the final tensor $\boldsymbol{D}$ follows
trivially. We leave this exercise for the interested reader. We present
the final solution in terms of the $3+1$-components and the original
metric.
\begin{align}
    A\indices{^\mu_0} &= e\indices{^\mu_{\bar0}} = n^\mu =
    \frac{1}{\alpha}(1, -\beta^i)^T,\\ B\indices{^i_1} &=
    e\indices{^i_{\bar1}} = \tilde{n}^i =
    \frac{1}{\tilde\alpha}(1,\tilde{\beta}^I)^T,\\ C\indices{^I_2} &=
    e\indices{^I_{\bar2}} = \frac{1}{A}(1, B),\\ D\indices{^3_3} &=
    e\indices{^3_{\bar3}} = \frac{1}{\sqrt{g_{33}}}\,.
\end{align}
The four-velocity of the Eulerian observer is well known, and the other
components are given by,
\begin{align}
    \tilde{\alpha} &= \sqrt{g_{11} + \frac{ (g_{12})^2g_{33} +
        (g_{13})^2g_{22} -
        2g_{12}g_{13}g_{23})}{(g_{23})^2-g_{22}g_{33}}},\\ \tilde{\beta}^I
    &= \frac{1}{(g_{23})^2-g_{22}g_{33}}\left(
    \begin{array}{c}
        g_{12}g_{33} - g_{13}g_{23}\\ g_{13}g_{22} - g_{12}g_{23}
    \end{array}
    \right),\\ A &= \sqrt{g_{22} - \frac{(g_{23})^2}{g_{33}}},\\ B &=
    -\frac{g_{23}}{g_{33}}\,.
\end{align}
The resulting tetrad from this procedure is always comoving with the
Eulerian observer. As usual, it can be boosted to obtain any other
tetrad. For example, in order to obtain a stationary tetrad $\bar
e\indices{^\mu_\nu}$, the boost matrix would be constructed as follows
\begin{align}
    \Lambda\indices{^\mu_\nu} &= \left(\begin{array}{cc} \gamma & -\gamma
      u_j \\
      ~& \\
      -\gamma u^i & \delta\indices{^i_j} + (\gamma - 1) {u^i
        u_j}/{u^k u_k}
    \end{array}\right),\\
    \bar e\indices{^0_0} &= \frac{1}{\sqrt{-g_{00}}} =
    e\indices{^0_\alpha}\Lambda\indices{^\alpha_0} =
    e\indices{^0_0}\gamma \quad\Rightarrow\quad\gamma =
    \frac{1}{e\indices{^0_0}\sqrt{-g_{00}}},\\ \bar e\indices{^i_0} &=
    0^i = e\indices{^i_\alpha}\Lambda\indices{^\alpha_0} =
    \gamma(n^i-e\indices{^i_j}u^j) \quad\Rightarrow\quad u^j =
    e\indices{_i^j} n^i\,.
\end{align}
Note that the inverse tetrad $e\indices{_\mu^\nu}$ and sub-tetrad
$e\indices{_i^j}$ are easy to compute due to the lower triangular nature
of our ansatz.

Finally, we compare the performance of our new approach to the
Gramm-Schmidt process~\citep{Gentle2007}. We use the following setup for
the performance comparison to simulate a realistic scenario similar to
GRMHD simulations. We initialise a metric and pre-compute the
corresponding $3+1$-components on a numerical Cartesian grid of
$200\times200\times200$ points. For the metric, we use a Cartesian
Kerr-Schild metric with $M=1$ and $a=0.5$, ensuring that no metric
components are zero. We then calculate a tetrad on each grid cell for
both codes on a single core and measure the run time 1000 times for
statistics. Tab.~\ref{tab: performance} shows the results for four
different cases. The stationary and comoving Gramm-Schmidt solutions only
differ in the choice of the first seed vector, $s^\mu_0 = (1, 0^i)$
vs. $s^\mu_0 = n^\mu$, and are therefore almost identical. The stationary
and comoving tetrad separation solutions differ in the additional Lorentz
boost. For this, we need to invert the lower triangular three-by-three
matrix $e\indices{^i_j}$, which adds a noticeable amount of time but
still outperforms the Gramm-Schmidt method.

Comparing the two methods, we get a speedup of $1.58$ for the stationary
case and $12.59$ for the comoving tetrad when using the tetrad separation
algorithm. The comoving tetrad is more desirable for our use case,
making this new method ideal for constructing tetrads.
\begin{table}
    \centering
    \begin{tabular}{|r|r|r|r|r|r|}
        \hline
                    &GS st.&GS com.&Sep st.&Sep com.\\
        \hline
        Average[ms] & 789.2&  803.8&  509.3&    62.7\\
        Max[ms]     & 795.7&  812.9&  514.0&    67.7\\
        Min[ms]     & 784.8&  799.1&  501.2&    61.2\\
        Std dev[ms] &   2.3&    2.5&    2.6&     2.1\\
        \hline
    \end{tabular}
    \caption{Gramm-Schmidt (\textit{GS}) and Tetrad-Separation
      (\textit{Sep}) run times for stationery (\textit{st.})  and
      comoving (\textit{com.}) tetrads.}
    \label{tab: performance}
\end{table}

\section{Initial Data}
\label{Appendix: Initial Data}

The initial data is given as energy and flux density in the LF,
$E_{\text{ID}}$, $F^i_{\text{ID}}$, for every grid
point. To convert these into initial data for the individual intensities,
we first convert them to the local IF, $\bar E_{\text{ID}}$,
$\bar F^i_{\text{ID}}$. There does not exist a unique mapping
from the IF energy and flux density to the intensity distribution. As an
ansatz, we use the Kent distribution~\citep{Kent1982} with the normalized
flux direction $\vec e_F$ as its direction vector.

\begin{align}
    \vec e_F &= \vec {\bar{F}}_{\text{ID}} / |\vec
         {\bar{F}}_{\text{ID}}|,\\ I(\theta) &= \bar
           E_{\text{ID}} \exp[\sigma \vec e_F \vec n(\theta) - A] =
         \bar E_{\text{ID}} \exp[\sigma\cos\theta - A]\,.
\end{align}
Without loss of generality, we can compute the moment integrals in a
spherical coordinate system that is aligned with the flux direction,
meaning $\vec e_F$ is parallel to $\vec e_z$ the upward pointing unit
vector
\begin{align}
    \bar E_{\text{ID}} &= \frac{1}{4\pi} \int_0^{2\pi}\int_0^\pi
    I(\theta) \sin\theta d\theta d\phi\\ &= \frac{\bar
        E_{\text{ID}}}{2} \int_0^\pi \exp[\sigma\cos\theta - A]
    \sin\theta d\theta\\ &= \bar E_{\text{ID}} \exp[-A]
    \frac{\sinh(\sigma)}{\sigma},
\end{align}
from which we deduce
\begin{align}
A &=
    \ln\left(\frac{\sinh(\sigma)}{\sigma}\right),\\ \bar
      F^z_{\text{ID}} &= \frac{1}{4\pi} \int_0^{2\pi}\int_0^\pi
    I(\theta) \cos\theta\sin\theta d\theta d\phi\\ &= \frac{\bar
        E_{\text{ID}}}{2} \int_0^\pi \exp[\sigma\cos\theta - A]
    \cos\theta\sin\theta d\theta\\ &= \frac{\bar
        E_{\text{ID}}}{\sigma^2} (\sigma \cosh\sigma - \sinh\sigma)\,,
\end{align}
so that
\begin{align}
\label{eq: F of sigma}
\bar F^z_{\text{ID}} = \frac{\bar
    E_{\text{ID}}}{\sinh\sigma}\frac{\sigma \cosh\sigma -
  \sinh\sigma}{\sigma}\,.
\end{align}

The variable $A$ scales the distribution to the given energy density and
can be determined analytically if $\sigma$ is known. The variable
$\sigma$ determines how narrow the distribution is and, therefore, the
flux density. We are interested in a function for $\sigma(|\vec
{\bar{F}}_{\text{ID}}| / \bar E_{\text{ID}})$, but
Eq.~\eqref{eq: F of sigma} is not analytically invertible. However,
it is strictly monotonically growing and, therefore, straightforward to
invert numerically using a lookup table.

While the Kent distribution allows us to map a given energy and flux
density to an intensity population distribution, it does not give us
control over the pressure density $P^{ij}$. With the above approach, the
pressure tensor will always be diagonal in the coordinate system aligned
with the flux density. To be more precise, the pressure tensor in the
aligned system is given by
\begin{align}
    P^{xx} &= \frac{\bar E_{\text{ID}}}{4\pi}
    \int_0^{2\pi}\int_0^\pi \exp[\sigma\cos\theta - A]
    (\sin\theta\cos\phi)^2 \sin\theta n^i n^j d\theta d\phi\nonumber\\ &=
    \frac{\bar E_{\text{ID}}}{\sigma^2\sinh\sigma}
    (\sigma\cosh(\sigma) - \sinh(\sigma)), \\ P^{yy} &=
    \frac{\bar E_{\text{ID}}}{4\pi} \int_0^{2\pi}\int_0^\pi
    \exp[\sigma\cos\theta - A] (\sin\theta\sin\phi)^2 \sin\theta n^i n^j
    d\theta d\phi\nonumber\\ &= \frac{\bar
        E_{\text{ID}}}{\sigma^2\sinh\sigma} (\sigma\cosh(\sigma) -
    \sinh(\sigma)), \\ P^{zz} &= \frac{\bar E_{\text{ID}}}{4\pi}
    \int_0^{2\pi}\int_0^\pi \exp[\sigma\cos\theta - A] (\cos\theta)^2
    \sin\theta n^i n^j d\theta d\phi\nonumber\\ &= \frac{\bar
        E_{\text{ID}}}{\sigma^2\sinh\sigma} ((\sigma^2 + 2)\sinh(\sigma)
    - 2\sigma\cosh(\sigma))\,.
\end{align}

Not all energy and flux density combinations can be achieved depending on
the chosen stencil. Each stencil has a maximum $\sigma_{\text{max}}$ it
can handle, depending on the population count $\Ndir$ and the additional
ghost directions. Once $\sigma$ exceeds $\sigma_{\text{max}}$, the
distribution will be too sharp to be resolved by the stencil and seem
discontinuous. This leads to errors in the velocity interpolation
becoming too big. To find this $\sigma_{\text{max}}$ value, we calculate
the Kent distribution for $\sigma = 1$. We then interpolate the intensity
to a grid of $100\times200$ points in the region $0.1\pi$ around the
north pole of the stencil. If the relative maximal interpolation error
does not exceed $1\%$, the current $\sigma$ value is acceptable, and we
increase $\sigma$ by an adaptive stepsize. Otherwise, we repeat the
process with a smaller stepsize until we have reached a reasonable
estimate for the maximal allowed $\sigma$. As a result, there is a
maximum relative flux density ${\bar F_i \bar F^i}/{\bar E} \leq 1$ a
stencil can resolve. Tables \ref{tab: sigma max 2D} and \ref{tab: sigma
  max 3D} show the maximum relative flux density for multiple Fourier and
Lebedev stencils with different orders and refinement levels.

\begin{table}
    \centering
    \begin{tabular}{|l|r|r|r|}\hline
        $N_{\text{real}} + N_{\text{ghost}}$ & Flux max & $N_{\text{real}} + N_{\text{ghost}}$ &  Flux max\\\hline
        $ 20 + 0$ & $0.911348$ & $ 16 +  4$ & $0.980785$\\
        $ 50 + 0$ & $0.989806$ & $ 40 + 10$ & $0.995807$\\
        $100 + 0$ & $0.995896$ & $ 80 + 20$ & $0.998677$\\
        $200 + 0$ & $0.998210$ & $160 + 40$ & $0.999231$\\\hline
    \end{tabular}
    \caption{Maximal achievable relative flux density for multiple
      Fourier stencils (2D) with different direction configurations. The
      last three columns show the respective population counts.}
    \label{tab: sigma max 2D}
\end{table}
\begin{table}
    \centering
    \begin{tabular}{|l|r|r|r|r|r|r|}\hline
        $p$ & $0.00$ - $0.00$ & $0.15$ - $0.00$ &  $0.00$ - $0.15$ &  $\Ndir$ &  $\Ndir$ &  $\Ndir$  \\\hline
        $11$ & $0.119301$ & $0.320176$ & $0.401914$ & $ 50$ & $ 54$ & $ 58$\\
        $15$ & $0.151213$ & $0.407415$ & $0.491836$ & $ 86$ & $ 94$ & $102$\\
        $17$ & $0.260644$ & $0.614494$ & $0.687429$ & $110$ & $126$ & $138$\\
        $19$ & $0.300239$ & $0.657584$ & $0.724383$ & $146$ & $166$ & $182$\\
        $21$ & $0.305549$ & $0.627482$ & $0.743574$ & $170$ & $190$ & $206$\\
        $23$ & $0.421269$ & $0.782529$ & $0.833982$ & $194$ & $218$ & $238$\\
        $29$ & $0.586264$ & $0.870299$ & $0.906942$ & $302$ & $338$ & $362$\\
        $31$ & $0.490054$ & $0.808158$ & $0.865103$ & $350$ & $390$ & $422$\\
        $35$ & $0.707619$ & $0.920281$ & $0.948240$ & $434$ & $490$ & $526$\\
        $41$ & $0.791457$ & $0.953830$ & $0.970163$ & $590$ & $662$ & $710$\\
        $47$ & $0.846724$ & $0.970273$ & $0.975960$ & $770$ & $862$ & $918$\\\hline
    \end{tabular}
    \caption{Maximal achievable relative flux density for multiple
      Lebedev stencils (3D) of order $p$. The last three columns show
      the respective population counts.}
    \label{tab: sigma max 3D}
\end{table}

Adaptive stencils achieve better results in beam tests even with a lower
$\Ndir$ because they can achieve higher relative flux values with fever
direction vectors. This is achieved by increasing the resolution in the
flux direction instead of homogeneously all around the sphere.

\begin{figure}
  \centering
  \includegraphics[width=0.48\textwidth]{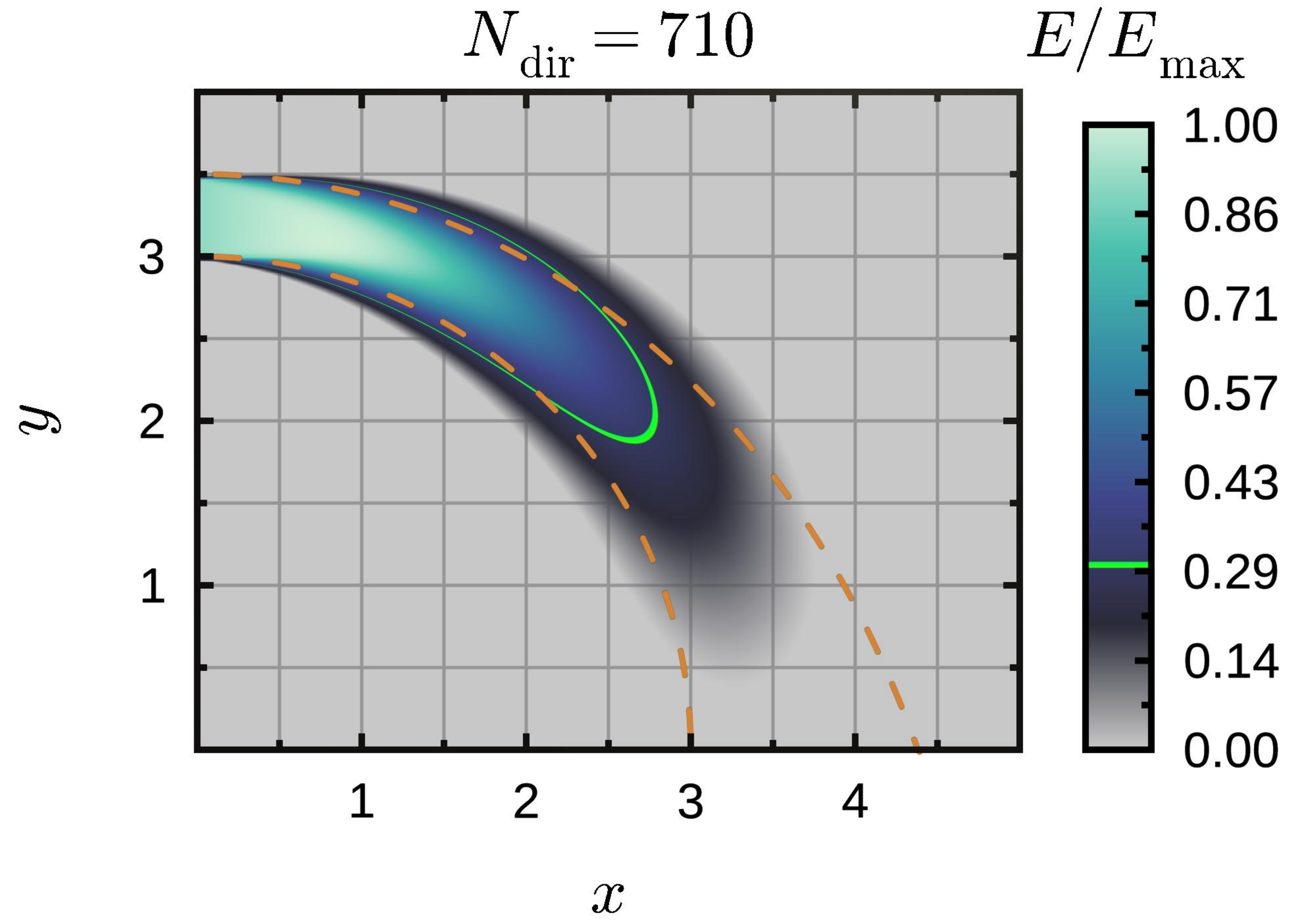}
    \caption{Normalised energy density for the 3D curved-beam test (see
      also Fig~\ref{fig: Curved Beam}) with maximal interpolation error
      of $5\%$ instead of $1\%$, resulting in $\sigma_{\text{max}} =
      84.75$ instead of $\sigma_{\text{max}} = 33.52$. The beam reaches
      its maximum energy density along its path instead of the very
      beginning, which is unphysical.}
    \label{fig: failed curved beam}
\end{figure}

In Fig.~\ref{fig: failed curved beam}, it is possible to note the effect
of a $\sigma > \sigma_{\text{max}}$ on a curved beam test. The energy
density along the beam increases and reaches its maximum in the middle of
the beam instead of the beginning. The maximum relative flux limitation
is only relevant in cases where the radiation field becomes highly
focused and traverses mono-directional, which might be an issue in some
extreme cases, \eg a black-hole jet emissions.

\bsp	
\label{lastpage}
\end{document}